\def\draftversion{false}

\RequirePackage{ifthen}
\ifthenelse{\equal{\draftversion}{true}}{
  \documentclass[aps,prx,10pt,galley,amsmath,amssymb,showpacs,citeautoscript,
                 superscriptaddress]{revtex4}
}{
  \documentclass[aps,prx,10pt,twocolumn,amsmath,amssymb,showpacs,citeautoscript,
                 superscriptaddress]{revtex4-1}
}

\ifthenelse{\equal{\draftversion}{true}}{
  \marginparwidth 2.7in
  \marginparsep 0.5in
  \newcounter{comm} 
  \def\commnext{\stepcounter{comm}}
  \def\commtext{{\bf\color{blue}[\arabic{comm}]}}
  \def\commmar{{\bf\color{blue}[\arabic{comm}]}}
  \def\dvm#1{\commnext\marginpar{\small DV\commmar: #1}\commtext}
  \def\cdm#1{\commnext\marginpar{\small CD\commmar: #1}\commtext}
  \def\cvm#1{\commnext\marginpar{\small CVdW\commmar: #1}\commtext}
  \def\mlab#1{\marginpar{\small\bf #1}}
  
}{
  \def\dvm#1{}
  \def\cdm#1{}
  \def\cvm#1{}
  \def\mlab#1{}
  
}

\usepackage[usenames]{color} 

\usepackage{ulem}

\usepackage{graphicx}
\usepackage{dcolumn}
\usepackage{bm}
\usepackage{epstopdf}
\begin{document}

\title{Correct implementation of polarization constants in wurtzite
materials and impact on III-nitrides}

\author{Cyrus E. Dreyer}
\affiliation{Materials Department, University of California,
Santa Barbara, CA 93106-5050}
\affiliation{Department of Physics and Astronomy, Rutgers
University, Piscataway, NJ 08845-0849}
\author{Anderson Janotti}
\altaffiliation{Current address: Materials Science and Engineering, University of Delaware, Newark, Delaware 19716-1501, USA}
\affiliation{Materials Department, University of California,
Santa Barbara, CA 93106-5050}
\author{Chris G. Van de Walle}
\affiliation{Materials Department, University of California,
Santa Barbara, CA 93106-5050}
\author{David Vanderbilt}
\affiliation{Department of Physics and Astronomy, Rutgers
University, Piscataway, NJ 08845-0849}

%
%

\date{\today}

\begin{abstract}

Accurate values for polarization discontinuities between
pyroelectric materials are critical for understanding and
designing the electronic properties of heterostructures. For
wurtzite materials, the zincblende structure has been used in the
literature as a reference to determine the effective spontaneous
polarization constants. We show that, because the zincblende
structure has a nonzero formal polarization, this method results in
a spurious contribution to the spontaneous polarization differences
between materials. In addition, we address the correct choice of
``improper'' versus ``proper'' piezoelectric constants. For the
technologically important III-nitride materials GaN, AlN, and
InN, we determine polarization discontinuities using a consistent
reference based on the layered hexagonal structure and the correct
choice of piezoelectric constants,
%
%
and discuss the results in light of available experimental data.

\end{abstract}


%
%

%
%
\maketitle

\section{Introduction}

Pyroelectric materials have emerged in a variety of electronic
and optoelectronic applications. Because of the symmetry of
their crystal structure these materials exhibit spontaneous
(SP) and piezoelectric (PZ) dipole moments \cite{Nye1957} which
manifest themselves as electric fields in heterostructure layers
and sheet charges at interfaces. In the technologically important
III-nitrides, which have the wurtzite (WZ) structure (space group
$P6_3mc$), polarization differences allow for strong carrier
confinement and the formation of a two-dimensional electron gas
(2DEG) with high density at AlGaN/GaN interfaces, exploited
in high electron mobility transistors (HEMTs). The effect of
polarization can also be detrimental, for example causing the
quantum-confined Stark effect in quantum wells of light-emitting
diodes (LEDs), which reduces radiative recombination rates and
shifts the emission wavelength. For both HEMTs and LEDs, accurate
values of the SP and PZ polarization constants are required for
a fundamental understanding as well as for device design.

Since experimental determination of the separate SP and PZ
contributions to the total polarization is very difficult,
calculated values of SP and PZ polarization constants are
widely used in simulations.  The PZ polarization constants are,
in principle, fairly straightforward to explicitly measure
or calculate \cite{Martin1972}.  However, the reported values
exhibit a considerable spread \cite{Feneberg2007}.  In addition,
the difference between so-called ``proper'' and ``improper'' PZ
constants \cite{Martin1972,Vanderbilt2000} is often overlooked,
even though it can give rise to significant quantitative changes
in the resulting polarization fields.  This difference is one
issue that is elucidated in the present paper.

The definition of SP polarization constants is even more subtle, and
they are typically not amenable to explicit experimental
determination, except in special cases \cite{Jonas2012}. The
calculation of SP polarization requires the choice of a reference
structure, which in the case of WZ semiconductors, has invariably been
chosen to be zincblende (ZB) \cite{BFV1997,BFV2001,Bechstedt2000}.  In
this work we will show that though ZB as a reference structure is
intuitively appealing, the SP polarization constants that result have
been misinterpreted, introducing a source of error into the predicted
values for bound sheet charge densities (and polarization fields).  We
also demonstrate that a proper choice of reference structure can
eliminate these problems, and we provide revised values that can be
directly inserted in current simulation tools.

While our theoretical considerations are general, we choose the
nitride semiconductors because they provide a suitable example to
illustrate the derivations, and because our findings have a
significant impact on this materials system of high (and still
increasing) technological importance.  In Sec.~\ref{sec:theory}
we review the underlying theory.  In Sec.~\ref{sec:spont} we address
the problems with choosing zincblende as a reference structure,
and propose a solution.  Sec.~\ref{improper} deals with
piezoelectric contributions, specifically the issue of proper
versus improper constants.  In Sec.~\ref{exp} we show that our
findings have important
consequences for nitride device structures
and compare with previous implementations and with experiment.
Section~\ref{sec:conc} concludes the paper.

\section{Calculating polarization constants in wurtzite}
\label{sec:theory}
For WZ films grown in the [0001] direction (\textit{i.e.}, the +$c$
direction), the polarization component $P_3$ is given by the sum
of the SP polarization at the wurtzite material's own
lattice parameters, $P_\text{SP}$, and the $z$ component of
the PZ polarization \cite{Nye1957}.
That is, for material $m$,
\begin{equation}
\label{totp}
\begin{split}
P_3^m=P_\text{SP}^m+(\epsilon^m_{1}+\epsilon^m_{2})e_{31}^m+\epsilon^m_{3}e_{33}^m,
\end{split}
\end{equation}
where (in Voigt notation) $\epsilon_i$ ($i$=1,2,3) is the
strain in the $i$ direction and $e_{3i}$ are the corresponding
piezoelectric constants
(specifically, the ``improper'' ones; see Section~\ref{improper}).
Henceforth we drop the subscript ``3''
from $P$ for simplicity; all unbolded quantities pertaining
to wurtzite are assumed to be in the $c$ direction.

The goal of this work is to derive the appropriate SP and PZ constants
that allow Eq.~(\ref{totp}) to be used in accurately determining
polarization differences at interfaces between different WZ materials.

\subsection{The modern theory of polarization}

Direct calculation of the polarization constants in
Eq.~(\ref{totp}) by first-principles electronic-structure methods
was enabled by the formulation of a rigorous theory of bulk
polarization, known as the modern theory of polarization (MTP)
\cite{KSV1993,Resta1994}. For a given structure $\lambda$, the
MTP allows calculation of the so-called ``formal'' polarization
\cite{KSV1993}:
\begin{align}
\label{pf}
{\bf P}_\text{f} & ={\bf P}_\text{ion}+{\bf P}_\text{el} \nonumber \\
&=  \frac{e}{\Omega}\sum_s Z_s^\text{ion}{\bf
R}_s^{(\lambda)}+\frac{ief}{8\pi^3}\sum^\text{occ}_j\int_\text{BZ}
d{\bf k}\langle u_{j,\bf k}^{(\lambda)}\vert \nabla_{\bf k}\vert
u_{j,\bf k}^{(\lambda)}\rangle,
\raisetag{3\baselineskip}
\end{align}
where $\Omega$ is the cell volume, $Z_s^\text{ion}$ is the charge of
the ion $s$ and ${\bf R}_\text{s}^{(\lambda)}$ is its position in the
$\lambda$ structure, $f$ is the spin degeneracy of the bands, the sum
$j$ runs over occupied bands, and $u_{j,\bf k}^{(\lambda)}$ are the
cell periodic parts of the Bloch wave functions. In Eq.~(\ref{pf}),
$\textbf{P}_{\text{el}}$ is the Berry phase taken over the valence-band
manifold \cite{KSV1993,Resta1994}. The formal polarization is
defined only modulo the ``quantum of polarization''
$e\textbf{R}/\Omega$, where \textbf{R} is any lattice constant and $e$
is the electron charge \cite{KSV1993,Resta1994}.

In the MTP, only {\it differences} between formal polarizations
of appropriate structures, $\lambda=0$ and $\lambda=1$, are
well defined:
\begin{equation}
\begin{split}
\label{delp}
\Delta{\bf P}={\bf P}_\text{f}^{(\lambda = 1)}-{\bf P}_\text{f}^{(\lambda = 0)}.
\end{split}
\end{equation}
The choice of the ``appropriate'' structures $\lambda = 0$
and $\lambda = 1$ rests on one of two possible considerations
to ensure that physical conclusions can be drawn from their
formal polarization differences.  First, if the two structures
are connected by an adiabatic, gap-preserving deformation path
\cite{KSV1993,Resta1994}, then their difference in polarization
[$\Delta{\bf P}$ in Eq.~(\ref{delp})] is given by the expression
\begin{equation}
\begin{split}
\label{polcur}
\Delta{\bf P}=\int^1_0 d\lambda \frac{\partial {\bf P}}{\partial \lambda},
\end{split}
\end{equation}
and corresponds to the zero-field adiabatic displacement
current. This quantity can, in principle, be determined
experimentally.  An obvious application is the calculation of
piezoelectric constants, which involves polarization differences
between structures with slightly different lattice constants
and/or internal structural parameters.

\subsection{Interface theorem}

The second consideration, as shown by Vanderbilt and King-Smith
\cite{VKS1993}, is that {\it if} an insulating interface can be
constructed between two structures, the difference in formal
polarization gives the bound charge, $\sigma_\text{b}$, that
builds up at the interface as a result of the continuity of the
displacement field over an interface with no free charge:
\begin{equation}
\label{disp}
\begin{split}
\sigma_\text{b}=({\bf P}_\text{f}^{\lambda=1} -{\bf
P}_\text{f}^{\lambda=0})\cdot {\bf \hat{n}}. 
\end{split}
\end{equation}
This is often referred to as the ``interface theorem.'' Since
there is no adiabatic path necessary between the two structures in
this consideration, $\lambda=0$ and $\lambda=1$ can be different
polymorphs of the same material (such as WZ and ZB structures of
GaN) or different materials altogether (such as GaN and AlN);
as long as they form an insulating interface, Eq.~(\ref{disp})
will give the bound charge accumulation at the interface.

From the interface theorem [Eq.~(\ref{disp})] and Eq.~(\ref{totp}),
the bound polarization charge at the interface between different
III-nitride materials ($m$ and $n$) is
\begin{equation}
\begin{split}
\sigma_\text{b} =\left[P_\text{SP}^m + e^m_{31}
 (\epsilon_1^m+\epsilon_2^m)+e^m_{33}\epsilon^m_3\right]\\
- \left[P_\text{SP}^n + e^n_{31}(\epsilon_1^n+\epsilon^n_2)
 +e^n_{33}\epsilon^n_3\right].
\end{split}
\end{equation}
As an example, we will take a realistic situation that occurs
in heterostructures, by assuming that material $n$ is strained
coherently to $m$ ($\epsilon_1^m=\epsilon_2^m=\epsilon_3^m=0$),
\textit{i.e.}, under plane stress ($\epsilon^n_1=\epsilon^n_2,
\epsilon^n_3= -2C^n_{13}/C^n_{33}\epsilon^n_1$, where $C_{ij}$
are the elastic constants). Therefore we have
\begin{equation}
\begin{split}
\label{bound}
\sigma_\text{b}&=(P_\text{SP}^m-P_\text{SP}^n )-
2\epsilon^n_1(e^n_{31}-e^n_{33}C^n_{13}/C^n_{33}) \\
&=\Delta P_\text{SP}^\text{int}-
2\epsilon^n_1(e^n_{31}-e^n_{33}C^n_{13}/C^n_{33}).
\end{split}
\end{equation}
Note that $\sigma_\text{b}$ is the charge density of electrons
at an interface for which material $n$ has been grown on top
of material $m$ in the $+c$ direction.

\section{Reference structure for spontaneous polarization}
\label{sec:spont}

\subsection{Effective spontaneous polarization constants}

We will first address the difference in spontaneous polarization
in Eq.~(\ref{bound}), $\Delta
P_\text{SP}^\text{int}$.
Strain effects will be taken into account separately in the
PZ part, so that
$\Delta P_\text{SP}^\text{int}$ is simply the
difference of formal polarizations of the respective
zero-strain structures,
\begin{equation}
\label{Pint}
\Delta P_\text{SP}^\text{int} =P^{m}_\text{f}\big |_{\epsilon=0}
  -P^{n}_\text{f}\big |_{\epsilon=0}.
\end{equation}

For purposes of Eq.~(\ref{totp}), we would like to define a SP
polarization constant that is a property of a single material. Simply
taking $P^{m}_\text{f}$ of Eq.~(\ref{Pint}) as $P^m_{\text{SP}}$ is
problematic, since formal polarization is multivalued, being only
well-defined modulo a quantum of polarization
$e\textbf{R}/\Omega$. Therefore, in every situation in which
Eq.~(\ref{totp}) is applied to determine $\sigma_b$ at an interface,
it must be confirmed that formal polarizations of the two materials
are taken on the same ``branch'' of $e\textbf{R}/\Omega$. A better approach is to take
$P^m_{\text{SP}}$ in Eq.~(\ref{totp}) as a so-called
``effective'' SP polarization, ${\bf P}_\text{eff}$, defined by Resta
and Vanderbilt \cite{Resta2007} to be the $\Delta{\bf P}$ in
Eq.~(\ref{delp}) that results as the system is taken from a
high-symmetry ``reference'' structure ($\lambda=0$) to the structure
of interest ($\lambda=1$).
That is,
\begin{equation}
\label{Peff} \textbf{P}_\textrm{eff}
=\textbf{P}_\textrm{f}^{(\lambda=1)}
-\textbf{P}_\textrm{f}^{(\lambda=0)} =\textbf{P}_\textrm{f}
-\textbf{P}_\textrm{f}^\textrm{ref} .
\end{equation}
%
Using
\textbf{P}$_{\text{eff}}$ to define the SP polarization of the
material removes the indeterminacy inherent to the formal
polarization.

The reference
structure is often chosen to be centrosymmetric, but it is important
to recognize that the formal polarization of centrosymmetric crystals
is not necessarily zero. This is because, as stated above,
$\textbf{P}_\text{f}$ is a multivalued vector field,
%
so it is
possible for a nonzero formal polarization to be unchanged (modulo
$e\textbf{R}/\Omega$) under the inversion operator. Nevertheless,
high symmetry puts restrictions on the possible values of
${\bf P}^{(\lambda=0)}_\text{f}$ \cite{VKS1993}.

While in principle effective polarization constants are still
{\it differences} in formal polarization between $\lambda=1$
and $\lambda=0$ (reference) structures, in practice they
can be used to compare spontaneous polarizations of different
materials to obtain $\Delta P_\text{SP}$ if such materials
share a reference structure with the same formal polarization.
Such a comparison
then correctly yields the interface charge density according to
the interface theorem of Ref.~\onlinecite{VKS1993}.
In such cases, $\Delta P_\text{SP}$ is just given by the difference
in \textit{effective} SP polarization of the materials,
\begin{equation}
\label{Pinttilde}
\Delta\widetilde{P}^\text{int}_\text{SP} =P^{m}_\text{eff}-P^{n}_\text{eff} .
\end{equation}

In the more general case that the reference formal polarizations do not match,
the correct change in SP polarization following from
Eqs.~(\ref{Pint}) and (\ref{Peff}) is
\begin{equation}
\label{Pint2}
\Delta P_\text{SP}^\text{int}=\Delta\widetilde{P}^\text{int}_\text{SP}
  +(P^{m,\text{ref}}_\text{f}-P^{n,\text{ref}}_\text{f}).
\end{equation}
That is, a correction term of the form
\begin{equation}
\label{Pintcorr}
\Delta P_\text{corr}^\text{ref}\equiv P^{m,\text{ref}}_\text{f}-P^{n,\text{ref}}_\text{f}
\end{equation}
has to be added to Eq.~(\ref{Pinttilde}). Unfortunately, this
correction term is not typically implemented in device simulation
packages (\textit{e.g.}, Ref.~\onlinecite{silense})
or used in the interpretation of experimental data (\textit{e.g.},
Ref.~\onlinecite{Ambacher2000}, which is considered a standard
reference in the field).

When the PZ terms are included as well, the total interface charge
given by Eq.~(\ref{bound}) becomes
\begin{equation}
\begin{split}
\label{boundcorr}
\sigma_\text{b}&= \Delta\widetilde{P}^\text{int}_\text{SP}
 +\Delta P_\text{corr}^\text{ref}
 - 2\epsilon^n_1(e^n_{31}-e^n_{33}C^n_{13}/C^n_{33}).
\end{split}
\end{equation}
Equation (\ref{boundcorr}) is a central result of the present work.

\subsection{Correction term for the effective spontaneous
polarization with the zincblende reference structure
\label{zbcorr}}
 
 As mentioned before, previous studies
\cite{BFV1997,BFV2001,Bechstedt2000} have exclusively used ZB (space
group $F\bar{4}3m$) as a reference structure for calculating the SP
polarization of the WZ. This structure is not centrosymmetric,
although it has sufficient symmetry to preclude any SP polarization
\cite{Nye1957}. The fact that an insulating (111) interface can be
constructed between the WZ and ZB polytypes \cite{Bechstedt2000}
makes
 it an appropriate reference structure. In fact, experimental
measurements have deduced the relative polarization between the WZ
and
 ZB phases of GaN \cite{Jonas2012}, which were found to be
consistent
 with the theoretical values in
Refs.~\onlinecite{BFV1997,BFV2001,Bechstedt2000}.
 
 However, there
is a subtlety with using ZB as a reference structure:
 it has a
nonzero formal polarization in the [111] direction,
$P_\text{f}^\text{ZB}$ (modulo $e\textbf{R}/\Omega$). Again, this is
consistent with the symmetry considerations because $P_\text{f}$ is a
multivalued vector quantity, and can be nonzero while still remaining
unchanged (modulo $e\textbf{R}/\Omega$) under the $F\bar{4}3m$
symmetry operations.  These symmetry operations dictate the possible
values of $P_\text{f}^\text{ZB}$, and therefore the resulting value
depends only on the lattice constant, not on the chemical species of
the atoms \cite{VKS1993} (see Section~S1 of the supplemental material
(SM) \cite{Supp} for more details on the formal polarization of
ZB). The ZB reference structures for the reported effective SP
polarization values for the III-nitrides were those with lattice
constants equal to the in-plane lattice constant of the corresponding
wurtzite material \cite{BFV1997,BFV2001,Bechstedt2000} (as confirmed
by our calculations), so $P_\text{f}^\text{ZB}$ will be
\textit{different} for GaN, AlN, and InN, and do not simply
constitute
 a constant shift of $P_\text{f}^\text{WZ}$ for all the
materials.
 Therefore, for the effective SP polarization constants
with the ZB
 reference to be implemented in Eq.~(\ref{bound}) to
determine the
 polarization difference between different WZ
materials, the correction
 term of Eq.~(\ref{Pintcorr}) is required,
as in Eq.~(\ref{boundcorr}).
 
 Consider the example of the
interface charge between InN and
 GaN. Although, as mentioned above,
the formal polarization of
 zincblende does not necessarily vanish,
the symmetry of the
 structure severely limits the possible
values. Specifically
 there are two possible values of the formal
polarization in the
 [111] direction that are consistent with the
symmetry: either
 $P_\text{f}$ vanishes, or it is equal to
$e\sqrt{3}/2a_n^2$
 (both modulo $e\sqrt{2}a_n/\Omega_n$), where
$a_n$ is the WZ in plane lattice constant
 of material $n$ and
$\Omega_n$ is the volume of the ZB primitive cell (see
Ref.~\onlinecite{VKS1993} or Section S1 of the SM \cite{Supp}). For
the III-nitrides it is the latter, giving
 a
correction term for GaN/InN:
\begin{equation}
\label{corr}
\begin{split}
\Delta P_\text{corr}^\text{(ZB ref)}&=P_\text{f}^\text{GaN,ZB}
  -P_\text{f}^\text{InN,ZB} \\
&=\frac{e\sqrt{3}}{2}\left(\frac{1}{(a_\text{GaN})^2}
 -\frac{1}{(a_\text{InN})^2}\right) \\
&=0.28 \text{ C/m}^2.
\end{split}
\end{equation}
When considering the SP polarization differences between WZ nitrides,
this represents a significant correction. In fact, as we will show in
Section~\ref{hex}, the correction is an order of magnitude larger than
the effective polarizations when they are calculated with the
zincblende reference \cite{BFV1997,BFV2001}.  As we shall see later in
Section~\ref{sec:compar}, this error is substantially reduced in practice
by an approximate error cancellation that occurs in connection with
the treatment of the PZ response.

There is nothing intrinsically wrong with using ZB as the reference
structure for defining WZ effective SP polarization; however if these
values are to be used to obtain polarization differences between
different WZ materials, the $\Delta P_\text{corr}^\text{ref}$ term
[Eq.~(\ref{Pintcorr}), or Eq.~(\ref{corr}) for the example of GaN/InN]
must be explicitly included in expressions such as Eq.~(\ref{Pint2})
or Eq.~(\ref{boundcorr}). To our knowledge, however, this has not
been properly implemented in the numerous previous evaluations of SP
polarization for nitride interfaces, and it would require changes in
the software for the many simulation tools that include modeling of
polarization fields in heterostructures.

\subsection{$P6_3/mmc$ hexagonal layered structure as an
alternative reference \label{hex}}

In order to avoid extensive changes in the simulation software,
and to enhance physical insight, we advocate another approach,
namely to determine effective SP polarization constants with
respect to a reference structure for which the formal polarization
is explicitly zero (so that $\Delta P^\text{int}_\text{SP}
=\Delta\widetilde{P}^\text{int}_\text{SP}$). A straightforward
choice for this reference structure is the layered hexagonal (H)
structure (space group $P6_3/mmc$), as was used for hexagonal
$P6_3mc$ $ABC$ materials \cite{Bennett2012}. This structure
is centrosymmetric, and we will show below with explicit
first-principles calculations that it remains insulating and
its formal polarization
vanishes.  The layered hexagonal structure can be obtained by an
adiabatic (gap preserving) increase of the internal structural
$u$ parameter from $u\approx 0.37-0.38$ of the WZ structure to
$u=0.5$. All that is required to avoid correction terms like
Eq.~(\ref{corr}) is to replace the effective SP polarization
constants currently used in the field (the ones referenced
to ZB \cite{BFV1997,BFV2001}) with those referenced to the H
reference structure.  We have explicitly verified that this leads
to expressions that are identical to those that would be obtained
for the ZB reference, provided the second term in Eq.~(\ref{Pint2})
or Eq.~(\ref{boundcorr}) is included.

The first-principles calculations of $P_\text{f}$ for the H, WZ,
and ZB structures of the III-nitrides were performed using density
functional theory with the screened hybrid functional of Heyd,
Scuseria, and Ernzerhof (HSE) \cite{HSE06} as implemented in the
{\sc Vasp} code \cite{vasp1}.  Hartree-Fock mixing parameters of
31\,\% for AlN and GaN, and 25\,\% for InN were used to correctly
describe the band gaps and structural parameters of each material.
Conventional functionals based on the local density approximation
(LDA) or generalized gradient approximation (GGA) predict InN to be
a metal, precluding the calculation of the polarization constants
if the $\Gamma$ point is included in the $k$-point mesh  (which is
required in {\sc Vasp}). Projector augmented wave potentials (PAW)
\cite{paw}, with the In and Ga $d$ electrons frozen in the core,
were used.  All calculations were performed on bulk primitive
cells, with a $6\times6\times8$ Monkhorst-Pack \cite{mp1976}
$k$-point mesh to sample the Brillouin zone, and a large energy
cutoff of 600 eV for the plane-wave basis set, chosen to ensure
convergence of the internal structural parameter $u$. The
calculated lattice parameters and band gaps, listed in Section~S2
of the SM \cite{Supp}, show good agreement
with experimental data.

\begin{figure}
\includegraphics[width=\columnwidth]{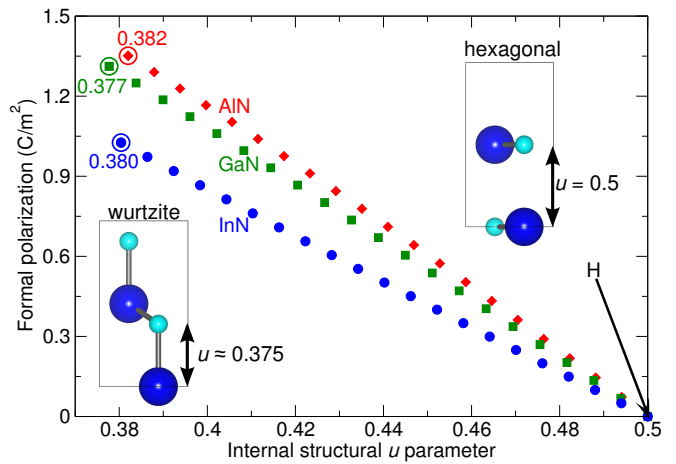}
\caption{\label{uint}
Formal polarization of InN, GaN, and AlN for structures as a
function of the internal structural parameter {\it u}, varying
between fully relaxed WZ (circled symbols, labelled with relaxed
$u$ value) and H ($u=0.5$), as shown schematically by ball and
stick models where smaller balls represent N atoms and larger
ones represent the cations. All other lattice parameters were
fixed at their relaxed WZ values.}
\end{figure}

We have calculated the electronic structure for structures
with increasing $u$, ranging from $u\approx 0.37$ to $u=0.5$
(Fig.~\ref{uint}), and confirmed that this path between WZ
and H is gap preserving. These calculations also show that
the formal polarization of the H structure is zero (modulo
$e\textbf{R}/\Omega$) for the III-nitrides (Fig.~\ref{uint}). We
remind the reader that this was not guaranteed, since ${\bf
P}_\text{f}$ can be nonzero and still consistent with inversion
symmetry, if the inversion operator changes ${\bf P}_\text{f}$
by a multiple of $e\textbf{R}/\Omega$. We have therefore verified
that the hexagonal phase is a reference structure for which there
is no spurious term in Eq.~(\ref{Pint2}).

In addition, by correcting for any discontinuities (in the amount
of a multiple of $e\textbf{R}/\Omega$) that may occur in the
calculations of formal polarizations along the path between WZ and
H, we have insured that we are comparing formal polarizations of
WZ GaN, AlN, InN on the same branch of $e\textbf{R}/\Omega$
\cite{Resta2007}.

\begin{table}
\caption{\label{sptab}
Effective spontaneous polarization constants in units of C/m$^2$
of wurtzite (WZ) GaN, AlN, and InN calculated using either the
hexagonal (H, space group $P6_3/mmc$) or zincblende (ZB, space
group $F\bar{4}3m$) reference structures. The lattice constant of
the ZB structure is chosen to match the in-plane lattice constant
of the WZ structure for the same material.   Results from previous
calculations \cite{BFV2001} that used the ZB reference are listed
for comparison.}

\begin{ruledtabular}
\begin{tabular}{cccc}
   & $P_\text{eff}^{\text{(H ref)}}$   &
     $P_\text{eff}^{\text{(ZB ref)}}$  &
     $P_\text{eff}^{\text{(ZB ref)}}$, prev.\footnotemark[1] \\
\hline
     GaN & $ 1.312$   & $-0.035$     & $-0.034$ \\
     AlN & $1.351$     & $-0.090$       & $-0.090$ \\
     InN & $1.026$    & $-0.053$       & $-0.042$ \\
\end{tabular}
\end{ruledtabular}
\footnotetext[1]{From Ref.~\onlinecite{BFV2001}.}
\end{table}

The calculated spontaneous polarization coefficients for the
WZ structure using either H or ZB as a reference are given in
Table~\ref{sptab}.  The results obtained by Bernardini {\it et
al.} \cite{BFV2001,BFV1997} are listed for comparison. The GGA
functional used in that work provides results that are very close
to those we obtained with HSE; the discrepancy is the largest
for InN, which is probably related to the fact, mentioned above,
that GGA predicts InN to be a metal.  Table~\ref{sptab} also
shows, however, that the choice of reference structure makes a
significant difference.
%
%
The magnitudes of the coefficients are much larger, and their signs
are different when H is used as the reference. We observe that it is
not just the absolute values, but also the relative differences
between the calculated polarization constants of the three materials
that differ from the previously reported values
\cite{BFV2001,BFV1997}.

The difference in sign of $P_\text{eff}^{\text{(H ref)}}$ compared to
$P_\text{eff}^{\text{(ZB ref)}}$ demonstrates that the conventional
wisdom that the SP polarization in WZ points in the -$c$ direction is
misleading. The formal polarization of WZ has no definite sign as this
would depend on the chosen branch. The effective SP depends on the
polarization difference, and therefore the sign will depend on the
sign and magnitude of the formal polarization of the reference
structure.

Even though the values reported in Table~\ref{sptab} for $P_{\text{eff}}^{\text{(H ref)}}$ and $P_{\text{eff}}^{\text{(ZB ref)}}$ clearly differ in sign, absolute magnitude, and relative differences between materials, we will show in Section~\ref{exp} that the final predictions based on both formulations are actually rather similar, because of the way the PZ contributions have been included in the previous work (\textit{cf.} Section~\ref{improper}).

\section{Improper versus proper piezoelectric constants
\label{improper}}

We now address the specifics of the PZ terms in
Eqs.~(\ref{bound}) and (\ref{boundcorr}). A complication that must be addressed is the
choice between improper and proper $e_{31}$ ($e_{33}$ has no such
complication) \cite{Vanderbilt2000,Martin1972}.

As we have done above, consider a thin layer of a WZ material
grown in the $c$ direction. If the layer is strained perpendicular
to the $c$ direction, the total bound charge on the $+c$ and $-c$
surfaces will change as a result of the polarization current,
or redistribution of charge, in the layer. If
metallic contacts on the $+c$ and $-c$ surfaces
are short-circuited
when the strain occurs, the current flow can be measured directly and
will give the proper PZ constant, denoted  $e_{31}^\text{prop}$
\cite{Vanderbilt2000,Martin1972}.

If the $+c$ and $-c$ faces are
in open-circuit boundary conditions,
the layer will
have a field across it due to the SP polarization, which will
be modified by the strain via two mechanisms. The first is the
same as in the proper case, as the strain will cause a flow of
polarization current. But in addition, since the field depends
on the charge \textit{density}, the change in the area
of the $c$-plane as a result of $\epsilon_1$
will dilute or concentrate the pre-strain bound
charge. For small strains the latter is given by the zero-strain
formal polarization \cite{Vanderbilt2000,Martin1972}. Taking both
of these mechanisms into account gives the improper PZ constant,
$e_{31}^\text{imp}$.

In the case of, \textit{e.g.},~Eq.~(\ref{boundcorr}), the PZ
constants correspond to the improper case, since their role
in the equation is to take into account the change in formal
polarization of material $n$ with strain, so that $\sigma_b$
corresponds to the bound charge at the coherent interface with
the in-plane lattice constant of material $m$. The change in
formal polarization with strain is an alternative definition of
the improper PZ constants \cite{Vanderbilt2000}.

From Refs.~\onlinecite{Martin1972} and \onlinecite{Vanderbilt2000},
the improper
PZ constant $e^{n,\text{imp}}_{31}$ is related to the proper
constant by
\begin{equation}
\label{imp}
e^{n,\text{imp}}_{31}={e}^{n,\text{prop}}_{31}
  -P_\text{f}^n\big|_{\epsilon=0},
\end{equation}
where $P_\text{f}^n\big|_{\epsilon=0}$ is the zero-strain formal
polarization of material $n$. There is no change to the $e_{33}$ PZ
constant. The proper PZ constant is a well-defined bulk quantity,
as it is related to the polarization current; however the improper
PZ constant is branch dependent \cite{Vanderbilt2000}. Here
also, defining polarization with respect to the H reference
proves useful. Since the formal polarization of the H
structure vanishes (Fig.~\ref{uint}), $P_\text{eff}^{n,\text{(H
ref)}}=P_\text{f}^{n,\text{WZ}}\big|_{\epsilon=0}$; this also
ensures that improper PZ constants for the different materials are
taken on the same branch, in the same way as this is confirmed
for the SP polarization constants. Therefore, consistent use of
the H reference structure allows us to write Eq.~(\ref{bound}) as
\begin{equation}
\begin{split}
\label{boundimp}
\sigma_\text{b}&=\Delta\widetilde{P}^\text{int,(H ref)}_\text{SP}\\
&- 2\epsilon^n_1\Big({e}^{n,\text{prop}}_{31}
-P_\text{eff}^{n,\text{(H ref)}}-e^{n,\text{prop}}_{33}C^n_{13}/C^n_{33}\Big),
\end{split}
\end{equation}
where $P_\text{eff}^{n,\text{(H ref)}}$ can be taken from Table \ref{sptab}.
\begin{table}[h]
\caption{\label{pztab}
Calculated piezoelectric polarization constants in units of C/m$^2$
compared with reported
%
values from the literature.
}
\begin{ruledtabular}
\begin{tabular}{c c c c c}
&	& proper	& improper 	& prev. reported\footnotemark[1] \\ \hline
GaN & e$_{31}$ 	& --0.551	& --1.863		&--0.22 to --0.55  \\
	& e$_{33}$ 	& 1.020 	& 1.020		&0.43 to 1.12 \\\hline
AlN 	& e$_{31}$ 	& --0.676 	& --2.027		&--0.38 to --0.81  \\
	& e$_{33}$ 	& 1.569 	& 1.569		&1.29 to 1.94 \\\hline
InN 	& e$_{31}$ 	&--0.604 	& --1.63		&--0.23 to --0.59  \\
	& e$_{33}$ 	&1.238 	& 1.238		& 0.39 to 1.09 \\
\end{tabular}
\end{ruledtabular}
\footnotetext[1]{From Ref.~\onlinecite{Feneberg2007} and references therein.}
\end{table}

Calculated proper PZ constants are given in the ``proper''
column of Table \ref{pztab}. Since the HSE hybrid functional
was used (and therefore density functional perturbation theory
was not implemented), finite differences were used to calculate
the derivatives with strain, following the procedure outlined
in  Eqs.~(4)-(6) in Ref.~\onlinecite{BFV1997}. Specifically,
improper PZ constants were calculated, and converted to proper
constants by adding $P_\text{f}^n\big|_{\epsilon=0}$ [see
Eq.~(\ref{imp})] as determined in the calculation. This removes
any dependence on the branch choice used in finite-difference
calculations \cite{Vanderbilt2000}. We then convert \textit{back}
to improper constants using $P_\text{eff}^{n,\text{(H ref)}}$
as discussed above, in order to ensure that the constants are
reported for the same branch for each material (``improper''
column in Table \ref{pztab}).

The WZ structure does have another nonzero
piezoelectric constant $e_{15}$, which couples a shear deformation in
a plane perpendicular to the $c$ plane ($\epsilon_{13}$ or
$\epsilon_{31}$) to the polarization in the $c$ plane. In this case,
there are two improper PZ constants, since a correction must be
included in the case of $\epsilon_{13}$ but not for $\epsilon_{31}$
\cite{Vanderbilt2000}. These elements do not enter in the situation we
consider in this work (plane stress conditions with the $c$ plane as
the growth plane), but may be important for growth on nonpolar or
semipolar planes  \cite{Romanov2006}.

It is important to comment on the PZ constants reported in the
literature \cite{Feneberg2007}.
When PZ polarization
constants have been implemented in simulations (\textit{e.g}.
Refs.~\onlinecite{silense,Ambacher2000,Feneberg2007}) it has
never been specified \textit{which} PZ constants are used for WZ
III-nitrides. However, by comparing our calculations of proper and
improper PZ constants (``proper'' and ``improper'' columns of Table
\ref{pztab}) with the  reported PZ constants in the literature
(``prev. reported'' of Table \ref{pztab}) we have found that the
reported constants are more likely to be the proper PZ constants.

From an experimental perspective, most of the experimental techniques
have measured \textit{total} polarization, and then deduced the PZ
constants in the nitrides using the SP constants from, \textit{e.g.},
Ref.~\onlinecite{BFV1997} using Eq.~\ref{totp}. As we will show in
Section \ref{exp}, observation of the effects of total polarization
can be misleading with regards to the differentiation between proper
and improper PZ constants, due to the error cancellation from the use
of the ZB reference in defining the SP polarization (discussed in
Section \ref{zbcorr}).

There have been direct measurements of the PZ constants, either by
probing the electromechanical coupling constants via surface acoustic
waves \cite{Oclock1973,Dubois1999}, or by using interferometry to
determine the strain caused by the application of a voltage
\cite{Muensit1998,Muensit1999,Guy1999,Lueng2000}. Both of these
techniques measure the {\it proper} constants, since neither is sensitive to
the change in surface charge density resulting from the deformation.
These reported values indeed agree well with our calculated values for the proper PZ constants,
both in sign and in  magnitude.

In previous work \cite{BFV2001},
$P_\text{eff}^{n,\text{(ZB ref)}}$ was used instead
of $P_\text{f}^n\big|_{\epsilon=0}$ in Eq.~(\ref{imp}) to
convert improper to proper $e_{31}$ PZ constants (\textit{cf.} Table
VI and V of Ref.~\onlinecite{BFV2001}).  Because of the
nonvanishing formal polarization of the ZB reference
structure, $P_\text{eff}^{n,\text{(ZB ref)}} \neq
P_\text{f}^{n,\text{WZ}}\big|_{\epsilon=0}$; instead, we
see from the discussion resulting in Eq.~(\ref{corr}) that
Eq.~(\ref{imp}) can be expressed as
\begin{equation}
\label{zbimp}
e^{n,\text{imp}}_{31}={e}^{n,\text{prop}}_{31}
 -\left(P_\text{eff}^{n,\text{ZB ref}}
 +\frac{e\sqrt{3}}{2a_n^2}\right),
\end{equation}
where $a_n$ is the equilibrium, in-plane lattice constant of the WZ material $n$. To
our knowledge the inclusion of the last term in Eq.~(\ref{zbimp})
has not been discussed in the literature.
Because of the small magnitude of $P_\text{eff}^{n,\text{(ZB
ref)}}$, neglecting the last term in Eq.~(\ref{zbimp}) led to
the conclusion in Ref.~\onlinecite{BFV2001} that the difference
between the proper and improper PZ constants is small, seemingly
rendering the distinction of no consequence.
Instead,
because of the large magnitude of $P_\text{eff}^{n,\text{(H ref)}}$
[and $e\sqrt{3}/2a_n^2$ in Eq.~(\ref{zbimp})],
the distinction between proper and improper PZ constants is actually
very significant.

\section{Comparison with reported experimental results \label{exp}}

\subsection{Correct expressions for total polarization for wurtzite materials}
\label{review}

Before discussing specific quantitative results for nitride
semiconductors, we briefly summarize the main points of the previous
sections and rigorously express the polarization of a given WZ
material [Eq.~(\ref{totp})]. Spontaneous polarization constants must
be defined with respect to a reference structure, and this choice of
reference structure must be taken into account when evaluating
polarization discontinuities at interfaces. We determined a correction
term [Eq.~(\ref{Pintcorr})] that is necessary when effective SP
polarization constants are used to determine the SP polarization
difference between materials at an interface. This correction term is
significant when the ZB reference structure is used [\textit{e.g.},
Eq.~(\ref{corr})], but is zero for the H reference.  Using H as a
reference is therefore more straightforward and is the approach we
advocate, with the SP constants $P_\text{eff}^{\text{(H ref)}}$ listed
in Table \ref{pztab}.

In addition, the improper PZ constants should be used to determine
interface bound charge and fields in heterostructure layers. These
can be obtained from the proper constant $e^\text{prop}_{31}$
by subtracting $P_\text{eff}^{\text{(H ref)}}$ (Table
\ref{pztab}). Therefore, in the notation of this paper,
Eq.~(\ref{totp}) is written rigorously as
\begin{equation}
\label{Ptotfin}
P= P_\text{eff}^{\text{(H ref)}}+(\epsilon_{1}+\epsilon_{2})
  \left(e^\text{prop}_{31}-P_\text{eff}^\text{(H ref)}\right)
  +\epsilon_{3}e^\text{prop}_{33}.
\end{equation}

\subsection{Calculation of sheet charges for III-nitrides}

\begin{figure}
\includegraphics[width=\columnwidth]{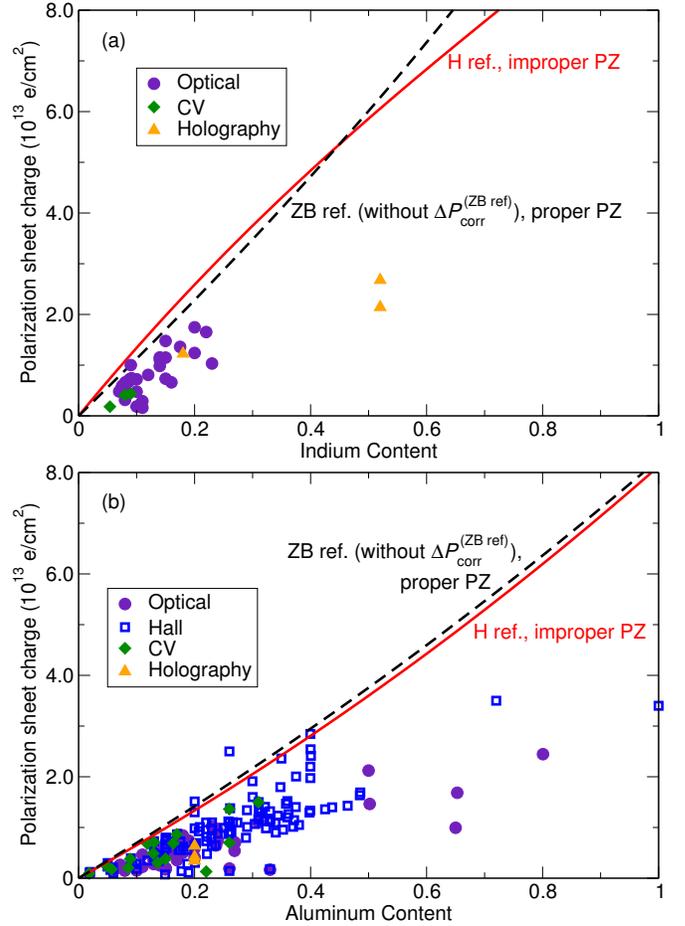}
\caption{\label{Both}
Absolute values for polarization sheet charges at the (a)
InGaN/GaN and (b) AlGaN/GaN interface as a function of alloy
content predicted from the spontaneous polarization constants
calculated using either the zincblende (ZB) reference structure
[without correction term, Eq.~(\ref{corr})] and the proper
piezoelectric constants (black dashed curve), or the hexagonal (H)
reference structure and improper piezoelectric constants (red solid
curve). Points are experimental values from the literature (see
Section S3 of the SM \cite{Supp} for references and values). }
\end{figure}

Because of the important impact of polarization on device
performance and design, a plethora of experimental studies
have been aimed at determining the effects of polarization at
GaN/InGaN and GaN/AlGaN heterostructures.  We have plotted these
reported results in Fig.~\ref{Both}, expressed as the magnitude
of polarization bound charge at the interface, as a function of
alloy content (a full list of references is provided in Section~S3 of
the SM \cite{Supp}).

For GaN grown in the $+c$ direction
with the InGaN (AlGaN) grown on top, the sign of the bound charge
at the interface will be negative (positive) \cite{Ambacher1999}.

In Fig.~\ref{Both}, the black dashed curves correspond to the
current practice in the field: sheet charges are predicted
based on (i) SP constants referenced to the ZB structure
($P_\text{eff}^{\text{(ZB ref)}}$ in Table \ref{sptab}) and
Eq.~(\ref{boundcorr}) \textit{without} the correction term $\Delta
P_\text{corr}^{\text{(ZB ref)}}$; and (ii) proper PZ constants
(``proper'' column in Table \ref{pztab}). Quantities for alloys
were obtained using linear interpolation.
For an explicit expression in terms of alloy content, see Eq.~(3)
in Section~S4 of the SM \cite{Supp}. Elastic constants were
taken from Ref.~\onlinecite{Vurgaftman2003}.

The red solid line in Fig.~\ref{Both} corresponds to
the implementation recommended in this work, \textit{i.e.,} using the
H reference structure and the improper PZ constants, as in
Eqs.~(\ref{boundimp}) and (\ref{Ptotfin}) [and Eq.~(5) in Section~S4
of the SM \cite{Supp}].


In view of the arguments given above, it may seem surprising
that the dashed black and solid red curves agree as well as they
do; we will return to this point in Section~\ref{sec:compar}.

\subsection{InGaN/GaN interfaces\label{ingan}}

For the InGaN/GaN system, most experimental studies have applied
optical techniques to determine the polarization fields in
GaN/InGaN/GaN quantum wells (QWs). This field can be probed
by varying QW width \cite{Hangleiter2003} or external biases
\cite{Renner2002} and measuring the change in the optical
properties of the QW (labeled ``optical'' in Fig.~\ref{Both}).
In addition, there have also been studies using time-resolved
PL to measure shifts due to screening of the polarization
field by photoexcited or electrically injected carriers
\cite{Turchinovich2003}. Other studies have been based on
electron holography \cite{Cherns1999}, where cross-sectional
transmission electron microscopy is conducted on InGaN/GaN
heterostructures to determine the depth-resolved electrostatic
potential in the growth direction, and capacitance-voltage
(CV) profiling of the fields \cite{Zhang2004}. When fields
are reported, we convert to bound charge for the purposes of
Fig.~\ref{Both}(a), assuming GaN/InGaN/GaN quantum well (with
thick barriers such that the electric field in the barriers is
presumed zero) using a simple parallel-plate capacitor model
($E=\sigma/\varepsilon_0\varepsilon_r$, using a relative dielectric
constant for GaN of 10 \cite{Barker1973} and for InN of 15
\cite{Madelung} and a linear interpolation for the dielectric
constant of InGaN). For specific values of the points in
Fig.~\ref{Both}, see Section~S3 of the SM \cite{Supp}.

The red curve in Fig.~\ref{Both}(a) is indeed in reasonable
agreement with the experimental observations, appearing to be an
upper bound of the data. The optical experiments usually rely
on Schr\"{o}dinger-Poisson simulations to determine the field
magnitude from the measured optical properties.  Uncertainties in
input parameters to these models such as well widths, compositions,
and composition profiles can result in quantitative differences. It
has been shown recently that taking into account the deviations from
an ideal QW structures when interpreting experimental observations can
account for the apparent discrepancy between the measurements and
theoretical prediction of polarization fields
\cite{Wu2012,Mcbride2014}. Such deviations are expected to be
significant for InGaN/GaN because of the large lattice mismatch and
the large difference in optimal growth temperatures for GaN and InGaN.

\subsection{AlGaN/GaN interfaces}

For the AlGaN/GaN system, there are two basic strategies for
experimentally determining polarization effects. The first is to
directly measure the polarization field in an AlGaN/GaN/AlGaN
(QW) structure with the same methods as used in the InGaN/GaN
case \cite{Grandjean1999,Leroux1999}. For the purposes of
Fig.~\ref{Both}(b) we have converted these fields to bound sheet
charge densities in an AlGaN/GaN/AlGaN quantum well (using a
relative dielectric constant for GaN of 10 \cite{Barker1973}).

The other strategy is to measure the density of the 2DEG at
the AlGaN/GaN interface in a HEMT structure (GaN channel, AlGaN
barrier); from this, the bound interface charge, $\sigma_\text{b}$,
can be derived \cite{Yu1997}.  The 2DEG density can be determined
either by Hall effect \cite{Ambacher2000} or CV \cite{Yu1997}
measurements.

The significant scatter in the experimental
data in Fig.~\ref{Both}(b) may have several origins. There are
experimental uncertainties that can influence fields, such as
incomplete strain relaxation in buffer layers \cite{Zhou2002}, and
differences in background
doping \cite{Kohler2011,Davidsson2004,Simon2001}.

As in the case of InGaN/GaN, the predicted sheet charges appear to be
an overestimation compared to the experimental observations. In the
case of optical measurements, Schr\"{o}dinger-Poisson
modeling is again typically used to interpret the measured properties,
and the same uncertainties and systematic errors may arise as
discussed in the case of InGaN/GaN
\cite{Arulkumaran2003,Mkhoyan2004}. For the cases where the
compensating 2DEG density is measured, the thickness of the AlGaN
layer and Schottky barrier height at the AlGaN surface will determine
whether the entire bound charge is compensated, which could be a
reason that the observations are slightly lower than predicted
theoretically \cite{Zhang2000}. Also, interface roughness and electron
traps due to dislocations and/or surface states have been proposed to
explain the reduced 2DEG density \cite{Ambacher2000}.

\subsection{Comparison of theoretical implementations}
\label{sec:compar}

The degree of agreement between results obtained based on the current
practice in the field (ZB reference, no correction term, proper PZ)
and our revised implementation (H reference, improper PZ constants)
merits some discussion.  It is clear from Table \ref{sptab} that the
SP polarization values determined with the H reference structure are
very different from those determined with the ZB reference; also, from
Table \ref{pztab}, the improper and proper $e_{31}$ coefficients are
very different. However, the similarity between the red solid and
black dashed curves in Fig.~\ref{Both} demonstrates that these two
large corrections cancel when we evaluate Eq.~(\ref{Ptotfin}).
I.e., the correct implementation (H reference, improper PZ constants)
results in values that are only slightly different from the current (incorrect) practice in the field
(ZB reference, no correction term, proper PZ).  We now demonstrate analytically
how this accidental agreement comes about.


If we take the difference between our revised implementation
(solid red curve in Fig.~\ref{Both}) and the
current practice in the field (dashed black
curve in Fig.~\ref{Both}) for a given $x$, we obtain the total
error in using the current practice in the field:
\begin{equation}
\label{diff}
\Delta P_{\text{error}} =x \Delta P_{\text{corr}}^{\text{(ZB ref)}}+2\epsilon_1(x)P_{\text{eff}}^{n, \text{(H ref)}}(x).
\end{equation}
A derivation of this expression is given in
Section~S5 of the SM \cite{Supp}. For
both AlGaN/GaN and InGaN/GaN, the two terms on the right-hand side
of Eq.~(\ref{diff}) have
opposite signs and a tendency to cancel.
To understand why, first note that $\epsilon_1(x)$ is approximately
linear in $x$, so that both terms can be regarded as being roughly
proportional to $\epsilon_1(x)$. In particular, linearizing
Eq.~(\ref{corr}) in $\epsilon_1$, we find
$x\Delta P_{\text{corr}}^{\text{(ZB ref)}} \simeq
  -2\epsilon_1(x) P_\text{f}^{m, \text{ZB}}$
[see also Eq.~(8) of the SM \cite{Supp}].  Thus
\begin{equation}
\begin{split}
\label{error}
\Delta P_{\text{error}} \simeq 2\epsilon_1(x) &\left(P_{\text{eff}}^{n, \text{(H ref)}}-P_\text{f}^{m, \text{ZB}}\right)
\\
=2\epsilon_1(x) &\left(P_{\text{f}}^{n, \text{WZ}}-P_\text{f}^{m, \text{ZB}}\right)
\\
=2\epsilon_1(x)&\left[P_{\text{eff}}^{n, \text{ZB ref}}+\left(P_\text{f}^{n, \text{ZB}}-P_\text{f}^{m, \text{ZB}}\right)\right],
\end{split}
\end{equation}
where in the second step we used the fact that the formal polarization
of the H structure vanishes. For the III-nitrides,
$\vert\epsilon_1\vert<0.1$, and we see from Table \ref{sptab} that $\vert
P_\text{eff}^{n, \text{ZB}}\vert<0.1$ C/m$^2$. The second term in the last line of Eq.~(\ref{error}), $P_\text{f}^{n, \text{ZB}}-P_\text{f}^{m, \text{ZB}}$, is related to the difference in in-plane lattice constants between material $n$ and material $m$ (Section \ref{zbcorr}). The largest value it will take for the materials considered in this study is $0.28$ C/m$^2$ for InN on GaN, as calculated in Eq.~(\ref{corr}), and it will be significantly smaller for lower alloy content and for the case of AlGaN on GaN.
The error is therefore the product of small factors, and thus
small in practice, significantly smaller than the errors in the SP and
PZ parts individually.

The small magnitude of $P_\text{eff}^{n, \text{ZB}}$ (Table~\ref{sptab}) demonstrates that
$P_\text{f}^{\text{WZ}}\sim P_\text{f}^{\text{ZB}}$ in these
materials (see Section S1 of the SM \cite{Supp}). The similarity between $P_\text{f}^{\text{ZB}}$ and
$P_\text{f}^{\text{WZ}}$ is not unexpected. Although ZB has sufficient
symmetry to preclude SP polarization, in the [111] direction the
structure only differs from the WZ $c$ direction by the stacking of
the cation/anion planes and a small deviation from the ideal WZ $u$
parameter and $c/a$ lattice constants. In WZ materials other than the III-nitrides, such
deviations could in principle be larger, and in that case $P_\text{eff}^{\text{ZB ref}}$ will be larger,
resulting in $\Delta P_{\text{error}}$ being more significant.

For AlGaN/GaN, the relatively modest difference in lattice constants
between AlN and GaN (and therefore modest $\epsilon_1$ values for
coherently strained alloy layers), and an almost exact cancellation
between $P_\text{f}^{\text{GaN,ZB}}$ and
$P_{\text{eff}}^{\text{AlGaN,(H ref)}}$ means that the difference
between implementations is small over the whole composition range
(\textit{cf.} Section~S5 of the SM \cite{Supp}). For InGaN/GaN, the
large lattice mismatch of InN and GaN and a less complete cancellation
of $P_\text{f}^{\text{GaN,ZB}}$ and $P_{\text{eff}}^{\text{InGaN,(H
ref)}}$ results in a significant deviation at higher In content; this
will be important for the prediction of polarization fields in
applications such as tunnel field-effect transistors based on thin,
high In-content interlayers \cite{Li2015}.

\begin{figure}
\includegraphics[width=\columnwidth]{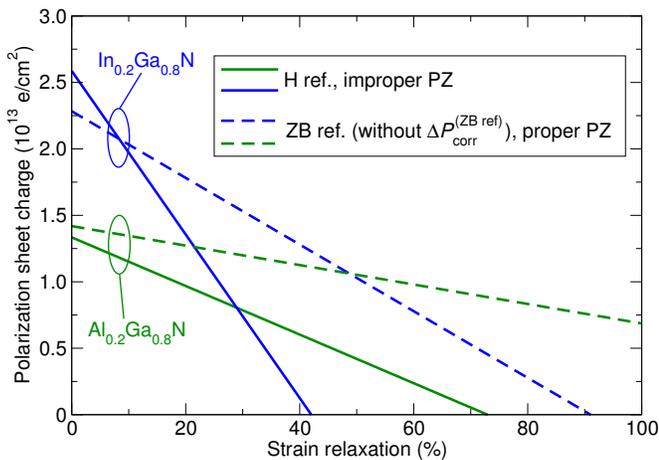}
\caption{\label{rel}
Absolute values for polarization sheet charges at the
In$_{0.2}$Ga$_{0.8}$N/GaN (blue) and Al$_{0.2}$Ga$_{0.8}$N/GaN
(green) interface as a function of percent strain relaxation.
0 \% relaxation corresponds to perfectly strained layers, 100 \% relaxation to
an unstrained overlayer at its bulk lattice constant.  Solid
curves are the correct implementation described in this work (H reference
for SP and improper PZ constants); dashed curves are the current
practice in the field (ZB reference for SP, without the correction
term, and proper PZ constants). }
\end{figure}

The two implementations differ significantly in the
\textit{relative} contributions of SP and PZ polarization. An
effect of this is illustrated by the case where there is strain
relaxation in the alloy layer. In Fig.~\ref{rel} the predicted
polarization bound charges for In$_{0.2}$Ga$_{0.8}$N/GaN (blue
curves) and Al$_{0.2}$Ga$_{0.8}$N/GaN (green curves) are shown
as a function
of strain relaxation of the layer, modeled by simply scaling
$\epsilon_1$. For both InGaN/GaN and AlGaN/GaN, the revised
implementation of this work predicts a much faster decrease
in bound charge at the interface than the current practice in
the field. Of course, strain relaxation is
associated with the presence of edge dislocations at
the interface,
which may themselves influence the interface bound charge;
this effect has not been taken into account
in either version of the implementation.

\section{Conclusions}
\label{sec:conc}

We have derived a rigorously correct implementation of
polarization constants in wurtzite materials, focusing on the
example of the III-nitrides. Our derivation has demonstrated the
impact of the choice of reference structure when calculating
spontaneous polarization constants using the modern theory of
polarization.  Insufficient care in using the values can result
in spurious contributions to the polarization discontinuities
at heterostructure interfaces.  We have provided new values
calculated with a consistent hexagonal (rather than zincblende)
reference structure.  In addition, we have demonstrated the
importance of choosing the correct piezoelectric constants
(improper), and provided values for these improper constants. These
revised values of the spontaneous and piezoelectric constants can
be directly used in simulations and to interpret experimental
observations. The revised implementation predicts a more rapid
decrease of polarization charge with strain relaxation for an alloy layer on GaN.

%
%

\acknowledgements{

Work by C. E. D. was supported by the U.S. Department of Energy
(DOE), Office of Science, Basic Energy Sciences (BES) under Award
DE-SC0010689.  Work by A. J. was supported by the Center for Low
Energy Systems Technology (LEAST), one of the six SRC STARnet
Centers, sponsored by MARCO and DARPA.
Work by D. V. was supported by ONR Grant N00014-12-1-1035.
Computational resources were provided by the CSC at the CNSI and
MRL (an NSF MRSEC, DMR-1121053) (NSF CNS-0960316) and by the National
Energy Research Scientific Computing Center, a DOE Office of
Science User Facility supported by the Office of Science of the
U.S. Department of Energy under Contract No. DE-AC02-05CH11231.
}

%

\begin{thebibliography}{76}%
\makeatletter
\providecommand \@ifxundefined [1]{%
 \@ifx{#1\undefined}
}%
\providecommand \@ifnum [1]{%
 \ifnum #1\expandafter \@firstoftwo
 \else \expandafter \@secondoftwo
 \fi
}%
\providecommand \@ifx [1]{%
 \ifx #1\expandafter \@firstoftwo
 \else \expandafter \@secondoftwo
 \fi
}%
\providecommand \natexlab [1]{#1}%
\providecommand \enquote  [1]{``#1''}%
\providecommand \bibnamefont  [1]{#1}%
\providecommand \bibfnamefont [1]{#1}%
\providecommand \citenamefont [1]{#1}%
\providecommand \href@noop [0]{\@secondoftwo}%
\providecommand \href [0]{\begingroup \@sanitize@url \@href}%
\providecommand \@href[1]{\@@startlink{#1}\@@href}%
\providecommand \@@href[1]{\endgroup#1\@@endlink}%
\providecommand \@sanitize@url [0]{\catcode `\\12\catcode `\$12\catcode
  `\&12\catcode `\#12\catcode `\^12\catcode `\_12\catcode `\%12\relax}%
\providecommand \@@startlink[1]{}%
\providecommand \@@endlink[0]{}%
\providecommand \url  [0]{\begingroup\@sanitize@url \@url }%
\providecommand \@url [1]{\endgroup\@href {#1}{\urlprefix }}%
\providecommand \urlprefix  [0]{URL }%
\providecommand \Eprint [0]{\href }%
\providecommand \doibase [0]{http://dx.doi.org/}%
\providecommand \selectlanguage [0]{\@gobble}%
\providecommand \bibinfo  [0]{\@secondoftwo}%
\providecommand \bibfield  [0]{\@secondoftwo}%
\providecommand \translation [1]{[#1]}%
\providecommand \BibitemOpen [0]{}%
\providecommand \bibitemStop [0]{}%
\providecommand \bibitemNoStop [0]{.\EOS\space}%
\providecommand \EOS [0]{\spacefactor3000\relax}%
\providecommand \BibitemShut  [1]{\csname bibitem#1\endcsname}%
\let\auto@bib@innerbib\@empty
\bibitem [{\citenamefont {Vanderbilt}\ and\ \citenamefont
  {King-Smith}(1993)}]{VKS1993}%
  \BibitemOpen
  \bibfield  {author} {\bibinfo {author} {\bibfnamefont {D.}~\bibnamefont
  {Vanderbilt}}\ and\ \bibinfo {author} {\bibfnamefont {R.~D.}\ \bibnamefont
  {King-Smith}},\ }\bibfield  {title} {\enquote {\bibinfo {title} {{Electric
  polarization as a bulk quantity and its relation to surface charge}},}\
  }\href {\doibase 10.1103/PhysRevB.48.4442} {\bibfield  {journal} {\bibinfo
  {journal} {Phys. Rev. B}\ }\textbf {\bibinfo {volume} {48}},\ \bibinfo
  {pages} {4442} (\bibinfo {year} {1993})}\BibitemShut {NoStop}%
\bibitem [{\citenamefont {Vurgaftman}\ and\ \citenamefont
  {Meyer}(2003)}]{Vurgaftman2003}%
  \BibitemOpen
  \bibfield  {author} {\bibinfo {author} {\bibfnamefont {I.}~\bibnamefont
  {Vurgaftman}}\ and\ \bibinfo {author} {\bibfnamefont {J.~R.}\ \bibnamefont
  {Meyer}},\ }\bibfield  {title} {\enquote {\bibinfo {title} {{Band parameters
  for nitrogen-containing semiconductors}},}\ }\href {\doibase
  10.1063/1.1600519} {\bibfield  {journal} {\bibinfo  {journal} {J. Appl.
  Phys.}\ }\textbf {\bibinfo {volume} {94}},\ \bibinfo {pages} {3675} (\bibinfo
  {year} {2003})}\BibitemShut {NoStop}%
\bibitem [{\citenamefont {Schulz}\ and\ \citenamefont
  {Thiemann}(1977)}]{Schulz1977}%
  \BibitemOpen
  \bibfield  {author} {\bibinfo {author} {\bibfnamefont {H.}~\bibnamefont
  {Schulz}}\ and\ \bibinfo {author} {\bibfnamefont {K.H.}\ \bibnamefont
  {Thiemann}},\ }\bibfield  {title} {\enquote {\bibinfo {title} {{Crystal
  structure refinement of AlN and GaN}},}\ }\href {\doibase
  10.1016/0038-1098(77)90959-0} {\bibfield  {journal} {\bibinfo  {journal}
  {Solid State Commun.}\ }\textbf {\bibinfo {volume} {23}},\ \bibinfo {pages}
  {815} (\bibinfo {year} {1977})}\BibitemShut {NoStop}%
\bibitem [{\citenamefont {Aumer}\ \emph {et~al.}(2001)\citenamefont {Aumer},
  \citenamefont {LeBoeuf}, \citenamefont {Moody},\ and\ \citenamefont
  {Bedair}}]{Aumer2001}%
  \BibitemOpen
  \bibfield  {author} {\bibinfo {author} {\bibfnamefont {M.~E.}\ \bibnamefont
  {Aumer}}, \bibinfo {author} {\bibfnamefont {S.~F.}\ \bibnamefont {LeBoeuf}},
  \bibinfo {author} {\bibfnamefont {B.~F.}\ \bibnamefont {Moody}}, \ and\
  \bibinfo {author} {\bibfnamefont {S.~M.}\ \bibnamefont {Bedair}},\ }\bibfield
   {title} {\enquote {\bibinfo {title} {{Strain-induced piezoelectric field
  effects on light emission energy and intensity from AlInGaN/InGaN quantum
  wells}},}\ }\href {\doibase 10.1063/1.1418453} {\bibfield  {journal}
  {\bibinfo  {journal} {Appl. Phys. Lett.}\ }\textbf {\bibinfo {volume} {79}},\
  \bibinfo {pages} {3803} (\bibinfo {year} {2001})}\BibitemShut {NoStop}%
\bibitem [{\citenamefont {Li}\ \emph {et~al.}(2002)\citenamefont {Li},
  \citenamefont {Xu}, \citenamefont {Xie}, \citenamefont {Tong}, \citenamefont
  {Zhang}, \citenamefont {Liu},\ and\ \citenamefont {Chua}}]{Li2002}%
  \BibitemOpen
  \bibfield  {author} {\bibinfo {author} {\bibfnamefont {Q.}~\bibnamefont
  {Li}}, \bibinfo {author} {\bibfnamefont {S.~J.}\ \bibnamefont {Xu}}, \bibinfo
  {author} {\bibfnamefont {M.~H.}\ \bibnamefont {Xie}}, \bibinfo {author}
  {\bibfnamefont {S.~Y.}\ \bibnamefont {Tong}}, \bibinfo {author}
  {\bibfnamefont {X.~H.}\ \bibnamefont {Zhang}}, \bibinfo {author}
  {\bibfnamefont {W.}~\bibnamefont {Liu}}, \ and\ \bibinfo {author}
  {\bibfnamefont {S.~J.}\ \bibnamefont {Chua}},\ }\bibfield  {title} {\enquote
  {\bibinfo {title} {{Strong screening effect of photo-generated carriers on
  piezoelectric field in In$_{0.13}$Ga$_{0.87}$N/In$_{0.03}$Ga$_{0.97}$N
  quantum wells}},}\ }\href {\doibase 10.1143/JJAP.41.L1093} {\bibfield
  {journal} {\bibinfo  {journal} {Jpn. J. Appl. Phys.}\ }\textbf {\bibinfo
  {volume} {41}},\ \bibinfo {pages} {L1093} (\bibinfo {year}
  {2002})}\BibitemShut {NoStop}%
\bibitem [{\citenamefont {Jho}\ \emph {et~al.}(2001)\citenamefont {Jho},
  \citenamefont {Yahng}, \citenamefont {Oh},\ and\ \citenamefont
  {Kim}}]{Jho2001}%
  \BibitemOpen
  \bibfield  {author} {\bibinfo {author} {\bibfnamefont {Y.~D.}\ \bibnamefont
  {Jho}}, \bibinfo {author} {\bibfnamefont {J.~S.}\ \bibnamefont {Yahng}},
  \bibinfo {author} {\bibfnamefont {E.}~\bibnamefont {Oh}}, \ and\ \bibinfo
  {author} {\bibfnamefont {D.~S.}\ \bibnamefont {Kim}},\ }\bibfield  {title}
  {\enquote {\bibinfo {title} {{Measurement of piezoelectric field and
  tunneling times in strongly biased InGaN/GaN quantum wells}},}\ }\href
  {\doibase 10.1063/1.1396315} {\bibfield  {journal} {\bibinfo  {journal}
  {Appl. Phys. Lett.}\ }\textbf {\bibinfo {volume} {79}},\ \bibinfo {pages}
  {1130} (\bibinfo {year} {2001})}\BibitemShut {NoStop}%
\bibitem [{\citenamefont {Hangleiter}\ \emph {et~al.}(2003)\citenamefont
  {Hangleiter}, \citenamefont {Hitzel}, \citenamefont {Lahmann},\ and\
  \citenamefont {Rossow}}]{Hangleiter2003}%
  \BibitemOpen
  \bibfield  {author} {\bibinfo {author} {\bibfnamefont {A.}~\bibnamefont
  {Hangleiter}}, \bibinfo {author} {\bibfnamefont {F.}~\bibnamefont {Hitzel}},
  \bibinfo {author} {\bibfnamefont {S.}~\bibnamefont {Lahmann}}, \ and\
  \bibinfo {author} {\bibfnamefont {U.}~\bibnamefont {Rossow}},\ }\bibfield
  {title} {\enquote {\bibinfo {title} {{Composition dependence of polarization
  fields in GaInN/GaN quantum wells}},}\ }\href {\doibase 10.1063/1.1601310}
  {\bibfield  {journal} {\bibinfo  {journal} {Appl. Phys. Lett.}\ }\textbf
  {\bibinfo {volume} {83}},\ \bibinfo {pages} {1169} (\bibinfo {year}
  {2003})}\BibitemShut {NoStop}%
\bibitem [{\citenamefont {Chichibu}\ \emph {et~al.}(1998)\citenamefont
  {Chichibu}, \citenamefont {Abare}, \citenamefont {Minsky}, \citenamefont
  {Keller}, \citenamefont {Fleischer}, \citenamefont {Bowers}, \citenamefont
  {Hu}, \citenamefont {Mishra}, \citenamefont {Coldren}, \citenamefont
  {DenBaars},\ and\ \citenamefont {Sota}}]{Chichibu1998}%
  \BibitemOpen
  \bibfield  {author} {\bibinfo {author} {\bibfnamefont {S.~F.}\ \bibnamefont
  {Chichibu}}, \bibinfo {author} {\bibfnamefont {A.~C.}\ \bibnamefont {Abare}},
  \bibinfo {author} {\bibfnamefont {M.~S.}\ \bibnamefont {Minsky}}, \bibinfo
  {author} {\bibfnamefont {S.}~\bibnamefont {Keller}}, \bibinfo {author}
  {\bibfnamefont {S.~B.}\ \bibnamefont {Fleischer}}, \bibinfo {author}
  {\bibfnamefont {J.~E.}\ \bibnamefont {Bowers}}, \bibinfo {author}
  {\bibfnamefont {E.}~\bibnamefont {Hu}}, \bibinfo {author} {\bibfnamefont
  {U.~K.}\ \bibnamefont {Mishra}}, \bibinfo {author} {\bibfnamefont {L.~A.}\
  \bibnamefont {Coldren}}, \bibinfo {author} {\bibfnamefont {S.~P.}\
  \bibnamefont {DenBaars}}, \ and\ \bibinfo {author} {\bibfnamefont
  {T.}~\bibnamefont {Sota}},\ }\bibfield  {title} {\enquote {\bibinfo {title}
  {{Effective band gap inhomogeneity and piezoelectric field in InGaN/GaN
  multiquantum well structures}},}\ }\href {\doibase 10.1063/1.122350}
  {\bibfield  {journal} {\bibinfo  {journal} {Appl. Phys. Lett.}\ }\textbf
  {\bibinfo {volume} {73}},\ \bibinfo {pages} {2006} (\bibinfo {year}
  {1998})}\BibitemShut {NoStop}%
\bibitem [{\citenamefont {Lefebvre}\ \emph {et~al.}(2001)\citenamefont
  {Lefebvre}, \citenamefont {Morel}, \citenamefont {Gallart}, \citenamefont
  {Taliercio}, \citenamefont {Allgre}, \citenamefont {Gil}, \citenamefont
  {Mathieu}, \citenamefont {Damilano}, \citenamefont {Grandjean},\ and\
  \citenamefont {Massies}}]{Lefebvre2001}%
  \BibitemOpen
  \bibfield  {author} {\bibinfo {author} {\bibfnamefont {P.}~\bibnamefont
  {Lefebvre}}, \bibinfo {author} {\bibfnamefont {A.}~\bibnamefont {Morel}},
  \bibinfo {author} {\bibfnamefont {M.}~\bibnamefont {Gallart}}, \bibinfo
  {author} {\bibfnamefont {T.}~\bibnamefont {Taliercio}}, \bibinfo {author}
  {\bibfnamefont {J.}~\bibnamefont {Allgre}}, \bibinfo {author} {\bibfnamefont
  {B.}~\bibnamefont {Gil}}, \bibinfo {author} {\bibfnamefont {H.}~\bibnamefont
  {Mathieu}}, \bibinfo {author} {\bibfnamefont {B.}~\bibnamefont {Damilano}},
  \bibinfo {author} {\bibfnamefont {N.}~\bibnamefont {Grandjean}}, \ and\
  \bibinfo {author} {\bibfnamefont {J.}~\bibnamefont {Massies}},\ }\bibfield
  {title} {\enquote {\bibinfo {title} {{High internal electric field in a
  graded-width InGaN/GaN quantum well: Accurate determination by time-resolved
  photoluminescence spectroscopy}},}\ }\href {\doibase 10.1063/1.1351517}
  {\bibfield  {journal} {\bibinfo  {journal} {Appl. Phys. Lett.}\ }\textbf
  {\bibinfo {volume} {78}},\ \bibinfo {pages} {1252} (\bibinfo {year}
  {2001})}\BibitemShut {NoStop}%
\bibitem [{\citenamefont {Kiesel}\ \emph {et~al.}(2001)\citenamefont {Kiesel},
  \citenamefont {Renner}, \citenamefont {Kneissl}, \citenamefont {{Van de
  Walle}}, \citenamefont {D\"{o}hler},\ and\ \citenamefont
  {Johnson}}]{Kiesel2001}%
  \BibitemOpen
  \bibfield  {author} {\bibinfo {author} {\bibfnamefont {P.}~\bibnamefont
  {Kiesel}}, \bibinfo {author} {\bibfnamefont {F.}~\bibnamefont {Renner}},
  \bibinfo {author} {\bibfnamefont {M.}~\bibnamefont {Kneissl}}, \bibinfo
  {author} {\bibfnamefont {C.~G.}\ \bibnamefont {{Van de Walle}}}, \bibinfo
  {author} {\bibfnamefont {G.~H.}\ \bibnamefont {D\"{o}hler}}, \ and\ \bibinfo
  {author} {\bibfnamefont {N.~M.}\ \bibnamefont {Johnson}},\ }\bibfield
  {title} {\enquote {\bibinfo {title} {{Electroabsorption Spectroscopy --
  Direct Determination of the Strong Piezoelectric Field in InGaN/GaN
  Heterostructure Diodes}},}\ }\href {\doibase
  10.1002/1521-396X(200111)188:1<131::AID-PSSA131>3.0.CO;2-C} {\bibfield
  {journal} {\bibinfo  {journal} {Phys. Stat. Sol. (a)}\ }\textbf {\bibinfo
  {volume} {188}},\ \bibinfo {pages} {131} (\bibinfo {year}
  {2001})}\BibitemShut {NoStop}%
\bibitem [{\citenamefont {Kiesel}\ \emph {et~al.}(2002)\citenamefont {Kiesel},
  \citenamefont {Renner}, \citenamefont {Kneissl}, \citenamefont {{Van de
  Walle}}, \citenamefont {D\"{o}hler},\ and\ \citenamefont
  {Johnson}}]{Kiesel2002}%
  \BibitemOpen
  \bibfield  {author} {\bibinfo {author} {\bibfnamefont {P.}~\bibnamefont
  {Kiesel}}, \bibinfo {author} {\bibfnamefont {F.}~\bibnamefont {Renner}},
  \bibinfo {author} {\bibfnamefont {M.}~\bibnamefont {Kneissl}}, \bibinfo
  {author} {\bibfnamefont {C.~G.}\ \bibnamefont {{Van de Walle}}}, \bibinfo
  {author} {\bibfnamefont {G.~H.}\ \bibnamefont {D\"{o}hler}}, \ and\ \bibinfo
  {author} {\bibfnamefont {N.~M.}\ \bibnamefont {Johnson}},\ }\bibfield
  {title} {\enquote {\bibinfo {title} {{Quantitative Analysis of Absorption and
  Field-Induced Absorption Changes in InGaN/GaN Quantum Wells}},}\ }\href
  {\doibase 10.1002/1521-3951(200212)234:3<742::AID-PSSB742>3.0.CO;2-G}
  {\bibfield  {journal} {\bibinfo  {journal} {Phys. Stat. Sol. (b)}\ }\textbf
  {\bibinfo {volume} {234}},\ \bibinfo {pages} {742--745} (\bibinfo {year}
  {2002})}\BibitemShut {NoStop}%
\bibitem [{\citenamefont {Renner}\ \emph {et~al.}(2002)\citenamefont {Renner},
  \citenamefont {Kiesel}, \citenamefont {D\"{o}hler}, \citenamefont {Kneissl},
  \citenamefont {{Van de Walle}},\ and\ \citenamefont {Johnson}}]{Renner2002}%
  \BibitemOpen
  \bibfield  {author} {\bibinfo {author} {\bibfnamefont {F.}~\bibnamefont
  {Renner}}, \bibinfo {author} {\bibfnamefont {P.}~\bibnamefont {Kiesel}},
  \bibinfo {author} {\bibfnamefont {G.~H.}\ \bibnamefont {D\"{o}hler}},
  \bibinfo {author} {\bibfnamefont {M.}~\bibnamefont {Kneissl}}, \bibinfo
  {author} {\bibfnamefont {C.~G.}\ \bibnamefont {{Van de Walle}}}, \ and\
  \bibinfo {author} {\bibfnamefont {N.~M.}\ \bibnamefont {Johnson}},\
  }\bibfield  {title} {\enquote {\bibinfo {title} {{Quantitative analysis of
  the polarization fields and absorption changes in InGaN/GaN quantum wells
  with electroabsorption spectroscopy}},}\ }\href {\doibase 10.1063/1.1493229}
  {\bibfield  {journal} {\bibinfo  {journal} {Appl. Phys. Lett.}\ }\textbf
  {\bibinfo {volume} {81}},\ \bibinfo {pages} {490} (\bibinfo {year}
  {2002})}\BibitemShut {NoStop}%
\bibitem [{\citenamefont {Kaplar}\ \emph {et~al.}(2004)\citenamefont {Kaplar},
  \citenamefont {Kurtz}, \citenamefont {Koleske},\ and\ \citenamefont
  {Fischer}}]{Kaplar2004}%
  \BibitemOpen
  \bibfield  {author} {\bibinfo {author} {\bibfnamefont {R.~J.}\ \bibnamefont
  {Kaplar}}, \bibinfo {author} {\bibfnamefont {S.~R.}\ \bibnamefont {Kurtz}},
  \bibinfo {author} {\bibfnamefont {D.~D.}\ \bibnamefont {Koleske}}, \ and\
  \bibinfo {author} {\bibfnamefont {A.~J.}\ \bibnamefont {Fischer}},\
  }\bibfield  {title} {\enquote {\bibinfo {title} {{Electroreflectance studies
  of Stark shifts and polarization-induced electric fields in InGaN/GaN single
  quantum wells}},}\ }\href {\doibase 10.1063/1.1690100} {\bibfield  {journal}
  {\bibinfo  {journal} {J. Appl. Phys.}\ }\textbf {\bibinfo {volume} {95}},\
  \bibinfo {pages} {4905} (\bibinfo {year} {2004})}\BibitemShut {NoStop}%
\bibitem [{\citenamefont {Kim}\ \emph {et~al.}(2010)\citenamefont {Kim},
  \citenamefont {Oh}, \citenamefont {Baek}, \citenamefont {Lee}, \citenamefont
  {Jung}, \citenamefont {Song}, \citenamefont {Kim}, \citenamefont {Ahn},
  \citenamefont {Yanqun},\ and\ \citenamefont {Song}}]{Kim2010}%
  \BibitemOpen
  \bibfield  {author} {\bibinfo {author} {\bibfnamefont {S.-M.}\ \bibnamefont
  {Kim}}, \bibinfo {author} {\bibfnamefont {H.~S.}\ \bibnamefont {Oh}},
  \bibinfo {author} {\bibfnamefont {J.~H.}\ \bibnamefont {Baek}}, \bibinfo
  {author} {\bibfnamefont {K.-H.}\ \bibnamefont {Lee}}, \bibinfo {author}
  {\bibfnamefont {G.~Y.}\ \bibnamefont {Jung}}, \bibinfo {author}
  {\bibfnamefont {J.-H.}\ \bibnamefont {Song}}, \bibinfo {author}
  {\bibfnamefont {H.-J.}\ \bibnamefont {Kim}}, \bibinfo {author} {\bibfnamefont
  {B.-J.}\ \bibnamefont {Ahn}}, \bibinfo {author} {\bibfnamefont
  {D.}~\bibnamefont {Yanqun}}, \ and\ \bibinfo {author} {\bibfnamefont {J.-H.}\
  \bibnamefont {Song}},\ }\bibfield  {title} {\enquote {\bibinfo {title}
  {{Effects of patterned sapphire substrates on piezoelectric field in
  blue-emitting InGaN multiple quantum wells}},}\ }\href {\doibase
  10.1109/LED.2010.2051406} {\bibfield  {journal} {\bibinfo  {journal} {IEEE
  Electr. Device Lett.}\ }\textbf {\bibinfo {volume} {31}},\ \bibinfo {pages}
  {842} (\bibinfo {year} {2010})}\BibitemShut {NoStop}%
\bibitem [{\citenamefont {Franssen}\ \emph {et~al.}(2004)\citenamefont
  {Franssen}, \citenamefont {Perlin},\ and\ \citenamefont
  {Suski}}]{Franssen2004}%
  \BibitemOpen
  \bibfield  {author} {\bibinfo {author} {\bibfnamefont {G.}~\bibnamefont
  {Franssen}}, \bibinfo {author} {\bibfnamefont {P.}~\bibnamefont {Perlin}}, \
  and\ \bibinfo {author} {\bibfnamefont {T.}~\bibnamefont {Suski}},\ }\bibfield
   {title} {\enquote {\bibinfo {title} {{Photocurrent spectroscopy as a tool
  for determining piezoelectric fields in In$_x$Ga$_{1-x}$N/GaN multiple
  quantum well light emitting diodes}},}\ }\href {\doibase
  10.1103/PhysRevB.69.045310} {\bibfield  {journal} {\bibinfo  {journal} {Phys.
  Rev. B}\ }\textbf {\bibinfo {volume} {69}},\ \bibinfo {pages} {045310}
  (\bibinfo {year} {2004})}\BibitemShut {NoStop}%
\bibitem [{\citenamefont {Brown}\ \emph {et~al.}(2005)\citenamefont {Brown},
  \citenamefont {Pope}, \citenamefont {Smowton}, \citenamefont {Blood},
  \citenamefont {Thomson}, \citenamefont {Chow}, \citenamefont {Bour},\ and\
  \citenamefont {Kneissl}}]{Brown2005}%
  \BibitemOpen
  \bibfield  {author} {\bibinfo {author} {\bibfnamefont {I.~H.}\ \bibnamefont
  {Brown}}, \bibinfo {author} {\bibfnamefont {I.~A.}\ \bibnamefont {Pope}},
  \bibinfo {author} {\bibfnamefont {P.~M.}\ \bibnamefont {Smowton}}, \bibinfo
  {author} {\bibfnamefont {P.}~\bibnamefont {Blood}}, \bibinfo {author}
  {\bibfnamefont {J.~D.}\ \bibnamefont {Thomson}}, \bibinfo {author}
  {\bibfnamefont {W.~W.}\ \bibnamefont {Chow}}, \bibinfo {author}
  {\bibfnamefont {D.~P.}\ \bibnamefont {Bour}}, \ and\ \bibinfo {author}
  {\bibfnamefont {M.}~\bibnamefont {Kneissl}},\ }\bibfield  {title} {\enquote
  {\bibinfo {title} {{Determination of the piezoelectric field in InGaN quantum
  wells}},}\ }\href {\doibase 10.1063/1.1896446} {\bibfield  {journal}
  {\bibinfo  {journal} {Appl. Phys. Lett.}\ }\textbf {\bibinfo {volume} {86}},\
  \bibinfo {pages} {131108} (\bibinfo {year} {2005})}\BibitemShut {NoStop}%
\bibitem [{\citenamefont {Thomsen}\ \emph {et~al.}(2011)\citenamefont
  {Thomsen}, \citenamefont {J\"{o}nen}, \citenamefont {Rossow},\ and\
  \citenamefont {Hangleiter}}]{Thomsen2011}%
  \BibitemOpen
  \bibfield  {author} {\bibinfo {author} {\bibfnamefont {M.}~\bibnamefont
  {Thomsen}}, \bibinfo {author} {\bibfnamefont {H.}~\bibnamefont {J\"{o}nen}},
  \bibinfo {author} {\bibfnamefont {U.}~\bibnamefont {Rossow}}, \ and\ \bibinfo
  {author} {\bibfnamefont {A.}~\bibnamefont {Hangleiter}},\ }\bibfield  {title}
  {\enquote {\bibinfo {title} {{Effects of spontaneous polarization on
  GaInN/GaN quantum well structures}},}\ }\href {\doibase 10.1063/1.3600221}
  {\bibfield  {journal} {\bibinfo  {journal} {J. Appl. Phys.}\ }\textbf
  {\bibinfo {volume} {109}},\ \bibinfo {pages} {123710} (\bibinfo {year}
  {2011})}\BibitemShut {NoStop}%
\bibitem [{\citenamefont {Turchinovich}\ \emph {et~al.}(2003)\citenamefont
  {Turchinovich}, \citenamefont {{Uhd Jepsen}}, \citenamefont {Monozon},
  \citenamefont {Koch}, \citenamefont {Lahmann}, \citenamefont {Rossow},\ and\
  \citenamefont {Hangleiter}}]{Turchinovich2003}%
  \BibitemOpen
  \bibfield  {author} {\bibinfo {author} {\bibfnamefont {D.}~\bibnamefont
  {Turchinovich}}, \bibinfo {author} {\bibfnamefont {P.}~\bibnamefont {{Uhd
  Jepsen}}}, \bibinfo {author} {\bibfnamefont {B.}~\bibnamefont {Monozon}},
  \bibinfo {author} {\bibfnamefont {M.}~\bibnamefont {Koch}}, \bibinfo {author}
  {\bibfnamefont {S.}~\bibnamefont {Lahmann}}, \bibinfo {author} {\bibfnamefont
  {U.}~\bibnamefont {Rossow}}, \ and\ \bibinfo {author} {\bibfnamefont
  {A.}~\bibnamefont {Hangleiter}},\ }\bibfield  {title} {\enquote {\bibinfo
  {title} {{Ultrafast polarization dynamics in biased quantum wells under
  strong femtosecond optical excitation}},}\ }\href {\doibase
  10.1103/PhysRevB.68.241307} {\bibfield  {journal} {\bibinfo  {journal} {Phys.
  Rev. B}\ }\textbf {\bibinfo {volume} {68}},\ \bibinfo {pages} {241307}
  (\bibinfo {year} {2003})}\BibitemShut {NoStop}%
\bibitem [{\citenamefont {Gfr\"{o}rer}\ \emph {et~al.}(1999)\citenamefont
  {Gfr\"{o}rer}, \citenamefont {Gemmer}, \citenamefont {Off}, \citenamefont
  {Im}, \citenamefont {Scholz},\ and\ \citenamefont
  {Hangleiter}}]{Gfrorer1999}%
  \BibitemOpen
  \bibfield  {author} {\bibinfo {author} {\bibfnamefont {O.}~\bibnamefont
  {Gfr\"{o}rer}}, \bibinfo {author} {\bibfnamefont {C.}~\bibnamefont {Gemmer}},
  \bibinfo {author} {\bibfnamefont {J.}~\bibnamefont {Off}}, \bibinfo {author}
  {\bibfnamefont {J.~S.}\ \bibnamefont {Im}}, \bibinfo {author} {\bibfnamefont
  {F.}~\bibnamefont {Scholz}}, \ and\ \bibinfo {author} {\bibfnamefont
  {A.}~\bibnamefont {Hangleiter}},\ }\bibfield  {title} {\enquote {\bibinfo
  {title} {{Direct Observation of Pyroelectric Fields in InGaN/GaN and
  AlGaN/GaN Heterostructures}},}\ }\href {\doibase
  10.1002/(SICI)1521-3951(199911)216:1<405::AID-PSSB405>3.0.CO;2-\#} {\bibfield
   {journal} {\bibinfo  {journal} {Phys. Stat. Sol. (b)}\ }\textbf {\bibinfo
  {volume} {216}},\ \bibinfo {pages} {405--408} (\bibinfo {year}
  {1999})}\BibitemShut {NoStop}%
\bibitem [{\citenamefont {Berkowicz}\ \emph {et~al.}(1999)\citenamefont
  {Berkowicz}, \citenamefont {Gershoni}, \citenamefont {Bahir}, \citenamefont
  {Abare}, \citenamefont {DenBaars},\ and\ \citenamefont
  {Coldren}}]{Berkowicz1999}%
  \BibitemOpen
  \bibfield  {author} {\bibinfo {author} {\bibfnamefont {E.}~\bibnamefont
  {Berkowicz}}, \bibinfo {author} {\bibfnamefont {D.}~\bibnamefont {Gershoni}},
  \bibinfo {author} {\bibfnamefont {G.}~\bibnamefont {Bahir}}, \bibinfo
  {author} {\bibfnamefont {A.~C.}\ \bibnamefont {Abare}}, \bibinfo {author}
  {\bibfnamefont {S.~P.}\ \bibnamefont {DenBaars}}, \ and\ \bibinfo {author}
  {\bibfnamefont {L.~A.}\ \bibnamefont {Coldren}},\ }\bibfield  {title}
  {\enquote {\bibinfo {title} {{Optical spectroscopy of InGaN/GaN quantum
  wells}},}\ }\href {\doibase
  10.1002/(SICI)1521-3951(199911)216:1<291::AID-PSSB291>3.0.CO;2-O} {\bibfield
  {journal} {\bibinfo  {journal} {Phys. Stat. Sol. (b)}\ }\textbf {\bibinfo
  {volume} {216}},\ \bibinfo {pages} {291--300} (\bibinfo {year}
  {1999})}\BibitemShut {NoStop}%
\bibitem [{\citenamefont {Park}\ \emph {et~al.}(2012)\citenamefont {Park},
  \citenamefont {Lee},\ and\ \citenamefont {Jang}}]{Park2012}%
  \BibitemOpen
  \bibfield  {author} {\bibinfo {author} {\bibfnamefont {S.~I.}\ \bibnamefont
  {Park}}, \bibinfo {author} {\bibfnamefont {J.~I.}\ \bibnamefont {Lee}}, \
  and\ \bibinfo {author} {\bibfnamefont {D.~H.}\ \bibnamefont {Jang}},\
  }\bibfield  {title} {\enquote {\bibinfo {title} {{Measurement of Internal
  Electric Field in GaN-Based Light-Emitting Diodes}},}\ }\href {\doibase
  10.1109/JQE.2012.2186610} {\bibfield  {journal} {\bibinfo  {journal} {IEEE J.
  Quantum Elect.}\ }\textbf {\bibinfo {volume} {48}},\ \bibinfo {pages} {500}
  (\bibinfo {year} {2012})}\BibitemShut {NoStop}%
\bibitem [{\citenamefont {Takeuchi}\ \emph {et~al.}(1998)\citenamefont
  {Takeuchi}, \citenamefont {Wetzel}, \citenamefont {Yamaguchi}, \citenamefont
  {Sakai}, \citenamefont {Amano}, \citenamefont {Akasaki}, \citenamefont
  {Kaneko}, \citenamefont {Nakagawa}, \citenamefont {Yamaoka},\ and\
  \citenamefont {Yamada}}]{Takeuchi1998}%
  \BibitemOpen
  \bibfield  {author} {\bibinfo {author} {\bibfnamefont {T.}~\bibnamefont
  {Takeuchi}}, \bibinfo {author} {\bibfnamefont {C.}~\bibnamefont {Wetzel}},
  \bibinfo {author} {\bibfnamefont {S.}~\bibnamefont {Yamaguchi}}, \bibinfo
  {author} {\bibfnamefont {H.}~\bibnamefont {Sakai}}, \bibinfo {author}
  {\bibfnamefont {H.}~\bibnamefont {Amano}}, \bibinfo {author} {\bibfnamefont
  {I.}~\bibnamefont {Akasaki}}, \bibinfo {author} {\bibfnamefont
  {Y.}~\bibnamefont {Kaneko}}, \bibinfo {author} {\bibfnamefont
  {S.}~\bibnamefont {Nakagawa}}, \bibinfo {author} {\bibfnamefont
  {Y.}~\bibnamefont {Yamaoka}}, \ and\ \bibinfo {author} {\bibfnamefont
  {N.}~\bibnamefont {Yamada}},\ }\bibfield  {title} {\enquote {\bibinfo {title}
  {{Determination of piezoelectric fields in strained GaInN quantum wells using
  the quantum-confined Stark effect}},}\ }\href {\doibase 10.1063/1.122247}
  {\bibfield  {journal} {\bibinfo  {journal} {Appl. Phys. Lett.}\ }\textbf
  {\bibinfo {volume} {73}},\ \bibinfo {pages} {1691} (\bibinfo {year}
  {1998})}\BibitemShut {NoStop}%
\bibitem [{\citenamefont {Lai}\ \emph {et~al.}(2002)\citenamefont {Lai},
  \citenamefont {Hsu}, \citenamefont {Chang}, \citenamefont {Tseng},
  \citenamefont {Lee}, \citenamefont {Chuo},\ and\ \citenamefont
  {Chyi}}]{Lai2002}%
  \BibitemOpen
  \bibfield  {author} {\bibinfo {author} {\bibfnamefont {C.~Y.}\ \bibnamefont
  {Lai}}, \bibinfo {author} {\bibfnamefont {T.~M.}\ \bibnamefont {Hsu}},
  \bibinfo {author} {\bibfnamefont {W.-H.}\ \bibnamefont {Chang}}, \bibinfo
  {author} {\bibfnamefont {K.-U.}\ \bibnamefont {Tseng}}, \bibinfo {author}
  {\bibfnamefont {C.-M.}\ \bibnamefont {Lee}}, \bibinfo {author} {\bibfnamefont
  {C.-C.}\ \bibnamefont {Chuo}}, \ and\ \bibinfo {author} {\bibfnamefont
  {J.-I.}\ \bibnamefont {Chyi}},\ }\bibfield  {title} {\enquote {\bibinfo
  {title} {{Direct measurement of piezoelectric field in
  In$_{0.23}$Ga$_{0.77}$N/GaN multiple quantum wells by electrotransmission
  spectroscopy}},}\ }\href {\doibase 10.1063/1.1426237} {\bibfield  {journal}
  {\bibinfo  {journal} {J. Appl. Phys.}\ }\textbf {\bibinfo {volume} {91}},\
  \bibinfo {pages} {531} (\bibinfo {year} {2002})}\BibitemShut {NoStop}%
\bibitem [{\citenamefont {Cherns}\ \emph {et~al.}(1999)\citenamefont {Cherns},
  \citenamefont {Barnard},\ and\ \citenamefont {Ponce}}]{Cherns1999}%
  \BibitemOpen
  \bibfield  {author} {\bibinfo {author} {\bibfnamefont {D.}~\bibnamefont
  {Cherns}}, \bibinfo {author} {\bibfnamefont {J.}~\bibnamefont {Barnard}}, \
  and\ \bibinfo {author} {\bibfnamefont {F.~A.}\ \bibnamefont {Ponce}},\
  }\bibfield  {title} {\enquote {\bibinfo {title} {{Measurement of the
  piezoelectric field across strained InGaN/GaN layers by electron
  holography}},}\ }\href {\doibase 10.1016/S0038-1098(99)00130-1} {\bibfield
  {journal} {\bibinfo  {journal} {Solid State Commun.}\ }\textbf {\bibinfo
  {volume} {111}},\ \bibinfo {pages} {281} (\bibinfo {year}
  {1999})}\BibitemShut {NoStop}%
\bibitem [{\citenamefont {Barnard}\ and\ \citenamefont
  {Cherns}(2000)}]{Barnard2000}%
  \BibitemOpen
  \bibfield  {author} {\bibinfo {author} {\bibfnamefont {J.~S.}\ \bibnamefont
  {Barnard}}\ and\ \bibinfo {author} {\bibfnamefont {D.}~\bibnamefont
  {Cherns}},\ }\bibfield  {title} {\enquote {\bibinfo {title} {{Direct
  observation of piezoelectric fields in GaN/ InGaN/GaN strained quantum
  wells}},}\ }\href {http://www.ncbi.nlm.nih.gov/pubmed/11108051} {\bibfield
  {journal} {\bibinfo  {journal} {J. Electron Microsc.}\ }\textbf {\bibinfo
  {volume} {49}},\ \bibinfo {pages} {281} (\bibinfo {year} {2000})}\BibitemShut
  {NoStop}%
\bibitem [{\citenamefont {Jia}\ \emph {et~al.}(2001)\citenamefont {Jia},
  \citenamefont {Yu}, \citenamefont {Keogh}, \citenamefont {Asbeck},
  \citenamefont {Miraglia}, \citenamefont {Roskowski},\ and\ \citenamefont
  {Davis}}]{Jia2001}%
  \BibitemOpen
  \bibfield  {author} {\bibinfo {author} {\bibfnamefont {L.}~\bibnamefont
  {Jia}}, \bibinfo {author} {\bibfnamefont {E.~T.}\ \bibnamefont {Yu}},
  \bibinfo {author} {\bibfnamefont {D.}~\bibnamefont {Keogh}}, \bibinfo
  {author} {\bibfnamefont {P.~M.}\ \bibnamefont {Asbeck}}, \bibinfo {author}
  {\bibfnamefont {P.}~\bibnamefont {Miraglia}}, \bibinfo {author}
  {\bibfnamefont {A.}~\bibnamefont {Roskowski}}, \ and\ \bibinfo {author}
  {\bibfnamefont {R.~F.}\ \bibnamefont {Davis}},\ }\bibfield  {title} {\enquote
  {\bibinfo {title} {{Polarization charges and polarization-induced barriers in
  Al$_x$Ga$_{1-x}$N/GaN and In$_y$Ga$_{1-y}$N/GaN heterostructures}},}\ }\href
  {\doibase 10.1063/1.1412594} {\bibfield  {journal} {\bibinfo  {journal}
  {Appl. Phys. Lett.}\ }\textbf {\bibinfo {volume} {79}},\ \bibinfo {pages}
  {2916} (\bibinfo {year} {2001})}\BibitemShut {NoStop}%
\bibitem [{\citenamefont {Zhang}\ \emph {et~al.}(2004)\citenamefont {Zhang},
  \citenamefont {Liu},\ and\ \citenamefont {Chua}}]{Zhang2004}%
  \BibitemOpen
  \bibfield  {author} {\bibinfo {author} {\bibfnamefont {X.~H.}\ \bibnamefont
  {Zhang}}, \bibinfo {author} {\bibfnamefont {W.}~\bibnamefont {Liu}}, \ and\
  \bibinfo {author} {\bibfnamefont {S.~J.}\ \bibnamefont {Chua}},\ }\bibfield
  {title} {\enquote {\bibinfo {title} {{Optical transitions in InGaN/GaN
  quantum wells: effects of the piezoelectric field}},}\ }\href {\doibase
  10.1016/j.jcrysgro.2004.04.084} {\bibfield  {journal} {\bibinfo  {journal}
  {J. Cryst. Growth}\ }\textbf {\bibinfo {volume} {268}},\ \bibinfo {pages}
  {521} (\bibinfo {year} {2004})}\BibitemShut {NoStop}%
\bibitem [{\citenamefont {Zhang}\ \emph {et~al.}(2000)\citenamefont {Zhang},
  \citenamefont {Smorchkova}, \citenamefont {Elsass}, \citenamefont {Keller},
  \citenamefont {Ibbetson}, \citenamefont {Denbaars}, \citenamefont {Mishra},\
  and\ \citenamefont {Singh}}]{Zhang2000}%
  \BibitemOpen
  \bibfield  {author} {\bibinfo {author} {\bibfnamefont {Y.}~\bibnamefont
  {Zhang}}, \bibinfo {author} {\bibfnamefont {I.~P.}\ \bibnamefont
  {Smorchkova}}, \bibinfo {author} {\bibfnamefont {C.~R.}\ \bibnamefont
  {Elsass}}, \bibinfo {author} {\bibfnamefont {S.}~\bibnamefont {Keller}},
  \bibinfo {author} {\bibfnamefont {J.~P.}\ \bibnamefont {Ibbetson}}, \bibinfo
  {author} {\bibfnamefont {S.}~\bibnamefont {Denbaars}}, \bibinfo {author}
  {\bibfnamefont {U.~K.}\ \bibnamefont {Mishra}}, \ and\ \bibinfo {author}
  {\bibfnamefont {J.}~\bibnamefont {Singh}},\ }\bibfield  {title} {\enquote
  {\bibinfo {title} {{Charge control and mobility in AlGaN/GaN transistors:
  Experimental and theoretical studies}},}\ }\href {\doibase 10.1063/1.373483}
  {\bibfield  {journal} {\bibinfo  {journal} {J. of Appl. Phys.}\ }\textbf
  {\bibinfo {volume} {87}},\ \bibinfo {pages} {7981} (\bibinfo {year}
  {2000})}\BibitemShut {NoStop}%
\bibitem [{\citenamefont {Davidsson}\ \emph {et~al.}(2004)\citenamefont
  {Davidsson}, \citenamefont {Gurusinghe}, \citenamefont {Andersson},\ and\
  \citenamefont {Zirath}}]{Davidsson2004}%
  \BibitemOpen
  \bibfield  {author} {\bibinfo {author} {\bibfnamefont {S.~K.}\ \bibnamefont
  {Davidsson}}, \bibinfo {author} {\bibfnamefont {M.}~\bibnamefont
  {Gurusinghe}}, \bibinfo {author} {\bibfnamefont {T.~G.}\ \bibnamefont
  {Andersson}}, \ and\ \bibinfo {author} {\bibfnamefont {H.}~\bibnamefont
  {Zirath}},\ }\bibfield  {title} {\enquote {\bibinfo {title} {{The influence
  of composition and unintentional doping on the two-dimensional electron gas
  density in AlGaN/GaN heterostructures}},}\ }\href {\doibase
  10.1007/s11664-004-0199-2} {\bibfield  {journal} {\bibinfo  {journal} {J. of
  Electron. Mater.}\ }\textbf {\bibinfo {volume} {33}},\ \bibinfo {pages} {440}
  (\bibinfo {year} {2004})}\BibitemShut {NoStop}%
\bibitem [{\citenamefont {Mart\'{i}nez-Criado}\ \emph
  {et~al.}(2001)\citenamefont {Mart\'{i}nez-Criado}, \citenamefont {Cros},
  \citenamefont {Cantarero}, \citenamefont {Ambacher}, \citenamefont {Miskys},
  \citenamefont {Dimitrov}, \citenamefont {Stutzmann}, \citenamefont {Smart},\
  and\ \citenamefont {Shealy}}]{Martnez-Criado2001}%
  \BibitemOpen
  \bibfield  {author} {\bibinfo {author} {\bibfnamefont {G.}~\bibnamefont
  {Mart\'{i}nez-Criado}}, \bibinfo {author} {\bibfnamefont {A.}~\bibnamefont
  {Cros}}, \bibinfo {author} {\bibfnamefont {A.}~\bibnamefont {Cantarero}},
  \bibinfo {author} {\bibfnamefont {O.}~\bibnamefont {Ambacher}}, \bibinfo
  {author} {\bibfnamefont {C.~R.}\ \bibnamefont {Miskys}}, \bibinfo {author}
  {\bibfnamefont {R.}~\bibnamefont {Dimitrov}}, \bibinfo {author}
  {\bibfnamefont {M.}~\bibnamefont {Stutzmann}}, \bibinfo {author}
  {\bibfnamefont {J.}~\bibnamefont {Smart}}, \ and\ \bibinfo {author}
  {\bibfnamefont {J.~R.}\ \bibnamefont {Shealy}},\ }\bibfield  {title}
  {\enquote {\bibinfo {title} {{Residual strain effects on the two-dimensional
  electron gas concentration of AlGaN/GaN heterostructures}},}\ }\href
  {\doibase 10.1063/1.1408268} {\bibfield  {journal} {\bibinfo  {journal} {J.
  Appl. Phys.}\ }\textbf {\bibinfo {volume} {90}},\ \bibinfo {pages} {4735}
  (\bibinfo {year} {2001})}\BibitemShut {NoStop}%
\bibitem [{\citenamefont {Binari}\ \emph {et~al.}(1997)\citenamefont {Binari},
  \citenamefont {Redwing}, \citenamefont {Kelner},\ and\ \citenamefont
  {Kruppa}}]{Binari1997}%
  \BibitemOpen
  \bibfield  {author} {\bibinfo {author} {\bibfnamefont {S.~C.}\ \bibnamefont
  {Binari}}, \bibinfo {author} {\bibfnamefont {J.~M.}\ \bibnamefont {Redwing}},
  \bibinfo {author} {\bibfnamefont {G.}~\bibnamefont {Kelner}}, \ and\ \bibinfo
  {author} {\bibfnamefont {W.}~\bibnamefont {Kruppa}},\ }\bibfield  {title}
  {\enquote {\bibinfo {title} {{AlGaN/GaN HEMTs grown on SiC substrates}},}\
  }\href {\doibase 10.1049/el:19970122} {\bibfield  {journal} {\bibinfo
  {journal} {Electron. Lett.}\ }\textbf {\bibinfo {volume} {33}},\ \bibinfo
  {pages} {242} (\bibinfo {year} {1997})}\BibitemShut {NoStop}%
\bibitem [{\citenamefont {Wu}\ \emph {et~al.}(1996)\citenamefont {Wu},
  \citenamefont {Keller}, \citenamefont {Keller}, \citenamefont {Kapolnek},
  \citenamefont {Kozodoy}, \citenamefont {Denbaars},\ and\ \citenamefont
  {Mishra}}]{Wu1996}%
  \BibitemOpen
  \bibfield  {author} {\bibinfo {author} {\bibfnamefont {Y.-F.}\ \bibnamefont
  {Wu}}, \bibinfo {author} {\bibfnamefont {B.~P.}\ \bibnamefont {Keller}},
  \bibinfo {author} {\bibfnamefont {S.}~\bibnamefont {Keller}}, \bibinfo
  {author} {\bibfnamefont {D.}~\bibnamefont {Kapolnek}}, \bibinfo {author}
  {\bibfnamefont {P.}~\bibnamefont {Kozodoy}}, \bibinfo {author} {\bibfnamefont
  {S.~P.}\ \bibnamefont {Denbaars}}, \ and\ \bibinfo {author} {\bibfnamefont
  {U.~K.}\ \bibnamefont {Mishra}},\ }\bibfield  {title} {\enquote {\bibinfo
  {title} {{Very high breakdown voltage and large transconductance realized on
  GaN heterojunction field effect transistors}},}\ }\href {\doibase
  10.1063/1.117607} {\bibfield  {journal} {\bibinfo  {journal} {Appl. Phys.
  Lett.}\ }\textbf {\bibinfo {volume} {69}},\ \bibinfo {pages} {1438} (\bibinfo
  {year} {1996})}\BibitemShut {NoStop}%
\bibitem [{\citenamefont {Nguyen}\ \emph {et~al.}(1998)\citenamefont {Nguyen},
  \citenamefont {Nguyen}, \citenamefont {Le},\ and\ \citenamefont
  {Grider}}]{Nguyen1998}%
  \BibitemOpen
  \bibfield  {author} {\bibinfo {author} {\bibfnamefont {C.}~\bibnamefont
  {Nguyen}}, \bibinfo {author} {\bibfnamefont {N.~X.}\ \bibnamefont {Nguyen}},
  \bibinfo {author} {\bibfnamefont {M.}~\bibnamefont {Le}}, \ and\ \bibinfo
  {author} {\bibfnamefont {D.~E.}\ \bibnamefont {Grider}},\ }\bibfield  {title}
  {\enquote {\bibinfo {title} {{High performance GaN/AlGaN MODFETs grown by
  RF-assisted MBE}},}\ }\href {\doibase 10.1049/el:19980198} {\bibfield
  {journal} {\bibinfo  {journal} {Electron. Lett.}\ }\textbf {\bibinfo {volume}
  {34}},\ \bibinfo {pages} {309} (\bibinfo {year} {1998})}\BibitemShut
  {NoStop}%
\bibitem [{\citenamefont {Wong}\ \emph {et~al.}(1998)\citenamefont {Wong},
  \citenamefont {Cai}, \citenamefont {Li}, \citenamefont {Wang}, \citenamefont
  {Jiang},\ and\ \citenamefont {Chen}}]{Wong1998}%
  \BibitemOpen
  \bibfield  {author} {\bibinfo {author} {\bibfnamefont {L.~W.}\ \bibnamefont
  {Wong}}, \bibinfo {author} {\bibfnamefont {S.~J.}\ \bibnamefont {Cai}},
  \bibinfo {author} {\bibfnamefont {R.}~\bibnamefont {Li}}, \bibinfo {author}
  {\bibfnamefont {K.}~\bibnamefont {Wang}}, \bibinfo {author} {\bibfnamefont
  {H.~W.}\ \bibnamefont {Jiang}}, \ and\ \bibinfo {author} {\bibfnamefont
  {M.}~\bibnamefont {Chen}},\ }\bibfield  {title} {\enquote {\bibinfo {title}
  {{Magnetotransport study on the two-dimensional electron gas in AlGaN/GaN
  heterostructures}},}\ }\href {\doibase 10.1063/1.121954} {\bibfield
  {journal} {\bibinfo  {journal} {Appl. Phys. Lett.}\ }\textbf {\bibinfo
  {volume} {73}},\ \bibinfo {pages} {1391} (\bibinfo {year}
  {1998})}\BibitemShut {NoStop}%
\bibitem [{\citenamefont {K\"{o}hler}\ \emph {et~al.}(2011)\citenamefont
  {K\"{o}hler}, \citenamefont {M\"{u}ller}, \citenamefont {Waltereit},
  \citenamefont {Pletschen}, \citenamefont {Polyakov}, \citenamefont {Lim},
  \citenamefont {Kirste}, \citenamefont {Menner}, \citenamefont {Br\"{u}ckner},
  \citenamefont {Ambacher}, \citenamefont {Buchheim},\ and\ \citenamefont
  {Goldhahn}}]{Kohler2011}%
  \BibitemOpen
  \bibfield  {author} {\bibinfo {author} {\bibfnamefont {K.}~\bibnamefont
  {K\"{o}hler}}, \bibinfo {author} {\bibfnamefont {S.}~\bibnamefont
  {M\"{u}ller}}, \bibinfo {author} {\bibfnamefont {P.}~\bibnamefont
  {Waltereit}}, \bibinfo {author} {\bibfnamefont {W.}~\bibnamefont
  {Pletschen}}, \bibinfo {author} {\bibfnamefont {V.}~\bibnamefont {Polyakov}},
  \bibinfo {author} {\bibfnamefont {T.}~\bibnamefont {Lim}}, \bibinfo {author}
  {\bibfnamefont {L.}~\bibnamefont {Kirste}}, \bibinfo {author} {\bibfnamefont
  {H.~P.}\ \bibnamefont {Menner}}, \bibinfo {author} {\bibfnamefont
  {P.}~\bibnamefont {Br\"{u}ckner}}, \bibinfo {author} {\bibfnamefont
  {O.}~\bibnamefont {Ambacher}}, \bibinfo {author} {\bibfnamefont
  {C.}~\bibnamefont {Buchheim}}, \ and\ \bibinfo {author} {\bibfnamefont
  {R.}~\bibnamefont {Goldhahn}},\ }\bibfield  {title} {\enquote {\bibinfo
  {title} {{Electrical properties of Al$_x$Ga$_{1-x}$N/GaN heterostructures
  with low Al content}},}\ }\href {\doibase 10.1063/1.3553866} {\bibfield
  {journal} {\bibinfo  {journal} {J. Appl. Phys.}\ }\textbf {\bibinfo {volume}
  {109}},\ \bibinfo {pages} {053705} (\bibinfo {year} {2011})}\BibitemShut
  {NoStop}%
\bibitem [{\citenamefont {K\"{o}hler}\ \emph {et~al.}(2009)\citenamefont
  {K\"{o}hler}, \citenamefont {M\"{u}ller}, \citenamefont {Waltereit},
  \citenamefont {Kirste}, \citenamefont {Menner}, \citenamefont {Bronner},\
  and\ \citenamefont {Quay}}]{Kohler2009}%
  \BibitemOpen
  \bibfield  {author} {\bibinfo {author} {\bibfnamefont {K.}~\bibnamefont
  {K\"{o}hler}}, \bibinfo {author} {\bibfnamefont {S}~\bibnamefont
  {M\"{u}ller}}, \bibinfo {author} {\bibfnamefont {Patrick}\ \bibnamefont
  {Waltereit}}, \bibinfo {author} {\bibfnamefont {L.}~\bibnamefont {Kirste}},
  \bibinfo {author} {\bibfnamefont {H.~P.}\ \bibnamefont {Menner}}, \bibinfo
  {author} {\bibfnamefont {W.}~\bibnamefont {Bronner}}, \ and\ \bibinfo
  {author} {\bibfnamefont {R\"{u}diger}\ \bibnamefont {Quay}},\ }\bibfield
  {title} {\enquote {\bibinfo {title} {{Growth and electrical properties
  applications and materials science of Al$_x$Ga$_{1-x}$N/GaN heterostructures
  with different Al-content}},}\ }\href {\doibase 10.1002/pssa.200925168}
  {\bibfield  {journal} {\bibinfo  {journal} {Phys. Stat. Sol. (a)}\ }\textbf
  {\bibinfo {volume} {206}},\ \bibinfo {pages} {2652} (\bibinfo {year}
  {2009})}\BibitemShut {NoStop}%
\bibitem [{\citenamefont {Nakajima}\ \emph {et~al.}(2010)\citenamefont
  {Nakajima}, \citenamefont {Sumida}, \citenamefont {Dhyani}, \citenamefont
  {Kawai},\ and\ \citenamefont {Narayanan}}]{Nakajima2010}%
  \BibitemOpen
  \bibfield  {author} {\bibinfo {author} {\bibfnamefont {A.}~\bibnamefont
  {Nakajima}}, \bibinfo {author} {\bibfnamefont {Y.}~\bibnamefont {Sumida}},
  \bibinfo {author} {\bibfnamefont {M.~H.}\ \bibnamefont {Dhyani}}, \bibinfo
  {author} {\bibfnamefont {H.}~\bibnamefont {Kawai}}, \ and\ \bibinfo {author}
  {\bibfnamefont {E.~M.~S.}\ \bibnamefont {Narayanan}},\ }\bibfield  {title}
  {\enquote {\bibinfo {title} {{High Density Two-Dimensional Hole Gas Induced
  by Negative Polarization at GaN/AlGaN Heterointerface}},}\ }\href {\doibase
  10.1143/APEX.3.121004} {\bibfield  {journal} {\bibinfo  {journal} {Appl.
  Phys. Express}\ }\textbf {\bibinfo {volume} {3}},\ \bibinfo {pages} {121004}
  (\bibinfo {year} {2010})}\BibitemShut {NoStop}%
\bibitem [{\citenamefont {Burm}\ \emph {et~al.}(1996)\citenamefont {Burm},
  \citenamefont {Schaff}, \citenamefont {Eastman}, \citenamefont {Amano},\ and\
  \citenamefont {Akasaki}}]{Burm1996}%
  \BibitemOpen
  \bibfield  {author} {\bibinfo {author} {\bibfnamefont {J.}~\bibnamefont
  {Burm}}, \bibinfo {author} {\bibfnamefont {W.~J.}\ \bibnamefont {Schaff}},
  \bibinfo {author} {\bibfnamefont {L.~F.}\ \bibnamefont {Eastman}}, \bibinfo
  {author} {\bibfnamefont {H.}~\bibnamefont {Amano}}, \ and\ \bibinfo {author}
  {\bibfnamefont {I.}~\bibnamefont {Akasaki}},\ }\bibfield  {title} {\enquote
  {\bibinfo {title} {{75 $\text{\AA}$ GaN channel modulation doped field effect
  transistors}},}\ }\href {\doibase 10.1063/1.116345} {\bibfield  {journal}
  {\bibinfo  {journal} {Appl. Phys. Lett.}\ }\textbf {\bibinfo {volume} {68}},\
  \bibinfo {pages} {2849} (\bibinfo {year} {1996})}\BibitemShut {NoStop}%
\bibitem [{\citenamefont {Wang}\ \emph {et~al.}(1999)\citenamefont {Wang},
  \citenamefont {Ohno}, \citenamefont {Lachab}, \citenamefont {Nakagawa},
  \citenamefont {Shirahama}, \citenamefont {Sakai},\ and\ \citenamefont
  {Ohno}}]{Wang1999}%
  \BibitemOpen
  \bibfield  {author} {\bibinfo {author} {\bibfnamefont {T.}~\bibnamefont
  {Wang}}, \bibinfo {author} {\bibfnamefont {Y.}~\bibnamefont {Ohno}}, \bibinfo
  {author} {\bibfnamefont {M.}~\bibnamefont {Lachab}}, \bibinfo {author}
  {\bibfnamefont {D.}~\bibnamefont {Nakagawa}}, \bibinfo {author}
  {\bibfnamefont {T.}~\bibnamefont {Shirahama}}, \bibinfo {author}
  {\bibfnamefont {S.}~\bibnamefont {Sakai}}, \ and\ \bibinfo {author}
  {\bibfnamefont {H.}~\bibnamefont {Ohno}},\ }\bibfield  {title} {\enquote
  {\bibinfo {title} {{Electron mobility exceeding 10$^4$ cm$^2$/V s in an
  AlGaN/GaN heterostructure grown on a sapphire substrate}},}\ }\href {\doibase
  10.1063/1.124151} {\bibfield  {journal} {\bibinfo  {journal} {Appl. Phys.
  Lett.}\ }\textbf {\bibinfo {volume} {74}},\ \bibinfo {pages} {3531} (\bibinfo
  {year} {1999})}\BibitemShut {NoStop}%
\bibitem [{\citenamefont {Liu}\ \emph {et~al.}(2006)\citenamefont {Liu},
  \citenamefont {Kauser}, \citenamefont {Schroepfer}, \citenamefont {Ruden},
  \citenamefont {Xie}, \citenamefont {Moon}, \citenamefont {Onojima},
  \citenamefont {Morko\c{c}}, \citenamefont {Son},\ and\ \citenamefont
  {Nathan}}]{Liu2006}%
  \BibitemOpen
  \bibfield  {author} {\bibinfo {author} {\bibfnamefont {Y.}~\bibnamefont
  {Liu}}, \bibinfo {author} {\bibfnamefont {M.~Z.}\ \bibnamefont {Kauser}},
  \bibinfo {author} {\bibfnamefont {D.~D.}\ \bibnamefont {Schroepfer}},
  \bibinfo {author} {\bibfnamefont {P.~P.}\ \bibnamefont {Ruden}}, \bibinfo
  {author} {\bibfnamefont {J.}~\bibnamefont {Xie}}, \bibinfo {author}
  {\bibfnamefont {Y.~T.}\ \bibnamefont {Moon}}, \bibinfo {author}
  {\bibfnamefont {N.}~\bibnamefont {Onojima}}, \bibinfo {author} {\bibfnamefont
  {H.}~\bibnamefont {Morko\c{c}}}, \bibinfo {author} {\bibfnamefont {K.-A.}\
  \bibnamefont {Son}}, \ and\ \bibinfo {author} {\bibfnamefont {M.~I.}\
  \bibnamefont {Nathan}},\ }\bibfield  {title} {\enquote {\bibinfo {title}
  {{Effect of hydrostatic pressure on the current-voltage characteristics of
  GaN/AlGaN/GaN heterostructure devices}},}\ }\href {\doibase
  10.1063/1.2200742} {\bibfield  {journal} {\bibinfo  {journal} {J. Appl.
  Phys.}\ }\textbf {\bibinfo {volume} {99}},\ \bibinfo {pages} {113706}
  (\bibinfo {year} {2006})}\BibitemShut {NoStop}%
\bibitem [{\citenamefont {Ding}\ \emph {et~al.}(2010)\citenamefont {Ding},
  \citenamefont {Guo}, \citenamefont {Xing}, \citenamefont {Chen},
  \citenamefont {Xu}, \citenamefont {Jia}, \citenamefont {Zhou},\ and\
  \citenamefont {Chen}}]{Ding2010}%
  \BibitemOpen
  \bibfield  {author} {\bibinfo {author} {\bibfnamefont {G.~J.}\ \bibnamefont
  {Ding}}, \bibinfo {author} {\bibfnamefont {L.~W.}\ \bibnamefont {Guo}},
  \bibinfo {author} {\bibfnamefont {Z.~G.}\ \bibnamefont {Xing}}, \bibinfo
  {author} {\bibfnamefont {Y.}~\bibnamefont {Chen}}, \bibinfo {author}
  {\bibfnamefont {P.~Q.}\ \bibnamefont {Xu}}, \bibinfo {author} {\bibfnamefont
  {H.~Q.}\ \bibnamefont {Jia}}, \bibinfo {author} {\bibfnamefont {J.~M.}\
  \bibnamefont {Zhou}}, \ and\ \bibinfo {author} {\bibfnamefont
  {H.}~\bibnamefont {Chen}},\ }\bibfield  {title} {\enquote {\bibinfo {title}
  {{Characterization of different-Al-content AlGaN/GaN heterostructures on
  sapphire}},}\ }\href {\doibase 10.1007/s11433-010-0083-4} {\bibfield
  {journal} {\bibinfo  {journal} {Sci. China Phys. Mech. Astron.}\ }\textbf
  {\bibinfo {volume} {53}},\ \bibinfo {pages} {49--} (\bibinfo {year}
  {2010})}\BibitemShut {NoStop}%
\bibitem [{\citenamefont {Chen}\ \emph {et~al.}(2005)\citenamefont {Chen},
  \citenamefont {Chang}, \citenamefont {Su}, \citenamefont {Wang},
  \citenamefont {Kuo},\ and\ \citenamefont {Shei}}]{Chen2005}%
  \BibitemOpen
  \bibfield  {author} {\bibinfo {author} {\bibfnamefont {W.~S.}\ \bibnamefont
  {Chen}}, \bibinfo {author} {\bibfnamefont {S.~J.}\ \bibnamefont {Chang}},
  \bibinfo {author} {\bibfnamefont {Y.~K.}\ \bibnamefont {Su}}, \bibinfo
  {author} {\bibfnamefont {R.~L.}\ \bibnamefont {Wang}}, \bibinfo {author}
  {\bibfnamefont {C.~H.}\ \bibnamefont {Kuo}}, \ and\ \bibinfo {author}
  {\bibfnamefont {S.~C.}\ \bibnamefont {Shei}},\ }\bibfield  {title} {\enquote
  {\bibinfo {title} {{Al$_x$Ga$_{1-x}$N/GaN heterostructure field effect
  transistors with various Al mole fractions in AlGaN barrier}},}\ }\href
  {\doibase 10.1016/j.jcrysgro.2004.12.007} {\bibfield  {journal} {\bibinfo
  {journal} {J. Cryst. Growth}\ }\textbf {\bibinfo {volume} {275}},\ \bibinfo
  {pages} {398} (\bibinfo {year} {2005})}\BibitemShut {NoStop}%
\bibitem [{\citenamefont {Jeganathan}\ \emph {et~al.}(2003)\citenamefont
  {Jeganathan}, \citenamefont {Ide}, \citenamefont {Shimizu},\ and\
  \citenamefont {Okumura}}]{Jeganathan2003}%
  \BibitemOpen
  \bibfield  {author} {\bibinfo {author} {\bibfnamefont {K.}~\bibnamefont
  {Jeganathan}}, \bibinfo {author} {\bibfnamefont {T.}~\bibnamefont {Ide}},
  \bibinfo {author} {\bibfnamefont {M.}~\bibnamefont {Shimizu}}, \ and\
  \bibinfo {author} {\bibfnamefont {H.}~\bibnamefont {Okumura}},\ }\bibfield
  {title} {\enquote {\bibinfo {title} {{Two-dimensional electron gases induced
  by polarization charges in AlN/GaN heterostructure grown by plasma-assisted
  molecular-beam epitaxy}},}\ }\href {\doibase 10.1063/1.1599979} {\bibfield
  {journal} {\bibinfo  {journal} {J. Appl. Phys.}\ }\textbf {\bibinfo {volume}
  {94}},\ \bibinfo {pages} {3260} (\bibinfo {year} {2003})}\BibitemShut
  {NoStop}%
\bibitem [{\citenamefont {Gaska}\ \emph {et~al.}(1998)\citenamefont {Gaska},
  \citenamefont {Osinsky}, \citenamefont {Yang},\ and\ \citenamefont
  {Shur}}]{Gaska1998}%
  \BibitemOpen
  \bibfield  {author} {\bibinfo {author} {\bibfnamefont {R.}~\bibnamefont
  {Gaska}}, \bibinfo {author} {\bibfnamefont {A.}~\bibnamefont {Osinsky}},
  \bibinfo {author} {\bibfnamefont {J.~W.}\ \bibnamefont {Yang}}, \ and\
  \bibinfo {author} {\bibfnamefont {M.~S.}\ \bibnamefont {Shur}},\ }\bibfield
  {title} {\enquote {\bibinfo {title} {{Self-Heating in High-Power AlGaN-GaN
  HFETs}},}\ }\href {\doibase 10.1109/55.661174} {\bibfield  {journal}
  {\bibinfo  {journal} {IEEE Electr. Device Lett.}\ }\textbf {\bibinfo {volume}
  {19}},\ \bibinfo {pages} {89} (\bibinfo {year} {1998})}\BibitemShut {NoStop}%
\bibitem [{\citenamefont {Asbeck}\ \emph {et~al.}(1997)\citenamefont {Asbeck},
  \citenamefont {Yu}, \citenamefont {Lau}, \citenamefont {Sullivan},
  \citenamefont {{Van Hove}},\ and\ \citenamefont {Redwing}}]{Asbeck1997}%
  \BibitemOpen
  \bibfield  {author} {\bibinfo {author} {\bibfnamefont {P.~M.}\ \bibnamefont
  {Asbeck}}, \bibinfo {author} {\bibfnamefont {E.~T.}\ \bibnamefont {Yu}},
  \bibinfo {author} {\bibfnamefont {S.~S.}\ \bibnamefont {Lau}}, \bibinfo
  {author} {\bibfnamefont {G.~J.}\ \bibnamefont {Sullivan}}, \bibinfo {author}
  {\bibfnamefont {J.}~\bibnamefont {{Van Hove}}}, \ and\ \bibinfo {author}
  {\bibfnamefont {J.}~\bibnamefont {Redwing}},\ }\bibfield  {title} {\enquote
  {\bibinfo {title} {{Piezoelectric charge densities in AlGaN/GaN HFETs}},}\
  }\href {\doibase 10.1049/el:19970843} {\bibfield  {journal} {\bibinfo
  {journal} {Electron. Lett.}\ }\textbf {\bibinfo {volume} {33}},\ \bibinfo
  {pages} {1230} (\bibinfo {year} {1997})}\BibitemShut {NoStop}%
\bibitem [{\citenamefont {Wang}\ \emph {et~al.}(2006)\citenamefont {Wang},
  \citenamefont {Wu},\ and\ \citenamefont {Wu}}]{Wang2006}%
  \BibitemOpen
  \bibfield  {author} {\bibinfo {author} {\bibfnamefont {D.-P.}\ \bibnamefont
  {Wang}}, \bibinfo {author} {\bibfnamefont {C.-C.}\ \bibnamefont {Wu}}, \ and\
  \bibinfo {author} {\bibfnamefont {C.-C.}\ \bibnamefont {Wu}},\ }\bibfield
  {title} {\enquote {\bibinfo {title} {{Determination of polarization charge
  density on interface of AlGaN/GaN heterostructure by electroreflectance}},}\
  }\href {\doibase 10.1063/1.2360909} {\bibfield  {journal} {\bibinfo
  {journal} {Appl. Phys. Lett.}\ }\textbf {\bibinfo {volume} {89}},\ \bibinfo
  {pages} {161903} (\bibinfo {year} {2006})}\BibitemShut {NoStop}%
\bibitem [{\citenamefont {Winzer}\ \emph {et~al.}(2003)\citenamefont {Winzer},
  \citenamefont {Goldhahn}, \citenamefont {Buchheim}, \citenamefont {Ambacher},
  \citenamefont {Link}, \citenamefont {Stutzmann}, \citenamefont {Smorchkova},
  \citenamefont {Mishra},\ and\ \citenamefont {Speck}}]{Winzer2003}%
  \BibitemOpen
  \bibfield  {author} {\bibinfo {author} {\bibfnamefont {A.~T.}\ \bibnamefont
  {Winzer}}, \bibinfo {author} {\bibfnamefont {R.}~\bibnamefont {Goldhahn}},
  \bibinfo {author} {\bibfnamefont {C.}~\bibnamefont {Buchheim}}, \bibinfo
  {author} {\bibfnamefont {O.}~\bibnamefont {Ambacher}}, \bibinfo {author}
  {\bibfnamefont {A.}~\bibnamefont {Link}}, \bibinfo {author} {\bibfnamefont
  {M.}~\bibnamefont {Stutzmann}}, \bibinfo {author} {\bibfnamefont
  {Y.}~\bibnamefont {Smorchkova}}, \bibinfo {author} {\bibfnamefont {U.~K.}\
  \bibnamefont {Mishra}}, \ and\ \bibinfo {author} {\bibfnamefont {J.~S.}\
  \bibnamefont {Speck}},\ }\bibfield  {title} {\enquote {\bibinfo {title}
  {{Photoreflectance studies of N- and Ga-face AlGaN/GaN heterostructures
  confining a polarization induced 2DEG}},}\ }\href {\doibase
  10.1002/pssb.200303351} {\bibfield  {journal} {\bibinfo  {journal} {Phys.
  Stat. Sol. (b)}\ }\textbf {\bibinfo {volume} {240}},\ \bibinfo {pages} {380}
  (\bibinfo {year} {2003})}\BibitemShut {NoStop}%
\bibitem [{\citenamefont {Li}\ \emph {et~al.}(2010)\citenamefont {Li},
  \citenamefont {Cao}, \citenamefont {Xing},\ and\ \citenamefont
  {Jena}}]{Li2010}%
  \BibitemOpen
  \bibfield  {author} {\bibinfo {author} {\bibfnamefont {G.}~\bibnamefont
  {Li}}, \bibinfo {author} {\bibfnamefont {Y.}~\bibnamefont {Cao}}, \bibinfo
  {author} {\bibfnamefont {H.~G.}\ \bibnamefont {Xing}}, \ and\ \bibinfo
  {author} {\bibfnamefont {D.}~\bibnamefont {Jena}},\ }\bibfield  {title}
  {\enquote {\bibinfo {title} {{High mobility two-dimensional electron gases in
  nitride heterostructures with high Al composition AlGaN alloy barriers}},}\
  }\href {\doibase 10.1063/1.3523358} {\bibfield  {journal} {\bibinfo
  {journal} {Appl. Phys. Lett.}\ }\textbf {\bibinfo {volume} {97}},\ \bibinfo
  {pages} {222110} (\bibinfo {year} {2010})}\BibitemShut {NoStop}%
\bibitem [{\citenamefont {Arulkumaran}\ \emph {et~al.}(2003)\citenamefont
  {Arulkumaran}, \citenamefont {Egawa}, \citenamefont {Ishikawa},\ and\
  \citenamefont {Jimbo}}]{Arulkumaran2003}%
  \BibitemOpen
  \bibfield  {author} {\bibinfo {author} {\bibfnamefont {S.}~\bibnamefont
  {Arulkumaran}}, \bibinfo {author} {\bibfnamefont {T.}~\bibnamefont {Egawa}},
  \bibinfo {author} {\bibfnamefont {H.}~\bibnamefont {Ishikawa}}, \ and\
  \bibinfo {author} {\bibfnamefont {T.}~\bibnamefont {Jimbo}},\ }\bibfield
  {title} {\enquote {\bibinfo {title} {{Characterization of
  different-Al-content Al$_x$Ga$_{1-x}$N/GaN heterostructures and
  high-electron-mobility transistors on sapphire}},}\ }\href {\doibase
  10.1116/1.1556398} {\bibfield  {journal} {\bibinfo  {journal} {J. Vac. Sci.
  Technol. B}\ }\textbf {\bibinfo {volume} {21}},\ \bibinfo {pages} {888}
  (\bibinfo {year} {2003})}\BibitemShut {NoStop}%
\bibitem [{\citenamefont {Ambacher}\ \emph {et~al.}(2000)\citenamefont
  {Ambacher}, \citenamefont {Foutz}, \citenamefont {Smart}, \citenamefont
  {Shealy}, \citenamefont {Weimann}, \citenamefont {Chu}, \citenamefont
  {Murphy}, \citenamefont {Sierakowski}, \citenamefont {Schaff}, \citenamefont
  {Eastman}, \citenamefont {Dimitrov}, \citenamefont {Mitchell},\ and\
  \citenamefont {Stutzmann}}]{Ambacher2000}%
  \BibitemOpen
  \bibfield  {author} {\bibinfo {author} {\bibfnamefont {O.}~\bibnamefont
  {Ambacher}}, \bibinfo {author} {\bibfnamefont {B.}~\bibnamefont {Foutz}},
  \bibinfo {author} {\bibfnamefont {J.}~\bibnamefont {Smart}}, \bibinfo
  {author} {\bibfnamefont {J.~R.}\ \bibnamefont {Shealy}}, \bibinfo {author}
  {\bibfnamefont {N.~G.}\ \bibnamefont {Weimann}}, \bibinfo {author}
  {\bibfnamefont {K.}~\bibnamefont {Chu}}, \bibinfo {author} {\bibfnamefont
  {M.}~\bibnamefont {Murphy}}, \bibinfo {author} {\bibfnamefont {A.~J.}\
  \bibnamefont {Sierakowski}}, \bibinfo {author} {\bibfnamefont {W.~J.}\
  \bibnamefont {Schaff}}, \bibinfo {author} {\bibfnamefont {L.~F.}\
  \bibnamefont {Eastman}}, \bibinfo {author} {\bibfnamefont {R.}~\bibnamefont
  {Dimitrov}}, \bibinfo {author} {\bibfnamefont {A.}~\bibnamefont {Mitchell}},
  \ and\ \bibinfo {author} {\bibfnamefont {M.}~\bibnamefont {Stutzmann}},\
  }\bibfield  {title} {\enquote {\bibinfo {title} {{Two dimensional electron
  gases induced by spontaneous and piezoelectric polarization in undoped and
  doped AlGaN/GaN heterostructures}},}\ }\href {\doibase 10.1063/1.371866}
  {\bibfield  {journal} {\bibinfo  {journal} {J. Appl. Phys.}\ }\textbf
  {\bibinfo {volume} {87}},\ \bibinfo {pages} {334} (\bibinfo {year}
  {2000})}\BibitemShut {NoStop}%
\bibitem [{\citenamefont {Franssen}\ \emph {et~al.}(2006)\citenamefont
  {Franssen}, \citenamefont {Plesiewicz}, \citenamefont {Dmowski},
  \citenamefont {Prystawko}, \citenamefont {Suski}, \citenamefont
  {Krupczy\'{n}ski}, \citenamefont {Jachymek}, \citenamefont {Perlin},\ and\
  \citenamefont {Leszczy\'{n}ski}}]{Franssen2006}%
  \BibitemOpen
  \bibfield  {author} {\bibinfo {author} {\bibfnamefont {G.}~\bibnamefont
  {Franssen}}, \bibinfo {author} {\bibfnamefont {J.~A.}\ \bibnamefont
  {Plesiewicz}}, \bibinfo {author} {\bibfnamefont {L.~H.}\ \bibnamefont
  {Dmowski}}, \bibinfo {author} {\bibfnamefont {P.}~\bibnamefont {Prystawko}},
  \bibinfo {author} {\bibfnamefont {T.}~\bibnamefont {Suski}}, \bibinfo
  {author} {\bibfnamefont {W.}~\bibnamefont {Krupczy\'{n}ski}}, \bibinfo
  {author} {\bibfnamefont {R.}~\bibnamefont {Jachymek}}, \bibinfo {author}
  {\bibfnamefont {P.}~\bibnamefont {Perlin}}, \ and\ \bibinfo {author}
  {\bibfnamefont {M.}~\bibnamefont {Leszczy\'{n}ski}},\ }\bibfield  {title}
  {\enquote {\bibinfo {title} {{Hydrostatic pressure dependence of
  polarization-induced interface charge in AlGaN/GaN heterostructures
  determined by means of capacitance-voltage characterization}},}\ }\href
  {\doibase 10.1063/1.2392719} {\bibfield  {journal} {\bibinfo  {journal} {J.
  Appl. Phys.}\ }\textbf {\bibinfo {volume} {100}},\ \bibinfo {pages} {113712}
  (\bibinfo {year} {2006})}\BibitemShut {NoStop}%
\bibitem [{\citenamefont {Grandjean}\ \emph {et~al.}(1999)\citenamefont
  {Grandjean}, \citenamefont {Damilano}, \citenamefont {Dalmasso},
  \citenamefont {Leroux}, \citenamefont {LaŸgt},\ and\ \citenamefont
  {Massies}}]{Grandjean1999}%
  \BibitemOpen
  \bibfield  {author} {\bibinfo {author} {\bibfnamefont {N.}~\bibnamefont
  {Grandjean}}, \bibinfo {author} {\bibfnamefont {B.}~\bibnamefont {Damilano}},
  \bibinfo {author} {\bibfnamefont {S.}~\bibnamefont {Dalmasso}}, \bibinfo
  {author} {\bibfnamefont {M.}~\bibnamefont {Leroux}}, \bibinfo {author}
  {\bibfnamefont {M.}~\bibnamefont {LaŸgt}}, \ and\ \bibinfo {author}
  {\bibfnamefont {J.}~\bibnamefont {Massies}},\ }\bibfield  {title} {\enquote
  {\bibinfo {title} {{Built-in electric-field effects in wurtzite AlGaN/GaN
  quantum wells}},}\ }\href {\doibase 10.1063/1.371241} {\bibfield  {journal}
  {\bibinfo  {journal} {J. Appl. Phys.}\ }\textbf {\bibinfo {volume} {86}},\
  \bibinfo {pages} {3714} (\bibinfo {year} {1999})}\BibitemShut {NoStop}%
\bibitem [{\citenamefont {Cingolani}\ \emph {et~al.}(2000)\citenamefont
  {Cingolani}, \citenamefont {Botchkarev}, \citenamefont {Tang}, \citenamefont
  {Morko\c{c}}, \citenamefont {Traetta}, \citenamefont {Coli}, \citenamefont
  {Lomascolo}, \citenamefont {{Di Carlo}}, \citenamefont {{Della Sala}},\ and\
  \citenamefont {Lugli}}]{Cingolani2000}%
  \BibitemOpen
  \bibfield  {author} {\bibinfo {author} {\bibfnamefont {R.}~\bibnamefont
  {Cingolani}}, \bibinfo {author} {\bibfnamefont {A.}~\bibnamefont
  {Botchkarev}}, \bibinfo {author} {\bibfnamefont {H.}~\bibnamefont {Tang}},
  \bibinfo {author} {\bibfnamefont {H.}~\bibnamefont {Morko\c{c}}}, \bibinfo
  {author} {\bibfnamefont {G.}~\bibnamefont {Traetta}}, \bibinfo {author}
  {\bibfnamefont {G.}~\bibnamefont {Coli}}, \bibinfo {author} {\bibfnamefont
  {M.}~\bibnamefont {Lomascolo}}, \bibinfo {author} {\bibfnamefont
  {A.}~\bibnamefont {{Di Carlo}}}, \bibinfo {author} {\bibfnamefont
  {F.}~\bibnamefont {{Della Sala}}}, \ and\ \bibinfo {author} {\bibfnamefont
  {P}~\bibnamefont {Lugli}},\ }\bibfield  {title} {\enquote {\bibinfo {title}
  {{Spontaneous polarization and piezoelectric field in
  GaN/Al$_{0.15}$Ga$_{0.85}$N quantum wells: Impact on the optical spectra}},}\
  }\href {\doibase 10.1103/PhysRevB.61.2711} {\bibfield  {journal} {\bibinfo
  {journal} {Phys. Rev. B}\ }\textbf {\bibinfo {volume} {61}},\ \bibinfo
  {pages} {2711} (\bibinfo {year} {2000})}\BibitemShut {NoStop}%
\bibitem [{\citenamefont {Chen}\ \emph {et~al.}(2013)\citenamefont {Chen},
  \citenamefont {Li}, \citenamefont {Wang}, \citenamefont {Huang},
  \citenamefont {Rong}, \citenamefont {Sang}, \citenamefont {Xu}, \citenamefont
  {Tang}, \citenamefont {Qin}, \citenamefont {Sumiya}, \citenamefont {Chen},
  \citenamefont {Ge},\ and\ \citenamefont {Shen}}]{Chen2013}%
  \BibitemOpen
  \bibfield  {author} {\bibinfo {author} {\bibfnamefont {G.}~\bibnamefont
  {Chen}}, \bibinfo {author} {\bibfnamefont {Z.~L.}\ \bibnamefont {Li}},
  \bibinfo {author} {\bibfnamefont {X.~Q.}\ \bibnamefont {Wang}}, \bibinfo
  {author} {\bibfnamefont {C.~C.}\ \bibnamefont {Huang}}, \bibinfo {author}
  {\bibfnamefont {X.}~\bibnamefont {Rong}}, \bibinfo {author} {\bibfnamefont
  {L.~W.}\ \bibnamefont {Sang}}, \bibinfo {author} {\bibfnamefont {F.~J.}\
  \bibnamefont {Xu}}, \bibinfo {author} {\bibfnamefont {N.}~\bibnamefont
  {Tang}}, \bibinfo {author} {\bibfnamefont {Z.~X.}\ \bibnamefont {Qin}},
  \bibinfo {author} {\bibfnamefont {M.}~\bibnamefont {Sumiya}}, \bibinfo
  {author} {\bibfnamefont {Y.~H.}\ \bibnamefont {Chen}}, \bibinfo {author}
  {\bibfnamefont {W.~K.}\ \bibnamefont {Ge}}, \ and\ \bibinfo {author}
  {\bibfnamefont {B.}~\bibnamefont {Shen}},\ }\bibfield  {title} {\enquote
  {\bibinfo {title} {{Effect of polarization on intersubband transition in
  AlGaN/GaN multiple quantum wells}},}\ }\href {\doibase 10.1063/1.4807131}
  {\bibfield  {journal} {\bibinfo  {journal} {Appl. Phys. Lett.}\ }\textbf
  {\bibinfo {volume} {102}},\ \bibinfo {pages} {192109} (\bibinfo {year}
  {2013})}\BibitemShut {NoStop}%
\bibitem [{\citenamefont {McAleese}\ \emph {et~al.}(2006)\citenamefont
  {McAleese}, \citenamefont {Costa}, \citenamefont {Graham}, \citenamefont
  {Xiu}, \citenamefont {Barnard}, \citenamefont {Kappers}, \citenamefont
  {Dawson}, \citenamefont {Godfrey},\ and\ \citenamefont
  {Humphreys}}]{McAleese2006}%
  \BibitemOpen
  \bibfield  {author} {\bibinfo {author} {\bibfnamefont {C.}~\bibnamefont
  {McAleese}}, \bibinfo {author} {\bibfnamefont {P.~M. F.~J.}\ \bibnamefont
  {Costa}}, \bibinfo {author} {\bibfnamefont {D.~M.}\ \bibnamefont {Graham}},
  \bibinfo {author} {\bibfnamefont {H.}~\bibnamefont {Xiu}}, \bibinfo {author}
  {\bibfnamefont {J.~S.}\ \bibnamefont {Barnard}}, \bibinfo {author}
  {\bibfnamefont {M.~J.}\ \bibnamefont {Kappers}}, \bibinfo {author}
  {\bibfnamefont {P.}~\bibnamefont {Dawson}}, \bibinfo {author} {\bibfnamefont
  {M.~J.}\ \bibnamefont {Godfrey}}, \ and\ \bibinfo {author} {\bibfnamefont
  {C.~J.}\ \bibnamefont {Humphreys}},\ }\bibfield  {title} {\enquote {\bibinfo
  {title} {{Electric fields in AlGaN/GaN quantum well structures}},}\ }\href
  {\doibase 10.1002/pssb.200565382} {\bibfield  {journal} {\bibinfo  {journal}
  {Phys. Stat. Sol. (b)}\ }\textbf {\bibinfo {volume} {243}},\ \bibinfo {pages}
  {1551} (\bibinfo {year} {2006})}\BibitemShut {NoStop}%
\bibitem [{\citenamefont {Langer}\ \emph {et~al.}(1999)\citenamefont {Langer},
  \citenamefont {Simon}, \citenamefont {Ortiz}, \citenamefont {Pelekanos},
  \citenamefont {Barski}, \citenamefont {Andr\'{e}},\ and\ \citenamefont
  {Godlewski}}]{Langer1999}%
  \BibitemOpen
  \bibfield  {author} {\bibinfo {author} {\bibfnamefont {R.}~\bibnamefont
  {Langer}}, \bibinfo {author} {\bibfnamefont {J.}~\bibnamefont {Simon}},
  \bibinfo {author} {\bibfnamefont {V.}~\bibnamefont {Ortiz}}, \bibinfo
  {author} {\bibfnamefont {N.~T.}\ \bibnamefont {Pelekanos}}, \bibinfo {author}
  {\bibfnamefont {A.}~\bibnamefont {Barski}}, \bibinfo {author} {\bibfnamefont
  {R.}~\bibnamefont {Andr\'{e}}}, \ and\ \bibinfo {author} {\bibfnamefont
  {M.}~\bibnamefont {Godlewski}},\ }\bibfield  {title} {\enquote {\bibinfo
  {title} {{Giant electric fields in unstrained GaN single quantum wells}},}\
  }\href {\doibase 10.1063/1.124193} {\bibfield  {journal} {\bibinfo  {journal}
  {Appl. Phys. Lett.}\ }\textbf {\bibinfo {volume} {74}},\ \bibinfo {pages}
  {3827} (\bibinfo {year} {1999})}\BibitemShut {NoStop}%
\bibitem [{\citenamefont {Leroux}\ \emph {et~al.}(1998)\citenamefont {Leroux},
  \citenamefont {Grandjean}, \citenamefont {La\"{u}gt}, \citenamefont
  {Massies}, \citenamefont {Gil}, \citenamefont {Lefebvre},\ and\ \citenamefont
  {Bigenwald}}]{Leroux1998}%
  \BibitemOpen
  \bibfield  {author} {\bibinfo {author} {\bibfnamefont {M.}~\bibnamefont
  {Leroux}}, \bibinfo {author} {\bibfnamefont {N.}~\bibnamefont {Grandjean}},
  \bibinfo {author} {\bibfnamefont {M.}~\bibnamefont {La\"{u}gt}}, \bibinfo
  {author} {\bibfnamefont {J.}~\bibnamefont {Massies}}, \bibinfo {author}
  {\bibfnamefont {B.}~\bibnamefont {Gil}}, \bibinfo {author} {\bibfnamefont
  {P.}~\bibnamefont {Lefebvre}}, \ and\ \bibinfo {author} {\bibfnamefont
  {P.}~\bibnamefont {Bigenwald}},\ }\bibfield  {title} {\enquote {\bibinfo
  {title} {{Quantum confined Stark effect due to built-in internal polarization
  fields in (Al,Ga)N/GaN quantum wells}},}\ }\href {\doibase
  10.1103/PhysRevB.58.R13371} {\bibfield  {journal} {\bibinfo  {journal} {Phys.
  Rev. B}\ }\textbf {\bibinfo {volume} {58}},\ \bibinfo {pages} {R13371}
  (\bibinfo {year} {1998})}\BibitemShut {NoStop}%
\bibitem [{\citenamefont {Suzuki}\ and\ \citenamefont
  {Iizuka}(1999)}]{Suzuki1999}%
  \BibitemOpen
  \bibfield  {author} {\bibinfo {author} {\bibfnamefont {N.}~\bibnamefont
  {Suzuki}}\ and\ \bibinfo {author} {\bibfnamefont {N.}~\bibnamefont
  {Iizuka}},\ }\bibfield  {title} {\enquote {\bibinfo {title} {{Effect of
  Polarization Field on Intersubband Transition in AlGaN/GaN Quantum Wells}},}\
  }\href {\doibase 10.1143/JJAP.38.L363} {\bibfield  {journal} {\bibinfo
  {journal} {Jpn. J. Appl. Phys.}\ }\textbf {\bibinfo {volume} {38}},\ \bibinfo
  {pages} {L363} (\bibinfo {year} {1999})}\BibitemShut {NoStop}%
\bibitem [{\citenamefont {Esmaeili}\ \emph {et~al.}(2009)\citenamefont
  {Esmaeili}, \citenamefont {Gholami}, \citenamefont {Haratizadeh},
  \citenamefont {Monemar}, \citenamefont {Holtz}, \citenamefont {Kamiyama},
  \citenamefont {Amano},\ and\ \citenamefont {Akasaki}}]{Esmaeili2009}%
  \BibitemOpen
  \bibfield  {author} {\bibinfo {author} {\bibfnamefont {M.}~\bibnamefont
  {Esmaeili}}, \bibinfo {author} {\bibfnamefont {M.}~\bibnamefont {Gholami}},
  \bibinfo {author} {\bibfnamefont {H.}~\bibnamefont {Haratizadeh}}, \bibinfo
  {author} {\bibfnamefont {B.}~\bibnamefont {Monemar}}, \bibinfo {author}
  {\bibfnamefont {P.~O.}\ \bibnamefont {Holtz}}, \bibinfo {author}
  {\bibfnamefont {S.}~\bibnamefont {Kamiyama}}, \bibinfo {author}
  {\bibfnamefont {H.}~\bibnamefont {Amano}}, \ and\ \bibinfo {author}
  {\bibfnamefont {I.}~\bibnamefont {Akasaki}},\ }\bibfield  {title} {\enquote
  {\bibinfo {title} {{Experimental and theoretical investigations of optical
  properties of GaN/AlGaN MQW nanostructures. Impact of built-in polarization
  fields}},}\ }\href {\doibase 10.2478/s11772-009-0010-2} {\bibfield  {journal}
  {\bibinfo  {journal} {Opto-Electron. Rev.}\ }\textbf {\bibinfo {volume}
  {17}},\ \bibinfo {pages} {293} (\bibinfo {year} {2009})}\BibitemShut
  {NoStop}%
\bibitem [{\citenamefont {Esmaeili}\ \emph {et~al.}(2007)\citenamefont
  {Esmaeili}, \citenamefont {Sabooni}, \citenamefont {Haratizadeh},
  \citenamefont {Paskov}, \citenamefont {Monemar}, \citenamefont {Holz},
  \citenamefont {Kamiyama},\ and\ \citenamefont {Iwaya}}]{Esmaeili2007}%
  \BibitemOpen
  \bibfield  {author} {\bibinfo {author} {\bibfnamefont {M.}~\bibnamefont
  {Esmaeili}}, \bibinfo {author} {\bibfnamefont {M.}~\bibnamefont {Sabooni}},
  \bibinfo {author} {\bibfnamefont {H.}~\bibnamefont {Haratizadeh}}, \bibinfo
  {author} {\bibfnamefont {P.~P.}\ \bibnamefont {Paskov}}, \bibinfo {author}
  {\bibfnamefont {B.}~\bibnamefont {Monemar}}, \bibinfo {author} {\bibfnamefont
  {P.~O.}\ \bibnamefont {Holz}}, \bibinfo {author} {\bibfnamefont
  {S.}~\bibnamefont {Kamiyama}}, \ and\ \bibinfo {author} {\bibfnamefont
  {M.}~\bibnamefont {Iwaya}},\ }\bibfield  {title} {\enquote {\bibinfo {title}
  {{Optical properties of GaN/AlGaN QW nanostructures with different well and
  barrier widths}},}\ }\href {\doibase 10.1088/0953-8984/19/35/356218}
  {\bibfield  {journal} {\bibinfo  {journal} {J. Phys: Condens. Mat.}\ }\textbf
  {\bibinfo {volume} {19}},\ \bibinfo {pages} {356218} (\bibinfo {year}
  {2007})}\BibitemShut {NoStop}%
\bibitem [{\citenamefont {Pinos}\ \emph {et~al.}(2008)\citenamefont {Pinos},
  \citenamefont {Marcinkevi\u{c}ius}, \citenamefont {Liu}, \citenamefont
  {Shur}, \citenamefont {Kuok\u{s}tis}, \citenamefont {Tamulaitis},
  \citenamefont {Gaska}, \citenamefont {Yang},\ and\ \citenamefont
  {Sun}}]{Pinos2008}%
  \BibitemOpen
  \bibfield  {author} {\bibinfo {author} {\bibfnamefont {A.}~\bibnamefont
  {Pinos}}, \bibinfo {author} {\bibfnamefont {S.}~\bibnamefont
  {Marcinkevi\u{c}ius}}, \bibinfo {author} {\bibfnamefont {K.}~\bibnamefont
  {Liu}}, \bibinfo {author} {\bibfnamefont {M.~S.}\ \bibnamefont {Shur}},
  \bibinfo {author} {\bibfnamefont {E.}~\bibnamefont {Kuok\u{s}tis}}, \bibinfo
  {author} {\bibfnamefont {G.}~\bibnamefont {Tamulaitis}}, \bibinfo {author}
  {\bibfnamefont {R.}~\bibnamefont {Gaska}}, \bibinfo {author} {\bibfnamefont
  {J.}~\bibnamefont {Yang}}, \ and\ \bibinfo {author} {\bibfnamefont
  {W.}~\bibnamefont {Sun}},\ }\bibfield  {title} {\enquote {\bibinfo {title}
  {{Screening dynamics of intrinsic electric field in AlGaN quantum wells}},}\
  }\href {\doibase 10.1063/1.2857467} {\bibfield  {journal} {\bibinfo
  {journal} {Appl. Phys. Lett.}\ }\textbf {\bibinfo {volume} {92}},\ \bibinfo
  {pages} {061907} (\bibinfo {year} {2008})}\BibitemShut {NoStop}%
\bibitem [{\citenamefont {Marcinkevi\u{c}ius}\ \emph
  {et~al.}(2007)\citenamefont {Marcinkevi\u{c}ius}, \citenamefont {Pinos},
  \citenamefont {Liu}, \citenamefont {Veksler}, \citenamefont {Shur},
  \citenamefont {Zhang},\ and\ \citenamefont {Gaska}}]{Marcinkevicius2007}%
  \BibitemOpen
  \bibfield  {author} {\bibinfo {author} {\bibfnamefont {S.}~\bibnamefont
  {Marcinkevi\u{c}ius}}, \bibinfo {author} {\bibfnamefont {A.}~\bibnamefont
  {Pinos}}, \bibinfo {author} {\bibfnamefont {K.}~\bibnamefont {Liu}}, \bibinfo
  {author} {\bibfnamefont {D.}~\bibnamefont {Veksler}}, \bibinfo {author}
  {\bibfnamefont {M.~S.}\ \bibnamefont {Shur}}, \bibinfo {author}
  {\bibfnamefont {J.}~\bibnamefont {Zhang}}, \ and\ \bibinfo {author}
  {\bibfnamefont {R.}~\bibnamefont {Gaska}},\ }\bibfield  {title} {\enquote
  {\bibinfo {title} {{Intrinsic electric fields in AlGaN quantum wells}},}\
  }\href {\doibase 10.1063/1.2679864} {\bibfield  {journal} {\bibinfo
  {journal} {Appl. Phys. Lett.}\ }\textbf {\bibinfo {volume} {90}},\ \bibinfo
  {pages} {081914} (\bibinfo {year} {2007})}\BibitemShut {NoStop}%
\bibitem [{\citenamefont {Kuokstis}\ \emph {et~al.}(2002)\citenamefont
  {Kuokstis}, \citenamefont {Chen}, \citenamefont {Gaevski}, \citenamefont
  {Sun}, \citenamefont {Yang}, \citenamefont {Simin}, \citenamefont {{Asif
  Khan}}, \citenamefont {Maruska}, \citenamefont {Hill}, \citenamefont {Chou},
  \citenamefont {Gallagher},\ and\ \citenamefont {Chai}}]{Kuokstis2002}%
  \BibitemOpen
  \bibfield  {author} {\bibinfo {author} {\bibfnamefont {E.}~\bibnamefont
  {Kuokstis}}, \bibinfo {author} {\bibfnamefont {C.~Q.}\ \bibnamefont {Chen}},
  \bibinfo {author} {\bibfnamefont {M.~E.}\ \bibnamefont {Gaevski}}, \bibinfo
  {author} {\bibfnamefont {W.~H.}\ \bibnamefont {Sun}}, \bibinfo {author}
  {\bibfnamefont {J.~W.}\ \bibnamefont {Yang}}, \bibinfo {author}
  {\bibfnamefont {G.}~\bibnamefont {Simin}}, \bibinfo {author} {\bibfnamefont
  {M.}~\bibnamefont {{Asif Khan}}}, \bibinfo {author} {\bibfnamefont {H.~P.}\
  \bibnamefont {Maruska}}, \bibinfo {author} {\bibfnamefont {D.~W.}\
  \bibnamefont {Hill}}, \bibinfo {author} {\bibfnamefont {M.~C.}\ \bibnamefont
  {Chou}}, \bibinfo {author} {\bibfnamefont {J.~J.}\ \bibnamefont {Gallagher}},
  \ and\ \bibinfo {author} {\bibfnamefont {B.}~\bibnamefont {Chai}},\
  }\bibfield  {title} {\enquote {\bibinfo {title} {{Polarization effects in
  photoluminescence of $c$- and $m$-plane GaN/AlGaN multiple quantum wells}},}\
  }\href {\doibase 10.1063/1.1524298} {\bibfield  {journal} {\bibinfo
  {journal} {Appl. Phys. Lett.}\ }\textbf {\bibinfo {volume} {81}},\ \bibinfo
  {pages} {4130} (\bibinfo {year} {2002})}\BibitemShut {NoStop}%
\bibitem [{\citenamefont {Ng}\ \emph {et~al.}(2001)\citenamefont {Ng},
  \citenamefont {Harel}, \citenamefont {Chu},\ and\ \citenamefont
  {Cho}}]{Ng2001}%
  \BibitemOpen
  \bibfield  {author} {\bibinfo {author} {\bibfnamefont {H.~M.}\ \bibnamefont
  {Ng}}, \bibinfo {author} {\bibfnamefont {R.}~\bibnamefont {Harel}}, \bibinfo
  {author} {\bibfnamefont {S.~N.~G.}\ \bibnamefont {Chu}}, \ and\ \bibinfo
  {author} {\bibfnamefont {A.~Y.}\ \bibnamefont {Cho}},\ }\bibfield  {title}
  {\enquote {\bibinfo {title} {{The effect of built-in electric field in
  GaN/AlGaN quantum wells with high AIN mole fraction}},}\ }\href {\doibase
  10.1007/s11664-001-0006-2} {\bibfield  {journal} {\bibinfo  {journal} {J.
  Electron. Mater.}\ }\textbf {\bibinfo {volume} {30}},\ \bibinfo {pages} {134}
  (\bibinfo {year} {2001})}\BibitemShut {NoStop}%
\bibitem [{\citenamefont {Im}\ \emph {et~al.}(1998)\citenamefont {Im},
  \citenamefont {Kollmer}, \citenamefont {Off}, \citenamefont {Sohmer},
  \citenamefont {Scholz},\ and\ \citenamefont {Hangleiter}}]{Im1998}%
  \BibitemOpen
  \bibfield  {author} {\bibinfo {author} {\bibfnamefont {J.~S.}\ \bibnamefont
  {Im}}, \bibinfo {author} {\bibfnamefont {H.}~\bibnamefont {Kollmer}},
  \bibinfo {author} {\bibfnamefont {J.}~\bibnamefont {Off}}, \bibinfo {author}
  {\bibfnamefont {A.}~\bibnamefont {Sohmer}}, \bibinfo {author} {\bibfnamefont
  {F.}~\bibnamefont {Scholz}}, \ and\ \bibinfo {author} {\bibfnamefont
  {A.}~\bibnamefont {Hangleiter}},\ }\bibfield  {title} {\enquote {\bibinfo
  {title} {{Reduction of oscillator strength due to piezoelectric fields in
  GaN/Al$_x$Ga$_{1-x}$N quantum wells}},}\ }\href {\doibase
  10.1103/PhysRevB.57.R9435} {\bibfield  {journal} {\bibinfo  {journal} {Phys.
  Rev. B}\ }\textbf {\bibinfo {volume} {57}},\ \bibinfo {pages} {R9435}
  (\bibinfo {year} {1998})}\BibitemShut {NoStop}%
\bibitem [{\citenamefont {Simon}\ \emph {et~al.}(2000)\citenamefont {Simon},
  \citenamefont {Langer}, \citenamefont {Barski},\ and\ \citenamefont
  {Pelekanos}}]{Simon2000}%
  \BibitemOpen
  \bibfield  {author} {\bibinfo {author} {\bibfnamefont {J.}~\bibnamefont
  {Simon}}, \bibinfo {author} {\bibfnamefont {R.}~\bibnamefont {Langer}},
  \bibinfo {author} {\bibfnamefont {A.}~\bibnamefont {Barski}}, \ and\ \bibinfo
  {author} {\bibfnamefont {N.}~\bibnamefont {Pelekanos}},\ }\bibfield  {title}
  {\enquote {\bibinfo {title} {{Spontaneous polarization effects in
  GaN/Al$_x$Ga$_{1-x}$N quantum wells}},}\ }\href {\doibase
  10.1103/PhysRevB.61.7211} {\bibfield  {journal} {\bibinfo  {journal} {Phys.
  Rev. B}\ }\textbf {\bibinfo {volume} {61}},\ \bibinfo {pages} {7211--7214}
  (\bibinfo {year} {2000})}\BibitemShut {NoStop}%
\bibitem [{\citenamefont {Simon}\ \emph {et~al.}(2001)\citenamefont {Simon},
  \citenamefont {Langer}, \citenamefont {Barski}, \citenamefont {Zervos},\ and\
  \citenamefont {Pelekanos}}]{Simon2001}%
  \BibitemOpen
  \bibfield  {author} {\bibinfo {author} {\bibfnamefont {J.}~\bibnamefont
  {Simon}}, \bibinfo {author} {\bibfnamefont {R.}~\bibnamefont {Langer}},
  \bibinfo {author} {\bibfnamefont {A.}~\bibnamefont {Barski}}, \bibinfo
  {author} {\bibfnamefont {M.}~\bibnamefont {Zervos}}, \ and\ \bibinfo {author}
  {\bibfnamefont {N.~T.}\ \bibnamefont {Pelekanos}},\ }\bibfield  {title}
  {\enquote {\bibinfo {title} {{Residual doping effects on the amplitude of
  polarization-induced electric fields in GaN/AlGaN quantum wells}},}\ }\href
  {\doibase 10.1002/1521-396X(200112)188:2<867::AID-PSSA867>3.0.CO;2-K}
  {\bibfield  {journal} {\bibinfo  {journal} {Phys. Stat. Sol. (a)}\ }\textbf
  {\bibinfo {volume} {188}},\ \bibinfo {pages} {867} (\bibinfo {year}
  {2001})}\BibitemShut {NoStop}%
\bibitem [{\citenamefont {Gfr\"{o}rer}\ \emph {et~al.}(2011)\citenamefont
  {Gfr\"{o}rer}, \citenamefont {Schl\"{u}sener}, \citenamefont {H\"{a}rle},
  \citenamefont {Scholz},\ and\ \citenamefont {Hangleiter}}]{Gfrorer2011}%
  \BibitemOpen
  \bibfield  {author} {\bibinfo {author} {\bibfnamefont {O.}~\bibnamefont
  {Gfr\"{o}rer}}, \bibinfo {author} {\bibfnamefont {T.}~\bibnamefont
  {Schl\"{u}sener}}, \bibinfo {author} {\bibfnamefont {V.}~\bibnamefont
  {H\"{a}rle}}, \bibinfo {author} {\bibfnamefont {F.}~\bibnamefont {Scholz}}, \
  and\ \bibinfo {author} {\bibfnamefont {A.}~\bibnamefont {Hangleiter}},\
  }\bibfield  {title} {\enquote {\bibinfo {title} {{Mechanisms of Strain
  Reduction in GaN and AlGaN/GaN Epitaxial Layers}},}\ }\href {\doibase
  10.1557/PROC-449-429} {\bibfield  {journal} {\bibinfo  {journal} {MRS Proc.}\
  }\textbf {\bibinfo {volume} {449}},\ \bibinfo {pages} {429} (\bibinfo {year}
  {2011})}\BibitemShut {NoStop}%
\bibitem [{\citenamefont {Winzer}\ \emph {et~al.}(2005)\citenamefont {Winzer},
  \citenamefont {Goldhahn}, \citenamefont {Gobsch}, \citenamefont {Link},
  \citenamefont {Eickhoff}, \citenamefont {Rossow},\ and\ \citenamefont
  {Hangleiter}}]{Winzer2005}%
  \BibitemOpen
  \bibfield  {author} {\bibinfo {author} {\bibfnamefont {A.~T.}\ \bibnamefont
  {Winzer}}, \bibinfo {author} {\bibfnamefont {R.}~\bibnamefont {Goldhahn}},
  \bibinfo {author} {\bibfnamefont {G.}~\bibnamefont {Gobsch}}, \bibinfo
  {author} {\bibfnamefont {A.}~\bibnamefont {Link}}, \bibinfo {author}
  {\bibfnamefont {M.}~\bibnamefont {Eickhoff}}, \bibinfo {author}
  {\bibfnamefont {U.}~\bibnamefont {Rossow}}, \ and\ \bibinfo {author}
  {\bibfnamefont {A.}~\bibnamefont {Hangleiter}},\ }\bibfield  {title}
  {\enquote {\bibinfo {title} {{Determination of the polarization discontinuity
  at the AlGaN/GaN interface by electroreflectance spectroscopy}},}\ }\href
  {\doibase 10.1063/1.1923748} {\bibfield  {journal} {\bibinfo  {journal}
  {Appl. Phys. Lett.}\ }\textbf {\bibinfo {volume} {86}},\ \bibinfo {pages}
  {1--3} (\bibinfo {year} {2005})}\BibitemShut {NoStop}%
\bibitem [{\citenamefont {Drabi\'{n}ska}\ \emph {et~al.}(2003)\citenamefont
  {Drabi\'{n}ska}, \citenamefont {Korona}, \citenamefont {Bo\.{z}ek},
  \citenamefont {Baranowski}, \citenamefont {Paku\l~a}, \citenamefont
  {Tomaszewicz},\ and\ \citenamefont {Gronkowski}}]{Drabinska2002}%
  \BibitemOpen
  \bibfield  {author} {\bibinfo {author} {\bibfnamefont {A.}~\bibnamefont
  {Drabi\'{n}ska}}, \bibinfo {author} {\bibfnamefont {K.~P.}\ \bibnamefont
  {Korona}}, \bibinfo {author} {\bibfnamefont {R.}~\bibnamefont {Bo\.{z}ek}},
  \bibinfo {author} {\bibfnamefont {J.~M.}\ \bibnamefont {Baranowski}},
  \bibinfo {author} {\bibfnamefont {K.}~\bibnamefont {Paku\l~a}}, \bibinfo
  {author} {\bibfnamefont {T.}~\bibnamefont {Tomaszewicz}}, \ and\ \bibinfo
  {author} {\bibfnamefont {J.}~\bibnamefont {Gronkowski}},\ }\bibfield  {title}
  {\enquote {\bibinfo {title} {{Investigation of 2D Electron Gas on AlGaN/GaN
  Interface by Electroreflectance}},}\ }\href {\doibase 10.1002/pssc.200390055}
  {\bibfield  {journal} {\bibinfo  {journal} {Phys. Stat. Sol. (c)}\ }\textbf
  {\bibinfo {volume} {333}},\ \bibinfo {pages} {329} (\bibinfo {year}
  {2003})}\BibitemShut {NoStop}%
\bibitem [{\citenamefont {Kurtz}\ \emph {et~al.}(2004)\citenamefont {Kurtz},
  \citenamefont {Allerman}, \citenamefont {Koleske}, \citenamefont {Baca},\
  and\ \citenamefont {Briggs}}]{Kurtz2004}%
  \BibitemOpen
  \bibfield  {author} {\bibinfo {author} {\bibfnamefont {S.~R.}\ \bibnamefont
  {Kurtz}}, \bibinfo {author} {\bibfnamefont {A.~A.}\ \bibnamefont {Allerman}},
  \bibinfo {author} {\bibfnamefont {D.~D.}\ \bibnamefont {Koleske}}, \bibinfo
  {author} {\bibfnamefont {A.~G.}\ \bibnamefont {Baca}}, \ and\ \bibinfo
  {author} {\bibfnamefont {R.~D.}\ \bibnamefont {Briggs}},\ }\bibfield  {title}
  {\enquote {\bibinfo {title} {{Electronic properties of the AlGaN/GaN
  heterostructure and two-dimensional electron gas observed by
  electroreflectance}},}\ }\href {\doibase 10.1063/1.1639955} {\bibfield
  {journal} {\bibinfo  {journal} {J. Appl. Phys.}\ }\textbf {\bibinfo {volume}
  {95}},\ \bibinfo {pages} {1888} (\bibinfo {year} {2004})}\BibitemShut
  {NoStop}%
\bibitem [{\citenamefont {Zhou}\ \emph {et~al.}(2002)\citenamefont {Zhou},
  \citenamefont {Shen}, \citenamefont {Someya}, \citenamefont {Yu},
  \citenamefont {Liu}, \citenamefont {Zhou}, \citenamefont {Zhang},
  \citenamefont {Shi}, \citenamefont {Zheng},\ and\ \citenamefont
  {Arakawa}}]{Zhou2002}%
  \BibitemOpen
  \bibfield  {author} {\bibinfo {author} {\bibfnamefont {Y.-G.}\ \bibnamefont
  {Zhou}}, \bibinfo {author} {\bibfnamefont {B.}~\bibnamefont {Shen}}, \bibinfo
  {author} {\bibfnamefont {T.}~\bibnamefont {Someya}}, \bibinfo {author}
  {\bibfnamefont {H.-Q.}\ \bibnamefont {Yu}}, \bibinfo {author} {\bibfnamefont
  {J.}~\bibnamefont {Liu}}, \bibinfo {author} {\bibfnamefont {H.-M.}\
  \bibnamefont {Zhou}}, \bibinfo {author} {\bibfnamefont {R.}~\bibnamefont
  {Zhang}}, \bibinfo {author} {\bibfnamefont {Y.}~\bibnamefont {Shi}}, \bibinfo
  {author} {\bibfnamefont {Y.-D.}\ \bibnamefont {Zheng}}, \ and\ \bibinfo
  {author} {\bibfnamefont {Y.}~\bibnamefont {Arakawa}},\ }\bibfield  {title}
  {\enquote {\bibinfo {title} {{Investigation of the polarization-induced
  charges in modulation-doped Al$_x$Ga$_{1-x}$N/GaN heterostructures through
  capacitanceÐvoltage profiling and simulation}},}\ }\href {\doibase
  10.1143/JJAP.41.2531} {\bibfield  {journal} {\bibinfo  {journal} {Jpn. J.
  Appl. Phys.}\ }\textbf {\bibinfo {volume} {41}},\ \bibinfo {pages} {2531}
  (\bibinfo {year} {2002})}\BibitemShut {NoStop}%
\bibitem [{\citenamefont {Yu}\ \emph {et~al.}(1997)\citenamefont {Yu},
  \citenamefont {Sullivan}, \citenamefont {Asbeck}, \citenamefont {Wang},
  \citenamefont {Qiao},\ and\ \citenamefont {Lau}}]{Yu1997}%
  \BibitemOpen
  \bibfield  {author} {\bibinfo {author} {\bibfnamefont {E.~T.}\ \bibnamefont
  {Yu}}, \bibinfo {author} {\bibfnamefont {G.~J.}\ \bibnamefont {Sullivan}},
  \bibinfo {author} {\bibfnamefont {P.~M.}\ \bibnamefont {Asbeck}}, \bibinfo
  {author} {\bibfnamefont {C.~D.}\ \bibnamefont {Wang}}, \bibinfo {author}
  {\bibfnamefont {D.}~\bibnamefont {Qiao}}, \ and\ \bibinfo {author}
  {\bibfnamefont {S.~S.}\ \bibnamefont {Lau}},\ }\bibfield  {title} {\enquote
  {\bibinfo {title} {{Measurement of piezoelectrically induced charge in
  GaN/AlGaN heterostructure field-effect transistors}},}\ }\href {\doibase
  10.1063/1.120138} {\bibfield  {journal} {\bibinfo  {journal} {Appl. Phys.
  Lett.}\ }\textbf {\bibinfo {volume} {71}},\ \bibinfo {pages} {2794} (\bibinfo
  {year} {1997})}\BibitemShut {NoStop}%
\bibitem [{\citenamefont {Smorchkova}\ \emph {et~al.}(1999)\citenamefont
  {Smorchkova}, \citenamefont {Elsass}, \citenamefont {Ibbetson}, \citenamefont
  {Vetury}, \citenamefont {Heying}, \citenamefont {Fini}, \citenamefont {Haus},
  \citenamefont {DenBaars}, \citenamefont {Speck},\ and\ \citenamefont
  {Mishra}}]{Smorchkova1999}%
  \BibitemOpen
  \bibfield  {author} {\bibinfo {author} {\bibfnamefont {I.~P.}\ \bibnamefont
  {Smorchkova}}, \bibinfo {author} {\bibfnamefont {C.~R.}\ \bibnamefont
  {Elsass}}, \bibinfo {author} {\bibfnamefont {J.~P.}\ \bibnamefont
  {Ibbetson}}, \bibinfo {author} {\bibfnamefont {R.}~\bibnamefont {Vetury}},
  \bibinfo {author} {\bibfnamefont {B.}~\bibnamefont {Heying}}, \bibinfo
  {author} {\bibfnamefont {P.}~\bibnamefont {Fini}}, \bibinfo {author}
  {\bibfnamefont {E.}~\bibnamefont {Haus}}, \bibinfo {author} {\bibfnamefont
  {S.~P.}\ \bibnamefont {DenBaars}}, \bibinfo {author} {\bibfnamefont {J.~S.}\
  \bibnamefont {Speck}}, \ and\ \bibinfo {author} {\bibfnamefont {U.~K.}\
  \bibnamefont {Mishra}},\ }\bibfield  {title} {\enquote {\bibinfo {title}
  {{Polarization-induced charge and electron mobility in AlGaN/GaN
  heterostructures grown by plasma-assisted molecular-beam epitaxy}},}\ }\href
  {\doibase 10.1063/1.371396} {\bibfield  {journal} {\bibinfo  {journal} {J.
  Appl. Phys.}\ }\textbf {\bibinfo {volume} {86}},\ \bibinfo {pages} {4520}
  (\bibinfo {year} {1999})}\BibitemShut {NoStop}%
\bibitem [{\citenamefont {Miller}\ \emph {et~al.}(2002)\citenamefont {Miller},
  \citenamefont {Yu}, \citenamefont {Poblenz}, \citenamefont {Elsass},\ and\
  \citenamefont {Speck}}]{Miller2002}%
  \BibitemOpen
  \bibfield  {author} {\bibinfo {author} {\bibfnamefont {E.~J.}\ \bibnamefont
  {Miller}}, \bibinfo {author} {\bibfnamefont {E.~T.}\ \bibnamefont {Yu}},
  \bibinfo {author} {\bibfnamefont {C.}~\bibnamefont {Poblenz}}, \bibinfo
  {author} {\bibfnamefont {C.}~\bibnamefont {Elsass}}, \ and\ \bibinfo {author}
  {\bibfnamefont {J.~S.}\ \bibnamefont {Speck}},\ }\bibfield  {title} {\enquote
  {\bibinfo {title} {{Direct measurement of the polarization charge in
  AlGaN/GaN heterostructures using capacitanceÐvoltage carrier profiling}},}\
  }\href {\doibase 10.1063/1.1477275} {\bibfield  {journal} {\bibinfo
  {journal} {Appl. Phys. Lett.}\ }\textbf {\bibinfo {volume} {80}},\ \bibinfo
  {pages} {3551} (\bibinfo {year} {2002})}\BibitemShut {NoStop}%
\bibitem [{\citenamefont {Bernardini}\ \emph {et~al.}(1997)\citenamefont
  {Bernardini}, \citenamefont {Fiorentini},\ and\ \citenamefont
  {Vanderbilt}}]{BFV1997}%
  \BibitemOpen
  \bibfield  {author} {\bibinfo {author} {\bibfnamefont {F.}~\bibnamefont
  {Bernardini}}, \bibinfo {author} {\bibfnamefont {V.}~\bibnamefont
  {Fiorentini}}, \ and\ \bibinfo {author} {\bibfnamefont {D.}~\bibnamefont
  {Vanderbilt}},\ }\bibfield  {title} {\enquote {\bibinfo {title} {{Spontaneous
  polarization and piezoelectric constants of III-V nitrides}},}\ }\href
  {\doibase 10.1103/PhysRevB.56.R10024} {\bibfield  {journal} {\bibinfo
  {journal} {Phys. Rev. B}\ }\textbf {\bibinfo {volume} {56}},\ \bibinfo
  {pages} {R10024} (\bibinfo {year} {1997})}\BibitemShut {NoStop}%
\end{thebibliography}%


\begin{thebibliography}{20}
\bibitem{Nye1957}
J. F. Nye (1957). {\it Physical Properties of Crystals}
(pp. 300). London: Oxford University Press.



\bibitem{Martin1972}
R. M. Martin, \textit{Piezoelectricity}, Phys. Rev. B {\bf 5}, 1607 (1972).



\bibitem{Feneberg2007}
M. Feneberg and K. Thonke, \textit{Polarization fields of III-nitrides grown in different crystal orientations}, J. Phys.: Condens. Matter {\bf 19},
403201 (2007).


\bibitem{Vanderbilt2000}
D. Vanderbilt, \textit{Berry-phase theory of proper piezoelectric response}, J. Phys. Chem. Sol. {\bf 61}, 147 (2000).


\bibitem{Jonas2012}
J. L\"{a}hnemann, O. Brandt, U. Jahn,
C. Pf\"{u}ller, C. Roder, P. Dogan, F. Grosse, A. Belabbes,
F. Bechstedt, A. Trampert, and L. Geelhaar, \textit{Direct experimental determination of the spontaneous polarization of GaN}, Phys. Rev. B \textbf{86},
081302 (2012).


\bibitem{BFV1997}
F. Bernardini, V. Fiorentini, D. Vanderbilt, \textit{Spontaneous polarization and piezoelectric constants of III-V nitrides}, Phys. Rev. B {\bf 56},
R10024 (1997).



\bibitem{BFV2001}
F. Bernardini, V. Fiorentini, D. Vanderbilt, \textit{Accurate calculation of polarization-related quantities in semiconductors}, Phys. Rev. B {\bf 63},
193201 (2001).



\bibitem{Bechstedt2000}
F. Bechstedt, U. Grossner, and J. Furthm\"{u}ller, \textit{Dynamics and polarization of group-III nitride lattices: A first-principles study}, Phys. Rev. B
{\bf 62}, 8003 (2000).


\bibitem{KSV1993}
R. D. King-Smith and D. Vanderbilt, \textit{Theory of polarization of crystalline solids}, Phys. Rev. B {\bf 47}, 1651
(1993).



\bibitem{Resta1994}
R. Resta, \textit{Macroscopic polarization in crystalline dielectrics: the geometric phase approach}, Rev. Mod. Phys. {\bf 66}, 899 (1994).



\bibitem{VKS1993}
D. Vanderbilt and R. D. King-Smith, \textit{Electric polarization as a bulk quantity and its relation to surface charge}, Phys. Rev. B {\bf 48}, 4442 (1993).



\bibitem{Resta2007}
R. Resta and D. Vanderbilt, in \textit{Physics of Ferroelectrics:
A Modern Perspective}, edited by K. Rabe, Ch. H. Ahn, and
J.-M. Triscone (Springer-Verlag, Berlin, 2007).



\bibitem{silense}
SiLENSe Physics Summary v5.2 (2011).



\bibitem{Ambacher2000}
 O. Ambacher, B. Foutz, J. Smart,
J. R. Shealy, N. G. Weimann,
 K. Chu, M. Murphy, R. Dimitrov,
A. Mitchell, and M. Stutzmann, \textit{Two dimensional electron gases
induced by spontaneous and piezoelectric polarization in undoped and
doped AlGaN/GaN heterostructures},
 J. Appl. Phys. {\bf 87}, 334
(2000).
 
 

\bibitem{Supp}
See Supplemental Material at [URL will be inserted by publisher] for a
discussion of formal polarization in ZB, 
a table of structural properties and band gaps, a list of experimental
references, equations used to generate curves in Fig.~\ref{Both}, and
equations demonstrating the difference between implementations.


\bibitem{Bennett2012}
 J. W. Bennett, K. F. Garrity, K. M. Rabe, and
D. Vanderbilt, \textit{Hexagonal ABC Semiconductors as
Ferroelectrics},
 Phys. Rev. Lett. {\bf 109}, 167602 (2012).
 
  

\bibitem{HSE06}
J. Heyd, G. E. Scuseria, M. Ernzerhof, \textit{Hybrid functionals based on a screened Coulomb potential}, J. Chem. Phys. {\bf 118},
8207 (2003); {\it ibid.} {\bf 124}, 219906 (2006).



\bibitem{vasp1}
G. Kresse and J. Furthm$\ddot{\rm u}$ller, \textit{Efficient iterative schemes for \text{ab initio} total-energy calculations using a plane-wave basis set}, Phys. Rev. B. {\bf 54},
11169 (1996).



\bibitem{paw}
P. E. Bl\"{o}chl, \textit{Projector augmented-wave method}, Phys. Rev. B 50, 17953 (1994).



\bibitem{mp1976}
H.J. Monkhorst, J. D. Pack, \textit{Special points for Brillouin-zone integrations}, Phys. Rev. B {\bf 13}, 5188 (1976).



\bibitem{Romanov2006}
A. E. Romanov, T. J. Baker, S. Nakamura, and J. S. Speck,
\textit{Strain-induced polarization in wurtzite III-nitride semipolar
layers}, 
 J. Appl. Phys. \textbf{100}, 023522 (2006).
 
 
 

\bibitem{Oclock1973}
G. D. O'Clock and M. T. Duffy, \textit{Acoustic surface wave properties of epitaxially grown aluminum nitride and gallium nitride on sapphire}, Appl. Phys. Lett. \textbf{23}, 55 (1973).


\bibitem{Dubois1999}
M.-A. Dubois and P. Muralt, \textit{Properties of aluminum nitride thin films for piezoelectric transducers and microwave filter applications}, Appl. Phys. Lett. \textbf{74}, 3032 (1999).




\bibitem{Muensit1998}
S. Muensit and I. L. Guy, \textit{The piezoelectric coefficient of gallium nitride thin films}, Appl. Phys. Lett. \textbf{72}, 1896 (1998).


\bibitem{Muensit1999}
S. Muensit, E. M. Goldys, and I. L. Guy, \textit{Shear piezoelectric coefficients of gallium nitride and aluminum nitride}, Appl. Phys. Lett. \textbf{75}, 3965 (1999).


\bibitem{Guy1999}
I. L. Guy, S. Muensit and E. M. Goldys, \textit{Extensional piezoelectric coefficients of gallium nitride and aluminum nitride}, Appl. Phys. Lett. \textbf{75}, 4133 (1999).


\bibitem{Lueng2000}
C. M. Lueng, H. L. W. Chan, C. Surya, and C. L. Choy, \textit{Piezoelectric coefficient of aluminum nitride and gallium nitride}, J. Appl. Phys. \textbf{88}, 5360 (2000).


\bibitem{Ambacher1999}
O. Ambacher, R. Dimitrov, M. Stutzmann, B. E. Foutz, M. J. Murphy,
J. A. Smart, J. R. Shealy, N. G. Weimann, K. Chu, M. Chumbes,
B. Green, A. J. Sierakowski, W. J. Schaff, and L. F. Eastman,
\textit{Two-dimensional electron gases induced by spontaneous and piezoelectric polarization charges in N- and Ga-face AlGaN / GaN heterostructures}, Phys. status solidi (b) \textbf{216}, 381 (1999).



\bibitem{Vurgaftman2003}
I. Vurgaftman and J. R. Meyer, \textit{Band parameters for nitrogen-containing semiconductors}, J. Appl. Phys. \textbf{94}, 3675 (2003).



\bibitem{Hangleiter2003}
A. Hangleiter, F. Hitzel, S. Lahmann, and U. Rossow, \textit{Composition dependence of polarization fields in GaInN/GaN quantum wells},
Appl. Phys. Lett. \textbf{83}, 1169 (2003).



\bibitem{Renner2002}
F. Renner, P. Kiesel, G. H. D\"{o}hler, M. Kneissl, C. G. Van de
Walle, and N. M. Johnson, \textit{Quantitative analysis of the polarization fields and absorption changes in InGaN/GaN quantum wells with electroabsorption spectroscopy}, Appl. Phys. Lett. {\bf 81}, 490 (2002).



\bibitem{Turchinovich2003}
D. Turchinovich, P. Uhd Jepsen, B. S. Monozon, M. Koch, S. Lahmann,
U. Rossow, and A. Hangleiter, \textit{Ultrafast polarization dynamics in biased quantum wells under strong femtosecond optical excitation}, Phys. Rev. B {\bf 68}, 241307 (2003).



\bibitem{Cherns1999}
D. Cherns, J. Barnard, and F. A. Ponce, \textit{Measurement of the piezoelectric field across strained InGaN/GaN layers by electron holography}, Solid State Commun. {\bf 111}, 281 (1999).



\bibitem{Zhang2004}
H. Zhang, E. J. Miller, E. T. Yu, C. Poblenz, and J. S. Speck, \textit{Analysis of interface electronic structure in In$_x$Ga$_{1-x}$N/GaN heterostructures}, J. Vac. Sci. Technol. B \textbf{22}, 2169 (2004).



\bibitem{Barker1973}
A. S. Barker, M. Ilegems, \textit{Infrared Lattice Vibrations and Free-Electron Dispersion in GaN}, Phys. Rev. B {\bf 7}, 743 (1973).



\bibitem{Madelung}
O. Madelung, \textit{Semiconductors: Data Handbook} (Springer
Berlin Heidelberg, 2004).



\bibitem{Wu2012}
Y.-R. Wu, R. Shivaraman, K.-C. Wang, and J. S. Speck, \textit{Analyzing the physical properties of InGaN multiple quantum well light emitting diodes from nano scale structure},
Appl. Phys. Lett. {\bf 101}, 083505 (2012).



\bibitem{Mcbride2014}
P. M. McBride, Q. Yan, and C. G. Van de Walle, \textit{Effects of In profile on simulations of InGaN / GaN multi-quantum-well light-emitting diodes},
Appl. Phys. Lett. \textbf{105}, 083507 (2014).



\bibitem{Grandjean1999}
N. Grandjean, B. Damilano, S. Dalmasso, M. Leroux, M. La\"{u}gt,
and J. Massies, \textit{Built-in electric-field effects in wurtzite AlGaN/GaN quantum wells}, J. Appl. Phys. {\bf 86}, 3714 (1999).



\bibitem{Leroux1999}
M. Leroux, N. Grandjean, J. Massies, B. Gil, P. Lefebvre, and
P. Bigenwald, \textit{Barrier-width dependence of group-III nitrides quantum-well transition energies}, Phys. Rev. B \textbf{60}, 1496 (1999).



\bibitem{Yu1997}
E. T. Yu, G. J. Sullivan, P. M. Asbeck, C. D. Wang, D. Qiao,
and S. S. Lau, \textit{Measurement of piezoelectrically induced charge in GaN/AlGaN heterostructure field-effect transistors}, Appl. Phys. Lett. {\bf 71},2794 (1997).



\bibitem{Zhou2002}
Y. Zhou, B. Shen, T. Someya, H. Yu, J. Liu, H. Zhou, R. Zhang,
Y. Shi, Y. Zheng, and Y. Arakawa, \textit{Investigation of the polarization-induced charges in modulation-doped Al$_x$Ga$_{1-x}$N/GaN heterostructures through capacitance–voltage profiling and simulation}, Jpn. J. Appl. Phys. {\bf 41},
2531 (2002).


\bibitem{Kohler2011}
 K. K\"{o}hler, S. M\"{u}ller, P. Waltereit, W. Pletschen,
 V. Polyakov, T. Lim, L. Kirste, H. P. Menner, P. Br\"{u}ckner,
 O. Ambacher, C. Buchheim, and R. Goldhahn, \textit{Electrical properties of Al$_x$Ga$_{1-x}$N/GaN heterostructures with low Al content}, J. Appl. Phys. {\bf
 109}, 053705 (2011).

 

\bibitem{Davidsson2004}
 S. K. Davidsson, M. Gurusinghe, T. G. Andersson, H. Zirath, \textit{The influence of composition and unintentional doping on the two-dimensional electron gas density in AlGaN/GaN heterostructures},
 J. Electron. Mater. {\bf 33}, 440 (2004).

 

\bibitem{Simon2001}
 J. Simon, R. Langer, A. Barski, M. Zervos, and N. T. Pelekanos, \textit{Residual doping effects on the amplitude of polarization-induced electric fields in GaN/AlGaN quantum wells}, 
 Phys. Stat. Sol. (a) {\bf 188}, 867 (2001).



\bibitem{Arulkumaran2003}
S. Arulkumaran, T. Egawa, H. Ishikawa, and T. Jimbo, \textit{Characterization of different Al content Al$_x$Ga$_{1-x}$N heterostructures and high-electron-mobility transistors on sapphire}, J.
Vac. Sci. Technol. B {\bf 21}, 888 (2003).



\bibitem{Mkhoyan2004}
K. A. Mkhoyan, J. Silcox, Z. Yu, W. J. Schaff, L. F. Eastman,
\textit{Formation of a quasi-two-dimensional electron gas in GaN/Al$_x$Ga$_{1-x}$N heterostructures with diffuse interfaces}, J. Appl. Phys. {\bf 95}, 1843 (2004).



\bibitem{Zhang2000} 
Y. Zhang, I. P. Smorchkova, C. R. Elsass,
 S. Keller, J. P. Ibbetson, S. P. DenBaars, U. K. Mishra, J. Singh, \textit{Charge control and mobility in AlGaN/GaN transistors: Experimental and theoretical studies}, J. Appl. Phys. {\bf 87}, 7981 (2000).

\bibitem{Li2015}
W. Li, S. Sharmin, H. Ilatikhameneh, R. Rahman, Y. Lu, J. Wang,
X. Yan, A. Seabaugh, G. Klimeck, D. Jena, and P. Fay, \textit{Polarization-Engineered III-Nitride Heterojunction Tunnel Field-Effect Transistors}, IEEE
J. Explor. Solid-State Comput. Devices Circuits \textbf{1}, 28
(2015).

 
\end{thebibliography}
%

\end{document}


\title{Correct implementation of polarization constants in wurtzite materials and impact on III-nitrides}
\author{Cyrus E. Dreyer}
\affiliation{Materials Department, University of California,
Santa Barbara, CA 93106-5050}
\affiliation{Department of Physics and Astronomy, Rutgers
University, Piscataway, NJ 08845-0849}
\author{Anderson Janotti}
\altaffiliation{Current address: Materials Science and Engineering, University of Delaware, Newark, Delaware 19716-1501, USA}
\affiliation{Materials Department, University of California,
Santa Barbara, CA 93106-5050}
\author{Chris G. Van de Walle}
\affiliation{Materials Department, University of California,
Santa Barbara, CA 93106-5050}
\author{David Vanderbilt}
\affiliation{Department of Physics and Astronomy, Rutgers
University, Piscataway, NJ 08845-0849}
\date{\today }
\maketitle

{\Large Supplemental Material}

\section{Polarization in Zincblende}

There is an additional subtlety when calculating the formal polarization of the zincblende (ZB) structure that is related to the choice of unit cell. If the primitive unit cell is used for the calculation [with the origin chosen such that the Ga atom is at (-1/8,-1/8,-1/8) and the N at (1/8,1/8,1/8)], then the result for III-nitirides will be what was determined in Ref.~\onlinecite{VKS1993}, Section III E. Specifically, the electronic part of the polarization vanishes, so the contribution simply comes from ${\textbf{P}}_{\text{ion}}$ of Eq.~(2) of the main text:
\begin{equation}
\label{prim}
\textbf{P}_{\text{f}}^{\text{ZB}}=\frac{e\sqrt{2}a_{\text{WZ}}}{\Omega}\left(\frac{1}{4},\frac{1}{4},\frac{1}{4}\right),
\end{equation}
where the $a_{\text{WZ}}$ is the in plane lattice parameter of the wurzite (WZ) material (related to the ZB lattice constant by $\sqrt{2}$). The magnitude in the [111] direction is therefore $\sqrt{3}ea_{\text{WZ}}/2\sqrt{2}\Omega=e\sqrt{3}/2a_{\text{WZ}}^2$ [since the volume of the ZB primitive cell is $\Omega=(\sqrt{2}a_{\text{WZ}})^3/4$]. Though choosing a different origin of the cell may change the value by quanta of polarization, there are no lattice vectors $\textbf{R}$ of the primitive ZB cell [$\sqrt{2}a_{\text{WZ}}$(1/2,1/2,0), $\sqrt{2}a_{\text{WZ}}$(0,1/2,1/2) and $\sqrt{2}a_{\text{WZ}}$(1/2,0,1/2)] that will result in a quantum of polarization $e\textbf{R}/\Omega$ that will take $\textbf{P}_{\text{f}}^{\text{ZB}}$ to zero, and therefore ZB truly has a nonvanishing formal polarization. For the III-nitrides, we list the values of formal polarization for ZB along with WZ and the layered hexagonal (H) structure in Table \ref{ftab} below.

We note that if a conventional eight-atom cubic unit cell is used, the results are misleading. The cubic cell volume is four times that of the primitive cell, $\Omega_c=4\Omega$, and has four ``dipoles'' such as the one in Eq.~(\ref{prim}); therefore, the magnitude of the polarization vector in the [111] direction is four times larger. Equation~(\ref{prim}) now becomes:
\begin{equation}
\textbf{P}_{\text{f}}^{\text{ZB}}(\text{cubic cell})=\frac{e\sqrt{2}a_{\text{WZ}}}{\Omega_c}\left(1,1,1\right) \, .
\end{equation}
However, $\sqrt{2}a_{\text{WZ}}(1,1,1)$ \textit{is} now given by the sum of lattice vectors of the \textit{cubic cell} [$\sqrt{2}a_{\text{WZ}}(1,0,0)$, $\sqrt{2}a_{\text{wz}}(0,1,0)$, and $\sqrt{2}a_{\text{WZ}}(0,0,1)$]; therefore it appears that $\textbf{P}_{\text{f}}^{\text{ZB}}(\text{cubic cell})$ vanishes modulo a quantum of polarization.
The calculation using the primitive cell is the rigorous result, as it is the smallest possible unit cell. For conventional unit cells of larger size, the quantum of polarization becomes smaller compared to the magnitude of the polarization vector, and the true values for the formal polarization cannot be ascertained.

\begin{table}
\caption{\label{ftab}
Calculated formal polarizations, in units of C/m$^2$,
of wurtzite (WZ), zincblende (ZB) and layered hexagonal
(H) GaN, AlN, and InN, all at the relaxed WZ
in-plane lattice constant.}
\begin{ruledtabular}
\begin{tabular}{cccc}
   & $P_\text{f}^{\text{(WZ)}}$   &
     $P_\text{f}^{\text{(ZB)}}$  &
     $P_\text{f}^{\text{(H)}}$ \\
\hline
     GaN & $ 1.31$   & $1.35$     & $0$ \\
     AlN & $1.35$     & $1.44$       & $0$ \\
     InN & $1.03$    & $1.07$       & $0$ \\
\end{tabular}
\end{ruledtabular}
\end{table}


\section{Structural parameters and band gaps calculated with the HSE hybrid functional}
\hspace{1pc}

\begin{table}[h]
\caption{\label{strtab} Parameters for the III-nitrides calculated with HSE.  Experimental data are listed for comparison.}
\begin{ruledtabular}
\begin{tabular}{c c  c c c}
     		& Property	& HSE (this work) 	& Experiment\footnotemark[1]	\\ \hline
     GaN 	& $a$ (\AA)	& 3.205			& 3.189		\\
    		& $c$ (\AA)	& 5.200			& 5.185		\\	
		& $u$ 		& 0.377			& 0.377\footnotemark[2]		\\
		& $E_g$ (eV)	& 3.496	 		& 3.4-3.5		\\ \hline
     AlN 	& $a$ (\AA)	& 3.099			& 3.112		\\
     		& $c$ (\AA)	& 4.959			& 4.982		\\	
		& $u$ 		& 0.382			& 0.382\footnotemark[2]		\\
		& $E_g$ (eV)	& 6.044 			& 6.1-6.3		\\ \hline
     InN 	& $a$ (\AA)	& 3.587			& 3.545		\\
     		& $c$ (\AA)	& 5.762			& 5.703		\\	
		& $u$ 		& 0.380			& --		\\
		& $E_g$ (eV)	& 0.646 		 	& 0.6-0.8		\\
\end{tabular}
\end{ruledtabular}
\footnotetext[1]{From Ref.~\onlinecite{Vurgaftman2003} unless otherwise specified.}
\footnotetext[2]{From Ref.~\onlinecite{Schulz1977}.}
\end{table}

\clearpage

\section{Experimental determination of polarization from the literature}

The experimental data points in Fig.~2 of the main text were taken from various literature studies that were intended to determine the polarization constants in the InGaN/GaN or AlGaN/GaN systems. We list these references in Table \ref{AlGaNHall}, \ref{AlGaNoptical}, \ref{AlGaNCV}, and \ref{InGaN} along with the reported values. Whether the actual measurement was bound charges at an interface or the field in a quantum well, we have converted the reported values to a polarization sheet charge for the purposes of Fig.~2 using the procedure outlined in Sec.~V of the main text.

\begin{table}[h]
\caption{\label{InGaN} Experimental data for GaN/InGaN/GaN quantum wells from optical (if not specified), holography, and CV measurements. In the cases where fields are reported in the reference, the bound charge is determined from the model described in the main text, Sec.~V.}
\begin{ruledtabular}
 \begin{tabular}{ >{\centering\arraybackslash}m{2.5cm} | >{\centering\arraybackslash}m{5cm} | >{\centering\arraybackslash}m{5cm} |>{\centering\arraybackslash}m{5cm} }
 Reference 	& InGaN content & Field ($10^5$ V/cm) & Bound charge ($10^{12} e^-/\text{cm}^2$)		\\
\hline
\onlinecite{Aumer2001}	& 0.08 	& 6.0        	&3.1 \\ \hline
\onlinecite{Li2002}		& 0.10	&13.6	&7.2 \\ \hline
\onlinecite{Jho2001} 	&0.15 	& 21.0	& 11.5 \\ \hline
\onlinecite{Hangleiter2003}	& 0.12, 0.22	& 15.0, 29.0 	& 8.1, 16.5  \\ \hline
\onlinecite{Chichibu1998}	& 0.10 	&3.5	&1.9 \\ \hline
\onlinecite{Lefebvre2001}	& 0.18, 0.15, 0.20	& 24.5, 27.0, 22.0 & 13.6, 14.8, 12.4 \\ \hline
\onlinecite{Kiesel2001}, \onlinecite{Kiesel2002}, \onlinecite{Renner2002}	& 0.07, 0.08, 0.08, 0.08, 0.08, 0.08, 0.09, 0.09, 0.09, 0.09, 0.09 & 10.5, 11.4, 11.1, 11.1, 12.6, 12.3, 12.9, 13.4, 13.7, 14.0, 14.0	&5.5, 6.0, 5.8, 5.8, 6.6, 6.5, 6.8, 7.1, 7.2, 7.4, 7.4 \\ \hline
\onlinecite{Kaplar2004}	& 0.07 	& 9.3	& 4.8 \\ \hline
\onlinecite{Kim2010} 	& 0.15  	& 13.4	& 7.33 \\ \hline
\onlinecite{Franssen2004}  &0.08 	& 11.0	& 5.76 \\ \hline
\onlinecite{Brown2005}	& 0.09	& 19.0	&10.0 \\ \hline
\onlinecite{Thomsen2011}	& 0.11 	&	& 2.9 \\ \hline
\onlinecite{Turchinovich2003}	& 0.20	& 31.0	& 17.5 \\ \hline
\onlinecite{Gfrorer1999}		& 0.11 & 3.0 & 1.6\\ \hline
\onlinecite{Berkowicz1999}		& 0.10 & 9.0 & 4.8 \\ \hline
 \onlinecite{Park2012} & 0.14, 0.14, 0.14 & 18.1, 21.2, 20.4 & 9.8, 11.5, 11.1 \\ \hline
 \onlinecite{Takeuchi1998} & 0.16 & 12.0 & 6.6 \\ \hline
\onlinecite{Lai2002} 	& 0.23 	&18.0	&10.3 \\ \hline
\onlinecite{Lai2002} (holography)	& 0.18 	&22	&12.2 \\ \hline
\onlinecite{Cherns1999} (holography) &0.52 & 40.0 & 26.8 \\ \hline
\onlinecite{Barnard2000} (holography) & 0.52 & 32.0 &21.4 \\ \hline
\onlinecite{Jia2001} (CV) & 0.08 & &4.1\\ \hline
\onlinecite{Zhang2004} (CV) & 0.05, 0.09 & &1.8, 4.4\\
\end{tabular}
\end{ruledtabular}
\end{table}

\begin{table}[h]
\caption{\label{AlGaNHall} Experimental data for GaN/AlGaN interfaces from Hall-effect measurements. }
\begin{ruledtabular}
 \begin{tabular}{ >{\centering\arraybackslash}m{5cm} | >{\centering\arraybackslash}m{5cm} |>{\centering\arraybackslash}m{5cm} }
 Reference 	& AlGaN content & Bound charge ($10^{12} e^-/\text{cm}^2$)		\\
\hline
 \onlinecite{Zhang2000}	& 0.09, 0.13, 0.17, 0.26, 0.31, 0.13, 0.18, 0.22, 0.22, 0.26, 0.29, 0.29, 0.31 & 3.9, 4.9, 8.7, 13.8, 15.0, 6.8, 7.2, 8.2, 11.0, 10.2, 10.1, 13.4, 13.5	\\ \hline
 \onlinecite{Davidsson2004} & 0.20, 0.20, 0.30, 0.35, 0.40, 0.40, 0.40, 0.40, 0.40, 0.37 & 9.8, 15.1, 19.1, 23.6, 28.5, 25.4, 24.1, 22.0, 19.7, 20.1					\\ \hline
 \onlinecite{Martnez-Criado2001} & 0.33, 0.34, 0.38	& 13.2, 9.0, 10.5 					\\ \hline
 \onlinecite{Binari1997} & 0.15		 &6.0					\\ \hline
 \onlinecite{Wu1996} & 0.15			 &7.9					\\ \hline
 \onlinecite{Nguyen1998} & 0.3	0			&16.0					\\ \hline
 \onlinecite{Wong1998}	& 0.05, 0.15		 &2.3, 6.7				\\ \hline
 \onlinecite{Kohler2011}	& 0.02, 0.06, 0.09, 0.14, 0.02, 0.05, 0.13, 0.15, 0.19, 0.24, 0.29	& 1.1, 2.1, 3.0, 3.3. 1.2, 1.6, 4.4, 4.6, 5.6, 6.5, 8.0					\\	\hline
 \onlinecite{Kohler2009}	& 0.12, 0.14, 0.14, 0.17, 0.17, 0.20, 0.23, 0.24, 0.26, 0.30, 0.31, 0.34, 0.36, 0.37&	 3.5, 4.6, 4.3, 5.6, 5.6, 6.5, 7.9, 8.8, 9.1, 10.7, 11.6, 12.3, 12.6, 14.0					\\ \hline
 \onlinecite{Nakajima2010}	& 0.23			& 11.0				\\ \hline
 \onlinecite{Burm1996}	& 0.16			& 7.3					\\ \hline
 \onlinecite{Wang1999}	& 0.10, 0.13, 0.18	& 2.8, 4.1, 6.2			\\ \hline
 \onlinecite{Liu2006} & 0.10, 0.15, 0.20				&4.1, 6.3, 8.7 					\\ \hline
 \onlinecite{Ding2010}	& 0.13, 0.23, 0.26, 0.36 & 7.3, 9.5, 11.0, 15.2	\\ \hline
 \onlinecite{Chen2005}	& 0.22, 0.26, 0.32, 0.36 &	7.3, 9.0, 11.3, 12.0	\\ \hline
 \onlinecite{Jeganathan2003}	& 1.0				& 34					\\ \hline
 \onlinecite{Gaska1998}	& 0.20			& 13					\\\hline
 \onlinecite{Asbeck1997}		& 0.05, 0.15, 0.15, 0.15, 0.25, 0.35 &	 3.0, 8.0, 7.5, 5.9, 9.9, 18.0		\\ \hline
 \onlinecite{Wang2006}	& 0.23			& 9.8					\\ \hline
 \onlinecite{Winzer2003}	 &  0.06, 0.10, 0.26, 0.33	& 1.0, 1.9, 1.4, 1.8					\\ \hline
 \onlinecite{Li2010}	 & 0.72			& 35					\\ \hline
 \onlinecite{Arulkumaran2003}	 & 0.21, 0.21, 0.27, 0.33, 0.33, 0.40, 0.40, 0.49, 0.48 & 9.2, 10.1, 11.0, 11.5, 11.2, 12.9, 13.3, 16.9, 16.4 \\ \hline
 \onlinecite{Ambacher2000}	& 0.15, 0.19, 0.18, 0.19, 0.20, 0.20, 0.23, 0.22, 0.26, 0.27, 0.26, 0.31, 0.31, 0.32, 0.32, 0.34, 0.35, 0.37, 0.37, 0.36, 0.43, 0.44, 0.46 & 0.7, 1.1, 1.4, 2.8, 3.0, 6.9, 6.4, 4.8, 11.2, 11.0, 8.7, 12.2, 13.0, 14.1, 9.8, 11.0, 9.6, 10.1, 11.4, 14.7, 13.6, 14.0, 14.3 \\  \hline
 \onlinecite{Franssen2006}	& 0.26			& 25 \\	
\end{tabular}
\end{ruledtabular}
\end{table}

\begin{table}[h]
\caption{\label{AlGaNoptical} Experimental data for AlGaN/GaN/AlGaN quantum wells from optical and holography measurements. In the cases where fields are reported in the reference, the bound charge is determined from the model described in the main text, Sec.~V.}
\begin{ruledtabular}
 \begin{tabular}{ >{\centering\arraybackslash}m{2.5cm} | >{\centering\arraybackslash}m{5cm} | >{\centering\arraybackslash}m{5cm} |>{\centering\arraybackslash}m{5cm} }
 Reference 	& AlGaN content & Field ($10^5$ V/cm) & Bound charge ($10^{12} e^-/\text{cm}^2$)		\\
\hline
\onlinecite{Grandjean1999}	&0.08, 0.08, 0.13, 0.17, 0.13, 0.17, 0.27, 0.27 && 3.8, 3.0, 5.6, 7.2, 7.4, 9.4, 14.2, 10.9 	\\ \hline
 \onlinecite{Cingolani2000} & 0.15			&& 2.0 \\ \hline
 \onlinecite{Chen2013}	&	0.50			&42.7 & 21.2	\\	 \hline	
 \onlinecite{McAleese2006}	& 0.20, 0.20	& 12.8, 8.3 &6.4, 4.1	\\ \hline
 \onlinecite{McAleese2006} (Hologaphy) & 0.20, 0.20, 0.20 & 12.8, 8.4, 6.9 & 6.4, 4.2, 3.4 \\ \hline
 \onlinecite{Langer1999} 	& 0.24		& 15.0 & 7.5	\\	 \hline
 \onlinecite{Leroux1998}		& 0.17		&8.3 & 4.1	\\ \hline
 \onlinecite{Suzuki1999} 		& 0.65 		& 20 & 9.94	\\ \hline
 \onlinecite{Esmaeili2009}, \onlinecite{Esmaeili2007}	& 0.07 & 4.8  & 2.4	\\ \hline
 \onlinecite{Pinos2008}		& 0.14		& 5.1 & 2.5	\\ \hline
 \onlinecite{Marcinkevicius2007}& 0.18, 0.11, 0.15 &10.2, 9.3, 3.8 & 5.1, 4.6, 1.9\\ \hline
 \onlinecite{Kuokstis2002}	& 0.18			&12.3 & 6.1	\\ \hline
 \onlinecite{Ng2001}		& 0.20, 0.50, 0.65, 0.80 & 11.9, 29.5, 33.9, 49.2 &5.9, 14.6, 16.8, 24.5	\\ \hline
 \onlinecite{Im1998}		& 0.15			&3.5 & 1.74	\\ \hline
 \onlinecite{Leroux1998}		& 0.11			&4.5 & 2.24	\\ \hline
 \onlinecite{Simon2000}		&0.24, 0.18, 0.18, 0.15, 0.07, 0.18, 0.17, 0.16, 0.16 &13.0, 13.0, 13.2, 9.0, 4.1, 13.3, 10.0, 10.0, 10.2 &  \\ \hline
 \onlinecite{Simon2001}		& 0.07, 0.15, 0.17, 0.18, 0.24, 0.18, 0.16, 0.16, 0.17 & 5.3, 11.7, 17.2, 16.9, 19.5, 13.8, 10.6, 10.2, 10.1& 2.6, 5.8, 8.5, 8.4, 9.7, 6.9, 5.3, 5.1, 5.0	\\ \hline
 \onlinecite{Gfrorer2011}		& 0.15			&14 & 7.0	\\ \hline
 \onlinecite{Wang2006}		& 0.23			& &10.2	\\ \hline
 \onlinecite{Winzer2003}		& 0.06, 0.10, 0.26, 0.33 & 3.4, 3.1, 3.8, 3.4 &1.7, 1.5, 1.9, 1.7	\\ \hline
 \onlinecite{Winzer2005}		& 0.31			& &11.0	\\ \hline
 \onlinecite{Drabinska2002}	&				&&	\\ \hline
 \onlinecite{Kurtz2004}		& 0.19			& 2.5 &7.4	\\
     \end{tabular}
\end{ruledtabular}
\end{table}

\begin{table}[h]\centering
\caption{\label{AlGaNCV} Experimental data for GaN/AlGaN interfaces from CV measurements.  }
\begin{ruledtabular}
  \begin{tabular}{ >{\centering\arraybackslash}m{5cm} | >{\centering\arraybackslash}m{5cm} |>{\centering\arraybackslash}m{5cm} }
 Reference 	& AlGaN content & Bound charge ($10^{12} e^-/\text{cm}^2$)		\\
 \hline
 \onlinecite{Zhou2002}		& 0.22 &1.3 \\ \hline
\onlinecite{Yu1997}			& 0.15 & 3.8 \\ \hline
 \onlinecite{Ambacher2000}	& 0.33 &10, 12 \\ \hline
 \onlinecite{Smorchkova1999} &0.09, 0.13, 0.17, 0.26, 0.31 & 3.8, 4.9, 8.7, 13.6, 15.0 \\ \hline
 \onlinecite{Miller2002}		& 0.05, 0.12, 0.16 &  2.3, 6.8, 6.9\\ \hline
 \onlinecite{Jia2001}		& 0.13 & 7.1 \\ \hline
 \onlinecite{Kohler2011}		& 0.02, 0.06, 0.09, 0.14  & 0.9, 1.8, 2.0, 3.1  \\ \hline
 \onlinecite{Franssen2006}	& 0.26 & 7.0 \\
     \end{tabular}
\end{ruledtabular}
\end{table}

\clearpage

\section{Bound charges at nitride interfaces}

Here we present the specific equations used to generate Fig.~2 in the main text. As in the main text, we assume a coherent $c$ plane interface of GaN and the alloy (InGaN or AlGaN), with the alloy layer under biaxial stress. The current practice in the field (black dashed curve in Fig.~2 of the main text) is to use the effective spontaneous (SP) polarization constants with respect to the zincblende (ZB) reference, without the correction term [$\Delta P_\text{corr}^\text{ref}$ introduced in Eq.~(11) of the main text], and the proper piezoelectric (PZ) constants. (These values are usually taken from Ref.~\onlinecite{BFV1997}.) The resulting equation for Al$_x$Ga$_{1-x}$N/GaN is
\begin{equation}
\label{curr}
\begin{split}
\sigma_\text{b}^\text{(ZB ref),prop}(x)&=\Delta\widetilde{P}_\text{SP}^\text{int,(ZB ref)}x- 2\frac{(a_\text{AlN}-a_\text{GaN})x}{a_\text{AlN}x+a_\text{GaN}(1-x)}\Bigg\{e^\text{AlN,prop}_{31}x+e^\text{GaN,prop}_{31}(1-x)\\&-\big[e_{33}^\text{AlN,prop}x+e_{33}^\text{GaN,prop}(1-x)\Big]\frac{C_{13}^\text{AlN}x+C_{13}^\text{GaN}(1-x)}{C_{33}^\text{AlN}x+C_{33}^\text{GaN}(1-x)}\Bigg\},
\end{split}
\end{equation}
where $\Delta\widetilde{P}_\text{SP}^\text{int,(ZB ref)}$ is
\begin{equation}
\label{AlGaN}
\begin{split}
\Delta\widetilde{P}_\text{SP}^\text{int,(ZB ref)}=P_\text{eff}^{\text{GaN, (ZB ref)}}-P_\text{eff}^{\text{AlN, (ZB ref)}}.
\end{split}
\end{equation}
An identical set of equations are used for In$_x$Ga$_{1-x}$N/GaN, with InN instead of AlN .


The red solid curve in Fig.~2 corresponds to using the H ref (or ZB with the correction term) and the improper PZ constants:
\begin{equation}
\label{rev}
\begin{split}
\sigma_\text{b}^\text{(H ref), imp}(x)&=\Delta\widetilde{P}_\text{SP}^\text{int,(H ref)}x- 2\frac{(a_\text{AlN}-a_\text{GaN})x}{a_\text{AlN}x+a_\text{GaN}(1-x)}\Bigg\{\left(e^\text{AlN,prop}_{31}-P_\text{eff}^\text{AlN, (H ref)}\right)x\\&+\left(e^\text{GaN,prop}_{31}-P_\text{eff}^\text{GaN, (H ref)}\right)(1-x)-\big[e_{33}^\text{AlN,prop}x+e_{33}^\text{GaN,prop}(1-x)\Big]\frac{C_{13}^\text{AlN}x+C_{13}^\text{GaN}(1-x)}{C_{33}^\text{AlN}x+C_{33}^\text{GaN}(1-x)}\Bigg\} \\&=(\Delta\widetilde{P}_\text{SP}^\text{int,(ZB ref)}+\Delta P_\text{corr}^\text{(ZB ref)})x- 2\frac{(a_\text{AlN}-a_\text{GaN})x}{a_\text{AlN}x+a_\text{GaN}(1-x)}\Bigg\{\left(e^\text{AlN,prop}_{31}-P_\text{eff}^\text{AlN, (H ref)}\right)x\\&+\left(e^\text{GaN,prop}_{31}-P_\text{eff}^\text{GaN, (H ref)}\right)(1-x)-\big[e_{33}^\text{AlN,prop}x+e_{33}^\text{GaN,prop}(1-x)\Big]\frac{C_{13}^\text{AlN}x+C_{13}^\text{GaN}(1-x)}{C_{33}^\text{AlN}x+C_{33}^\text{GaN}(1-x)}\Bigg\},
\end{split}
\end{equation}
where
\begin{equation}
\label{corr}
\Delta P_\text{corr}^\text{(ZB ref)}= \frac{e\sqrt{3}}{2}\left(\frac{1}{(a_\text{GaN})^2}-\frac{1}{(a_\text{AlN})^2}\right),
\end{equation}
and similarly for InGaN/GaN.

\clearpage

\section{Difference between implementations}

The difference between the current practice in the field (ZB reference, no correction term, proper PZ)
and our revised implementation (H reference, improper PZ constants)
can be determined by taking the difference of Eq.~(\ref{curr}) and Eq.~(\ref{rev}). For the case of AlGaN/GaN:

\begin{equation}
\label{diffimp}
\begin{split}
\sigma_\text{b}^\text{(H ref), imp}(x)-\sigma_\text{b}^\text{(ZB ref),prop}(x)=&x\left[\Delta\widetilde{P}_\text{SP}^\text{int,(H ref)}-\Delta\widetilde{P}_\text{SP}^\text{int,(ZB ref)}\right]
\\
&-2\frac{(a_\text{AlN}-a_\text{GaN})x}{a_\text{AlN}x+a_\text{GaN}(1-x)}\left[-P_\text{eff}^\text{AlN, (H ref)}(x)-P_\text{eff}^\text{GaN, (H ref)}(1-x)\right]
\\
=&x\left[\left(\Delta\widetilde{P}_\text{SP}^\text{int,(ZB ref)}+\Delta P_\text{corr}^\text{(ZB ref)}\right)-\Delta\widetilde{P}_\text{SP}^\text{int,(ZB ref)}\right]
\\
&+2\frac{(a_\text{AlN}-a_\text{GaN})x}{a_\text{AlN}x+a_\text{GaN}(1-x)}\left[P_\text{eff}^\text{AlN, (H ref)}(x)+P_\text{eff}^\text{GaN, (H ref)}(1-x)\right]
\\
=&x\Delta P_\text{corr}^\text{(ZB ref)}+2\frac{(a_\text{AlN}-a_\text{GaN})x}{a_\text{AlN}x+a_\text{GaN}(1-x)}\left[P_\text{eff}^\text{AlN, (H ref)}(x)+P_\text{eff}^\text{GaN, (H ref)}(1-x)\right]
\\
=&x\Delta P_\text{corr}^\text{(ZB ref)}+2\varepsilon_1(x)P_\text{eff}^\text{AlGaN, (H ref)}(x).
\end{split}
\end{equation}
We can gain some more insight by linearizing the first term in Eq.~({\ref{diffimp}):
\begin{equation}
\label{lincorr}
\begin{split}
x\Delta P_\text{corr}^\text{(ZB ref)}&=x \frac{e\sqrt{3}}{2}\left(\frac{1}{(a_\text{GaN})^2}-\frac{1}{(a_\text{AlN})^2}\right)
\\
&= x\frac{e\sqrt{3}}{2}\frac{1}{(a_\text{GaN})^2}\left(1-\frac{(a_\text{GaN})^2}{(a_\text{AlN})^2}\right)
\\
&= xP_{\text{f}}^{\text{GaN, ZB}}\left(1-\frac{(a_\text{GaN})^2}{(a_\text{AlN})^2}\right)
\\
&\simeq 2 xP_{\text{f}}^{\text{GaN, ZB}}\left(1-\frac{a_\text{GaN}}{a_\text{AlN}}\right)
\\
&=-2 P_{\text{f}}^{\text{GaN, ZB}}\left(x\frac{a_\text{GaN}-a_\text{AlN}}{a_\text{AlN}}\right)
\\
&\simeq -2 P_{\text{f}}^{\text{GaN, ZB}}\varepsilon_1(x)
\end{split}
\end{equation}
So we see that the difference in implementations is
\begin{equation}
\label{diffimp2}
\begin{split}
\sigma_\text{b}^\text{(H ref), imp}(x)-\sigma_\text{b}^\text{(ZB ref),prop}(x)&\simeq 2 \varepsilon_1(x)\left[P_\text{eff}^\text{AlGaN, (H ref)}(x)- P_{\text{f}}^{\text{GaN, ZB}}\right]
\\
&=2 \varepsilon_1(x)\left[xP_\text{f}^{\text{AlN, WZ}}+(1-x)P_\text{f}^{\text{GaN, WZ}}- P_{\text{f}}^{\text{GaN, ZB}}\right]
\\
&=2 \varepsilon_1(x)\left[x\left(P_\text{f}^{\text{AlN, WZ}}-P_\text{f}^{\text{GaN, WZ}}\right)+\left(P_{\text{f}}^{\text{GaN, WZ}}-P_{\text{f}}^{\text{GaN, ZB}}\right)\right]
\\
&=2 \varepsilon_1(x)\left[x\left(P_\text{f}^{\text{AlN, WZ}}-P_\text{f}^{\text{GaN, WZ}}\right)+P_{\text{eff}}^{\text{GaN, (ZB ref)}}\right]
\end{split}
\end{equation}
Therefore, the difference is small for small strains, and/or when there is a large cancellation of the terms in the square brackets. We see from Table I of the main text that $P_\text{f}^{\text{AlN, WZ}}-P_\text{f}^{\text{GaN, WZ}}=0.039 $ C/m$^2\sim -P_{\text{eff}}^{\text{GaN, (ZB ref)}}$, hence the close agreement with between the black dashed and red solid curves in Fig.~2(b) of the main text (along with the relatively small magnitude of the strain). For the case of InGaN, $P_\text{f}^{\text{InN, WZ}}-P_\text{f}^{\text{GaN, WZ}}=-0.286$ C/m$^2$ which is the same sign as $P_{\text{eff}}^{\text{GaN, (ZB ref)}}$, hence the larger discrepancy between the black dashed and red solid curves in Fig.~2(a) of the main text (also combined with a larger strain between InN and GaN).

\bibliography{Supbib}